\documentclass[12pt]{article}
\input{LaTeX/preamble}
\usepackage{lineno}

\begin{document}

\title{Inference for bivariate extremes via a semi-parametric angular-radial model}
\author[1,2*]{Callum John Rowlandson Murphy-Barltrop}
\author[3]{Ed Mackay}
\author[4,5]{Philip Jonathan}
\affil[1]{Technische Universität Dresden, Institut Für Mathematische Stochastik, Helmholtzstraße 10, 01069 Dresden, Germany}
\affil[2]{Center for Scalable Data Analytics and Artificial Intelligence (ScaDS.AI), Dresden/Leipzig, Germany}
\affil[3]{Deparment of Engineering, University of Exeter, TR10 9FE, United Kingdom}
\affil[4]{Department of Mathematics and Statistics, Lancaster University LA1 4YF, United Kingdom}
\affil[5]{Shell Research Limited, London SE1 7NA, United Kingdom}
\affil[*]{Correspondence to: callum.murphy-barltrop@tu-dresden.de}
\date{\footnotesize \today}

\maketitle

\begin{abstract}
    The modelling of multivariate extreme events is important in a wide variety of applications, including flood risk analysis, metocean engineering and financial modelling. A wide variety of statistical techniques have been proposed in the literature; however, many such methods are limited in the forms of dependence they can capture, or make strong parametric assumptions about data structures. In this article, we introduce a novel inference framework for bivariate extremes based on a semi-parametric angular-radial model. This model overcomes the limitations of many existing approaches and provides a unified paradigm for assessing joint tail behaviour. Alongside inferential tools, we also introduce techniques for assessing uncertainty and goodness of fit. Our proposed technique is tested on simulated data sets alongside observed metocean time series', with results indicating generally good performance.  
\end{abstract}

\noindent%

\doublespacing

\section{Introduction} \label{sec:intro}

\subsection{Multivariate extreme value modelling} \label{subsec:mv_evt_modelling}

The modelling of multivariate extremes is an active area of research, with applications spanning many domains, including meteorology \citep{Chavez-Demoulin2005}, metocean engineering \citep{Jonathan2013, Vanem2022}, financial modelling \citep{Castro-Camilo2018} and flood risk assessment \citep{w13040474}. Typically, approaches in this research field are comprised of two steps: first, modelling the extremes of individual variables and transforming to common margins, followed by modelling of the dependence between the extremes of different variables. We refer to this dependence as the extremal dependence structure henceforth. 

This article discusses inference for multivariate extremes using an angular-radial model for the \textit{probability density function}, illustrated using examples in two dimensions. To place the proposed model in context, we first provide a brief synopsis of the existing literature for multivariate extremes.
%
Given a random vector $(X,Y) \in \mathbb{R}^2$ with marginal distributions functions $F_X$ and $F_Y$, the strength of dependence in the upper tail of $(X,Y)$ can be quantified in terms of the tail dependence coefficient, $\chi\in[0,1]$, defined as the limiting probability
\begin{align}
    \chi = \lim_{u\to1} \Pr(F_X(X)>u \mid F_Y(Y)>u),
\end{align}
when this limit exists \citep{Joe1997}. When $\chi>0$, the components of $(X,Y)$ are said to be \textit{asymptotically dependent} (AD) in the upper tail, and when $\chi=0$, they are said to be \textit{asymptotically independent} (AI). Much of the focus of recent work in multivariate extreme value theory has been related to developing a general framework for modelling joint extremes of $(X,Y)$ which is applicable to both AD and AI cases, and can be used to evaluate joint tail behaviour in the region where at least one variable is large.  

To discuss the approaches proposed to date and their associated limitations, it is helpful to categorise them in terms of whether they assume heavy- or light-tailed margins, and whether they consider the distribution or density function. Classical multivariate extreme value theory assumes heavy-tailed margins, and is based on the framework of multivariate regular variation \citep[MRV,][]{Resnick1987}. It addresses the case where $\chi>0$, and has been widely studied -- see \citet{Beirlant2004}, \citet{haan2006extreme} and  \citet{Resnick2007} for reviews. Under some regularity conditions, equivalent asymptotic descriptions of joint extremal behaviour can be obtained from either the density or distribution function \citep{haan1987}. 

In the MRV framework, any distribution with $\chi=0$ has the same asymptotic representation. To address this issue, \cite{Ledford1996, Ledford1997} proposed a method to characterise joint extremes for both AI and AD distributions in the region where both variables are large. Model forms for the Ledford-Tawn representation were proposed by \citet{Ramos2009}. The resulting framework also assumes heavy-tailed margins and is referred to as hidden regular variation \citep[HRV,][]{Resnick2002}. However, for AI distributions, a description of extremal behaviour in the region where both variables are large may not be the most useful, since extremes of both variables are unlikely to occur simultaneously. Moreover, for AI distributions with certain regularity conditions, the asymptotic representation in this framework is governed only by the properties of the distribution along the line $y=x$ \citep{Mackay2023}. To provide a more useful representation for AI distributions, applicable in the region where either variable is large, \citet{Wadsworth2013} introduced an asymptotic model for the joint survivor function on standard exponential margins. In contrast to the MRV framework, the resulting model provides a useful description of AI distributions, but all AD distributions have the same representation. 

More recently, there has been interest in modelling the limiting shapes of scaled sample clouds, or \textit{limit sets}. The study of limit sets has been around since the 1960s \citep{Fisher1969,Davis1988}, and recent works from \citet{Nolde2014} and \citet{Nolde2022} have shown that these sets are directly linked to several representations for multivariate extremes. For a given distribution, the limit set is obtained by evaluating the asymptotic behaviour of the joint density function on light tailed margins. Many recent approaches have focused on estimation of the limit set in order to approximate extremal dependence properties; see, for instance, \citet{Simpson2022, Wadsworth2024,Majumder2023} and \citet{Papastathopoulos2024}.   However, the limit set itself does not provide a full description of the asymptotic joint density or distribution, so is less useful from a practical modelling perspective. 

To understand the limitations of the methods discussed above, it is instructive to provide an illustration of the joint distribution and density functions on heavy- and light-tailed margins for AI and AD random vectors. All the methods discussed above have equivalent representations in angular-radial coordinates, so without loss of generality, we consider the angular-radial dependence. The first step for most methods for modelling multivariate extremes is to transform variables to common margins. Define 
\begin{align*}
    (X_P,Y_P) &= \left((1-F_X(X))^{-1},(1-F_Y(Y))^{-1}\right) \in [1,\infty)^2,\\
    (X_E,Y_E) &= \left(-\log(1-F_X(X)), -\log(1-F_Y(Y))\right)\in[0,\infty)^2,
\end{align*}
so that $(X_P,Y_P)$ and $(X_E,Y_E)$ have standard Pareto and exponential margins, respectively. Note that $(X_E,Y_E)=(\log(X_P),\log(Y_P))$, and that the dependence structure or copula of $(X,Y)$ remains unchanged by the marginal transformation \citep{Sklar1959}. Furthermore, the joint survivor function $\bar{F}_P(x,y)=\Pr(X_P>x,Y_P>y)$ is related to the joint survivor function of $(X_E,Y_E)$ by $\bar{F}_E(x,y) = \bar{F}_P (\exp(x), \exp(y))$. Moreover, if $(X_E,Y_E)$ has joint density function $f_E(x,y)$, then $(X_P,Y_P)$ has joint density $f_P(\exp(x),\exp(y)) = \exp(-r) f_E(x,y)$, where $r=x+y$.. 

\begin{figure}[ht]
    \centering
    \includegraphics[scale=0.5]{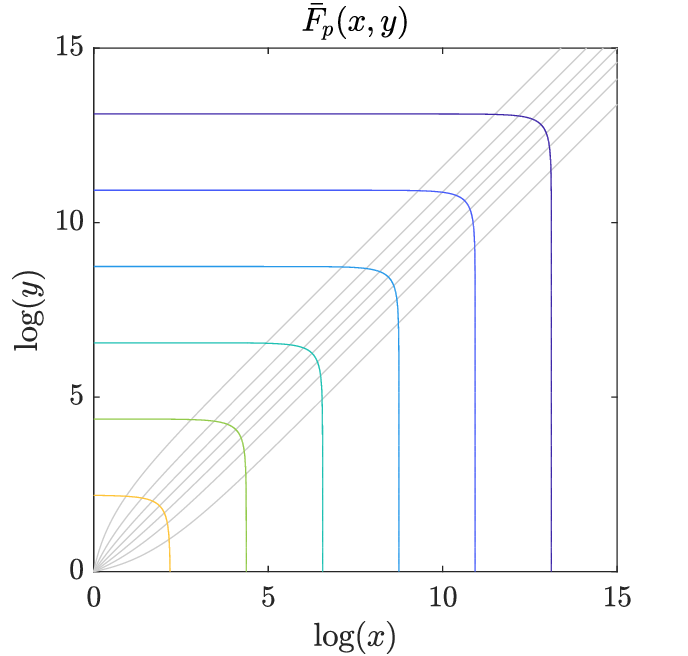}
    \includegraphics[scale=0.5]{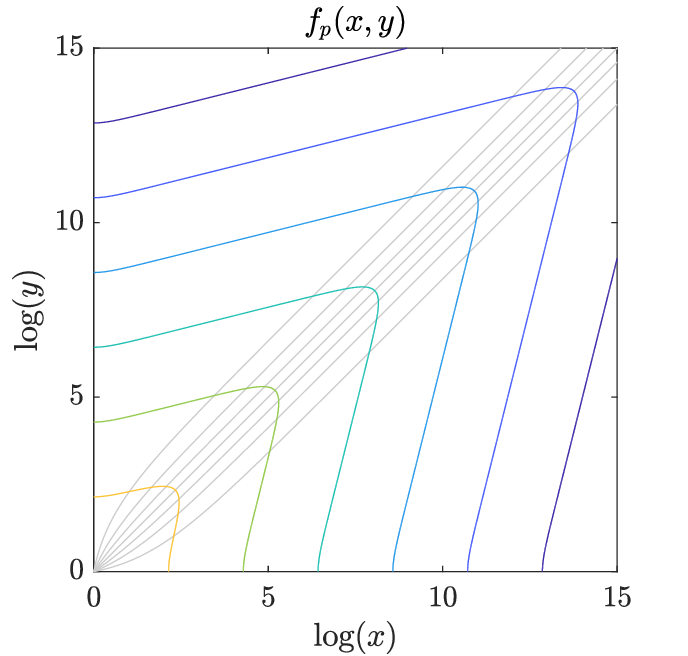}\\
    \vskip10pt
    \includegraphics[scale=0.5]{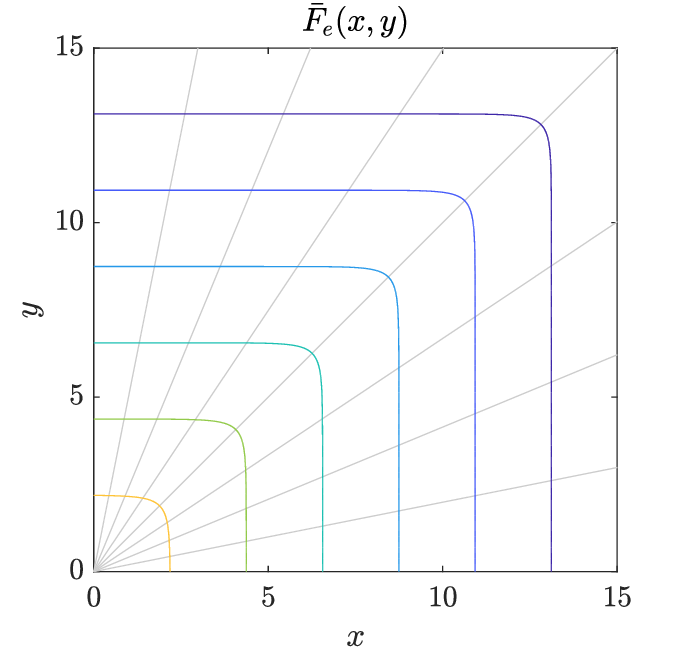}
    \includegraphics[scale=0.5]{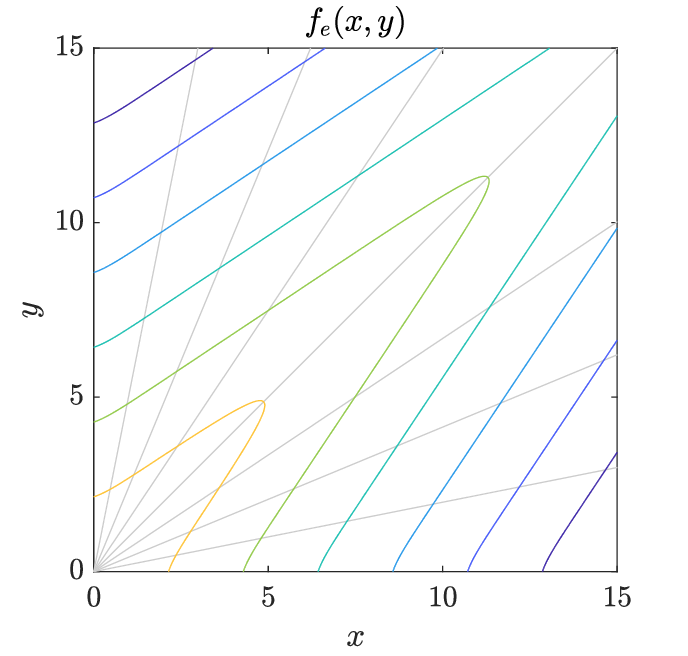}
    \caption{Representations of a Joe copula (with parameter $\alpha=3$) on standard Pareto (upper row) and standard exponential margins (lower row). Left plots: Contours of joint survivor function at equal logarithmic increments. Right plots: Contours of joint density function at equal logarithmic increments. Light grey lines show rays of constant angle on each margin.}
    \label{fig:Margins_onesided_Joe}
\end{figure}

Figure \ref{fig:Margins_onesided_Joe} shows the joint survivor and density functions for the AD Joe copula, as defined in the Supplementary Material, on standard Pareto and exponential margins. Rays of constant angle on each margin are also shown. On Pareto margins, with the axes shown on a logarithmic scale, lines of constant angle asymptote to lines with unit gradient. As such, the MRV framework provides a description of joint tail behaviour in the region close to the line $\log(y)=\log(x)$, i.e., where $X_P$ and $Y_P$ are of similar magnitudes. In this region, the contours of the joint density and survivor functions asymptote to a curve of constant shape, which describes the joint extremal behaviour in this region. In contrast, the angular-radial description appears different on exponential margins. For the joint survivor function, the contours of constant probability appear to asymptote towards the line $\max(x,y) = c$ for some constant $c$. \citet{Wadsworth2013} showed that is the case for all AD distributions. Informally, this is because for a distribution to be AD, the probability mass must be concentrated close to the line $y=x$, so when the density is integrated to obtain the survivor function, the dominant contribution comes from this region. In contrast to the joint survivor function, the angular-radial description of the joint density is not the same for all AD distributions on exponential margins. 

\begin{figure}[ht]
    \centering
    \includegraphics[scale=0.5]{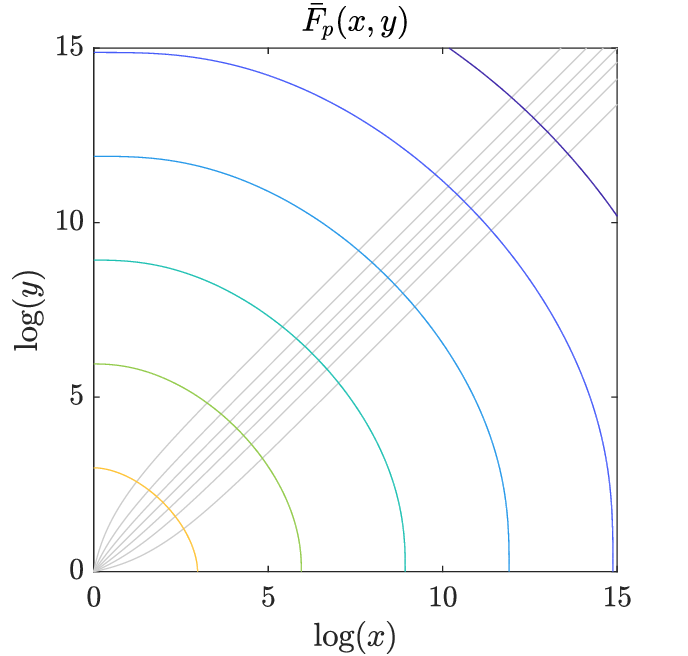}
    \includegraphics[scale=0.5]{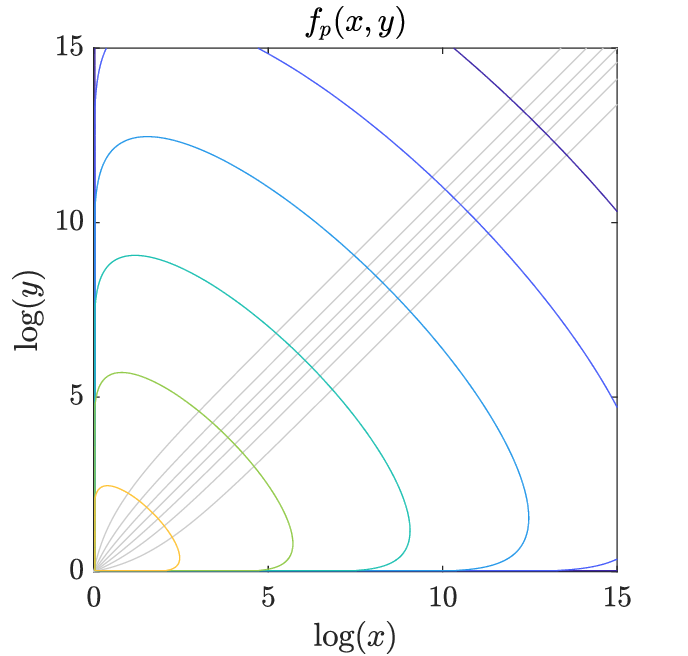}\\
    \vskip10pt
    \includegraphics[scale=0.5]{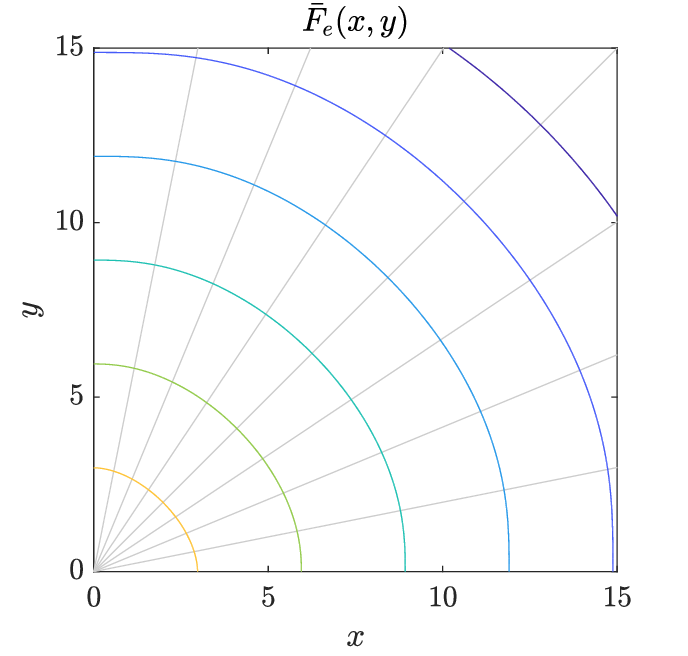}
    \includegraphics[scale=0.5]{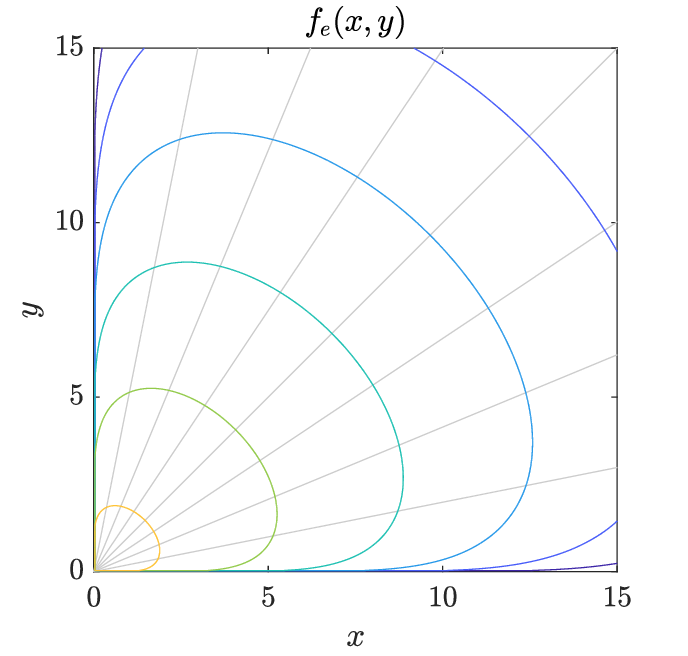}
    \caption{As previous figure, but for Gaussian copula with $\rho=0.5$.}    \label{fig:Margins_onesided_gaussian}
\end{figure}

Figure \ref{fig:Margins_onesided_gaussian} shows a similar set of plots for the AI Gaussian copula, also defined in the Supplementary Material. In this case, contours for both the joint density and joint survivor function are curved on both sets of margins. The angular-radial model on Pareto margins describes the section of the curves close to the line $y=x$, which asymptote to straight lines as $x+y\to\infty$. Therefore, the HRV description of the asymptotic behaviour is effectively a straight line approximation to a curve, and is only applicable in the region close to the line $y=x$; see \citet{Mackay2023} for details. In contrast, the angular-radial description of both the density and survivor functions on exponential margins provides a more useful description of asymptotic behaviour. That is, the representation on exponential margins is valid for the full angular range, whereas the representation on Pareto margins is only valid in the joint exceedance region where we are unlikely to observe the largest values of either variable for variables which are AI.

\begin{figure}[ht]
    \centering
    \includegraphics[scale=0.5]{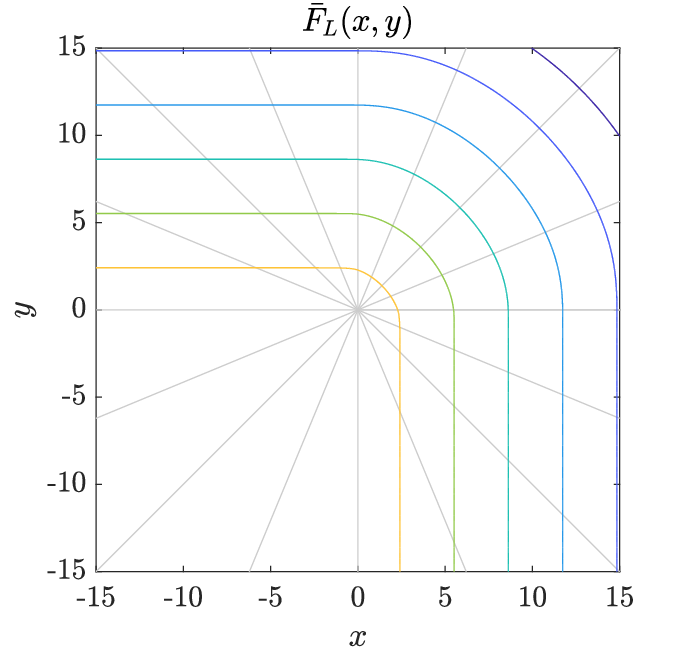}
    \includegraphics[scale=0.5]{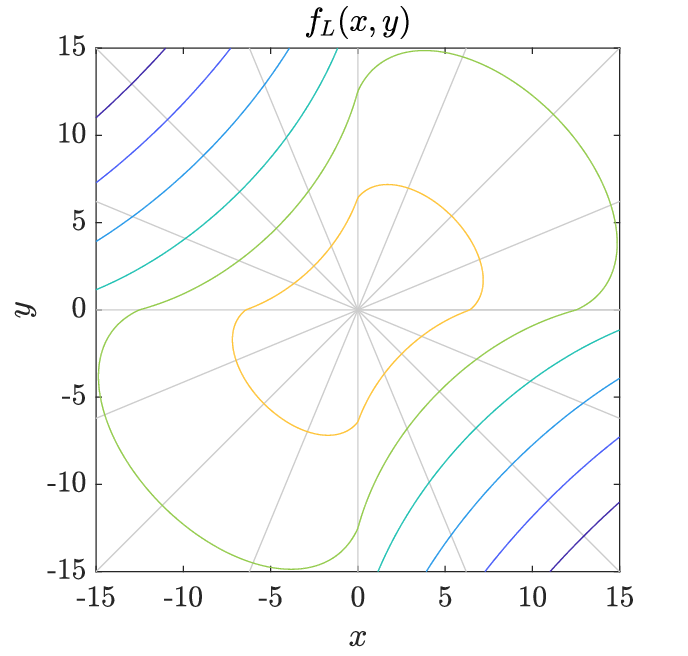}
    \caption{Gaussian copula on standard Laplace margins. Left: Contours of joint survivor function at equal logarithmic increments. Right: Contours of joint density function at equal logarithmic increments. Light grey lines show rays of constant angle.}
    \label{fig:Margins_twosided_gaussian}
\end{figure}

In some applications it is useful to describe the extremal behaviour of a random vector for both large and small values of certain variables; see Section \ref{subsec:motiv_examples}. In this case, it is more useful to work on symmetric two-sided margins, rather than one-sided margins. Figure \ref{fig:Margins_twosided_gaussian} shows the joint survivor and density functions for a Gaussian copula on standard Laplace margins. The angular-radial variation of the joint survivor function is useful in the first quadrant of the plane, but is less useful in the other quadrants. In the second and fourth quadrants, the contours of the joint survivor function asymptote to the corresponding marginal levels, providing no information about the asymptotic behaviour of the distribution in this region. In contrast, the joint density function provides useful asymptotic information in all regions of the plane. 

This motivates an intuitively-appealing angular-radial description of the joint density function, referred to as the semi-parametric angular-radial (SPAR) model \citep{mackay2022imex}, which we consider in detail in this article. 
%
%
A similar model was recently proposed by \citet{Papastathopoulos2024}, although the application was only considered for standard Laplace margins. However, the SPAR framework can be applied on any type of margin. \citet{Mackay2023} showed that on heavy-tailed margins, SPAR is consistent with the MRV/HRV frameworks, and on light-tailed margins, SPAR is consistent with limit set theory. However, the SPAR framework is more general than limit set theory, as it provides an explicit model for the density in extreme regions of the variable space. Moreover, there are distributions which have degenerate limit sets in some regions, for which there is still a useful SPAR representation. 

In the SPAR framework, variables are transformed to angular-radial coordinates, and it is assumed that the conditional radial distribution is in the domain of attraction of an extreme value distribution. This implies the radial tail conditioned on angle can be approximated by a non-stationary generalised Pareto (GP) distribution. The SPAR approach generalises the model proposed by \citet{Wadsworth2017}, in which angular and radial components are assumed to be independent. In the \citet{Wadsworth2017} model, the margins and angular-radial coordinate system are selected so that the assumption of independent angular and radial components is satisfied. The SPAR framework removes this requirement, providing a more flexible representation for multivariate extremes. 

While a strong theoretical foundation for the SPAR model is provided in \citet{Mackay2023}, inference for this model has not yet been demonstrated. Inference via this framework would offer advantages over many existing approaches, and a fitted SPAR model could be used to estimate extreme quantities commonly applied in practice, such as risk measures \citep{Murphy-Barltrop2023} and joint tail probabilities \citep{Keef2013a}.

The SPAR model reframes multivariate extreme value modelling as non-stationary peaks over threshold (POT) modelling with angular dependence. Many approaches have been proposed for non-stationary POT inference e.g. \citet{Randell2016, Youngman2019, Zanini2020}. In this paper, we introduce an `off-the-shelf' inference framework for the SPAR model. This framework, which utilises generalised additive models \citep[GAMs;][]{Wood2017} for capturing the relationship between radial and angular components, offers a high degree of flexibility and can capture a wide variety of extremal dependence structures, as demonstrated in Sections \ref{sec:sim_study} and \ref{sec:case_study}. Our approach offers utility across a wide range of applications and provides a convenient, practical framework for performing inference on multivariate extremes. Moreover, our inference framework is ready to use by practitioners; open-source software for fitting the SPAR model is available at \url{https://github.com/callumbarltrop/SPAR}. For ease of discussion and illustration, we restrict attention to the bivariate setting throughout, noting that the SPAR model is not limited to this setting.  



\subsection{Motivating examples} \label{subsec:motiv_examples}

\begin{figure}[ht]
    \centering
    \includegraphics[width=\textwidth]{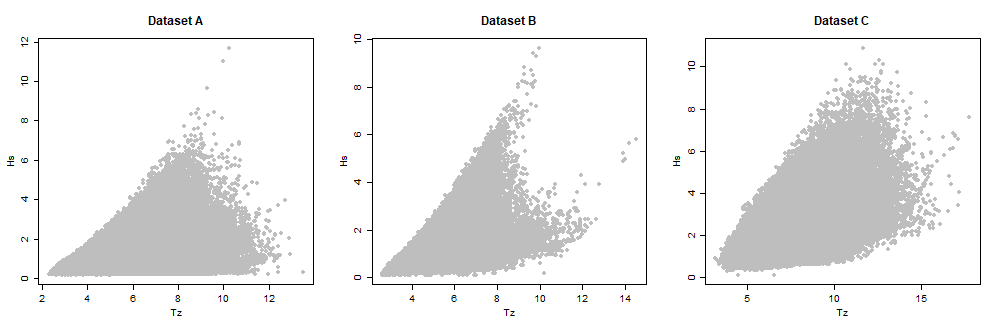}
    \caption{Metocean data sets A (left), B (centre) and C (right) comprised of hourly $T_z$ and $H_s$ observations.}
    \label{fig:metocean_dataset}
\end{figure}

To demonstrate the practical applicability of our proposed inference framework, we consider three bivariate metocean time series made up of zero-up-crossing period, $T_z$, and significant wave height, $H_s$, observations. We label these data sets as A, B and C, with each data set corresponding to a location off the coast of North America. data sets A and B were previously considered in a benchmarking exercise for environmental contours \citep{Haselsteiner2021}. Observations were recorded on an hourly basis over 40, 31 and 42 year time periods for data sets A, B and C, resulting in $n_A = 320740$, $n_B=241815$ and $n_C = 328247$ observations, respectively, once missing observations are taken into account. Exploratory analysis indicates the joint time series are approximately stationary over the observation period. Understanding the joint extremes of metocean variables is important in the field of ocean engineering for assessing the reliability of offshore structures. Wave loading on structures is dependent on both wave height and period, and the largest loads on a structure may not necessarily occur with the largest wave heights. Resonances in a structure may result in the largest responses occurring with either short- or long-period waves, meaning it is necessary to characterise the joint distribution in both of these ranges. These data sets are illustrated in Figure \ref{fig:metocean_dataset}. 

Metocean data sets of this type can often exhibit complex dependence structures, for which many multivariate models fail to account. For example, data set B exhibits clear asymmetry in its dependence structure. Moreover, as demonstrated in \citet{Haselsteiner2021}, many existing approaches for modelling metocean data sets perform poorly in practice, often misrepresenting the joint tail behaviour or not offering sufficient flexibility to capture the complex data structures. These shortcomings can have drastic consequences if fitted models are used to inform the design bases for offshore structures, as is common in practice.



This paper is structured as follows. In Section \ref{sec:SPAR_formulation}, we briefly introduce the SPAR model and outline our assumptions. In Section \ref{sec:angular_dens}, we introduce a technique to estimate the density of the angular component. In Section \ref{sec:cond_radial_dens}, we introduce a framework for estimating the density of the radial component, conditioned on a fixed angle. In Section \ref{sec:practical_tools}, we introduce tools for quantifying uncertainty and assessing goodness of fit when applying the SPAR model in practice. In Section \ref{sec:sim_study} and \ref{sec:case_study}, we apply the proposed framework to simulated and real data sets, respectively, illustrating the proposed framework can accurately capture a wide range of extremal dependence structures for both prescribed and unknown marginal distriutions. We conclude in Section \ref{sec:discussion} with a discussion and outlook on future work.  


\section{The SPAR Model} \label{sec:SPAR_formulation}

\subsection{Coordinate systems}

Let $(X,Y)$ denote a random vector in $\RR^2$ with continuous joint density function $f_{X,Y}$ and simply connected support containing the point $(0,0)$. The SPAR model for $f_{X,Y}$ requires a transformation from Cartesian to polar coordinates. Polar coordinates can be defined in various ways; see \citet{Mackay2023} for discussion. In this paper, we restrict attention to two particular angular-radial systems corresponding to the $L1$ and $L2$ norms, defined as $\norm{(x,y)}{p} := (|x|^p + |y|^p)^{1/p},$ $p=1,2$, for $(x,y) \in \RR^2$. We define $R_p := \norm{(X,Y)}{p}$, $p = 1,2$, and consider these variables as radial components of $(X,Y)$. Such definitions of radial variables are common in multivariate extreme value models \citep[e.g.,][]{DeHaan1998, Wadsworth2017}. When using the $L2$ norm to define the radial variable, the corresponding angular variable is usually defined as $\Theta = \text{atan2}(X,Y)$, where $\text{atan2}$ is the four-quadrant inverse tan function. The map between $(X,Y)$ and $(R_2,\Theta)$ is bijective on $\RR^2\setminus\{(0,0)\}$. When using the $L1$ norm to define radii, the angular variable is typically defined as $W:=X/\norm{(X,Y)}{1}$ \citep[e.g.,][chapter 5]{Resnick1987}. The random vector $(R_1,W)$ has a one-to-one correspondence with $(X,Y)$ in the upper half of the plane ($Y\ge 0$), but the use of the vector $(R_1,W)$ becomes ambiguous if we are interested in the full plane, since $W$ contains no information about the sign of $Y$.

With this in mind, we follow \citet{Mackay2023} and define the bijective angular functions $\mathcal{A}_p: \mathcal{U}_p \to (-2,2]$, where $\mathcal{U}_p:= \{ (u,v) \in \RR^2 \mid \norm{(u,v)}{p} = 1\}$ is the unit circle for the $L_p$ norm. For $p=1,2$, these are defined as 
\begin{align*}
    \mathcal{A}_1(u,v) &:= \varepsilon(v) (1-u)\\
    \mathcal{A}_2(u,v) &:= \frac{2}{\pi} \text{atan2}(u,v),
\end{align*}
where $\varepsilon(v) = 1$ for $v\geq 0$ and $-1$ otherwise, is the generalised signum function. The functions $\mathcal{A}_p(u,v)$ give a scaled measure of the distance along the unit circle $\mathcal{U}_p$ from the point $(1,0)$ to $(u,v)$, measured counter-clockwise.  

With angular functions established, we define the angular variables of $(X,Y)$ to be $Q_p := \mathcal{A}_p(X/R_p,Y/R_p)$, $p=1,2$. The corresponding radial-angular mapping $t:\RR^2 \setminus \{ (0,0)  \} \to (0,\infty) \times (-2,2]$ given by 
\begin{equation*}
    t(x,y):= \left(\norm{(x,y)}{p},\mathcal{A}_p\left(\frac{x}{\norm{(x,y)}{p}},\frac{y}{\norm{(x,y)}{p}}\right)\right), 
\end{equation*}
is bijective for $p = 1,2$. Consequently, we can recover $(X,Y)$ from its radial and angular components, i.e., $(X,Y) = R_p \mathcal{A}^{-1}_p(Q_p)$ for $p = 1,2$. We note that $Q_2=2\Theta/\pi$. However, we use the variable $Q_2$ here, in preference to $\Theta$, so that the angular range is the same for both $Q_1$ and $Q_2$. The joint density of $(R_p,Q_p)$ can be written in terms of the joint density of $(X,Y)$, 
\begin{equation*}
    f_{R_1,Q_1}(r_1,q_1) = r_1 f_{X,Y}(r_1 \mathcal{A}^{-1}_1(q_1)), \hspace{1em}f_{R_2,Q_2}(r_2,q_2) = \frac{\pi r_2}{2} f_{X,Y}(r_2 \mathcal{A}^{-1}_2(q_2)),
\end{equation*}
where the terms $r_1$ and $(\pi r_2)/2$ are the Jacobians of the respective transformations. For ease of notation, we henceforth drop the subscripts on the radial and angular components and simply let $(R,Q)$ denote one of the coordinate systems, with corresponding joint density function $f_{R,Q}$. 

\citet{Mackay2023} showed that the choice of coordinate system does not affect whether the SPAR model assumptions (discussed below) are satisfied. However, the coordinates may affect the inference, so in the examples presented in Sections \ref{sec:sim_study} and \ref{sec:case_study}, we consider both $L1$ and $L2$ polar coordinates.

\subsection{Conditional radial tail assumption} \label{subsec:gpd_tail_assumption}
Applying Bayes theorem, the joint density $f_{R,Q}$ can be written in the conditional form $f_{R,Q}(r,q) = f_{Q}(q)f_{R_{q}}(r\mid q),$ where $f_{Q}(q)$ denotes the marginal density of $Q$, $R_{q}:= R\mid~(Q~=~q),$ $q \in (-2,2]$ and $f_{R_{q}}(r\mid q)$ denotes the density of $R_{q}$, with corresponding distribution function $F_{R_q}(r\mid q)$. Viewed in this way, the modelling of joint extremes is reduced to the modelling of the angular density, $f_{Q}$, and the tail of the conditional density, $f_{R_{q}}$.


Given any $\gamma \in (0,1)$, define $u_{\gamma}:(-2,2] \to \RR_+$ as $u_{\gamma}(q) = \inf\{r \mid F_{R_q}(r \mid q) \geq \gamma \}$ for all $q \in (-2,2]$, implying $\Pr(R_{q} \leq u_{\gamma}(q)) = \gamma$. We refer to $u_{\gamma}(q), q \in (-2,2]$ as the threshold function henceforth. For the SPAR model, we assume that for all $q \in (-2,2]$, there exists a normalising function $c_{q}:\RR_+ \to \RR_+$ such that 
\begin{equation} \label{eqn:GPD_SPAR}
       \Pr \left( \frac{R_{q} - u_{\gamma}(q)}{c_{q}(u_{\gamma}(q))} \leq r \;  \Big\vert \; R_{q} > u_{\gamma}(q) \right) \to 1 - \left\{1 + \xi(q) r \right\}_+^{-1/\xi(q)}, \hspace{.5em} r > 0,
\end{equation}
as $\gamma \to 1^-$, with $\xi(q) \in \RR$. The right hand side of equation \eqref{eqn:GPD_SPAR} denotes the cumulative distribution function of a generalised Pareto (GP) distribution, and we term $\xi(q)$ the shape parameter function. The case $\xi(q) = 0$ can be interpreted as the limit of equation \eqref{eqn:GPD_SPAR} as $\xi(q) \to 0$.  Assumption \eqref{eqn:GPD_SPAR} is equivalent to the assumption that $R_q$ is in the domain of attraction of an extreme value distribution \citep{Balkema1974}. Given the wide range of univariate distributions satisfying this assumption, it is reasonable to expect the convergence of \eqref{eqn:GPD_SPAR} to hold in many cases for $R_{q}$ also. \citet{Mackay2023} showed that this assumption holds for a wide variety of theoretical examples. 

This convergence motivates a model for the upper tail of $R_q$. Assuming that equation \eqref{eqn:GPD_SPAR} approximately holds for some $\gamma < 1$ close to $1$, we have 
\begin{equation} \label{eqn:POT_SPAR}
       \Pr \left( R_{q} - u_{\gamma}(q) \leq r \;  \Big\vert \; R_{q} > u_{\gamma}(q) \right) \approx F_{GP}(r \mid \tau(q),\xi(q)) := 1 - \left\{1 + \frac{\xi(q) r}{\tau(q)} \right\}_+^{-1/\xi(q)}, \hspace{.5em} r > 0,
\end{equation}
for some $\tau(q) \in \RR_+$ which we refer to as the scale parameter function. The inclusion of the scale parameter removes the need to estimate the normalising function $c_q$, and this is equivalent to the standard peaks over threshold approximation used in univariate extreme value theory \citep{Davison1990}.

Given $q \in (-2,2]$ and $r \geq u_{\gamma}(q)$, assumption \eqref{eqn:POT_SPAR} implies that
\begin{align*}
    \bar{F}_{R_q}(r\mid q) &= \Pr(R_{q} > u_{\gamma}(q)) \left[ \Pr \left( R_{q} > r \;  \Big\vert \; R_{q} > u_{\gamma}(q) \right) \right],  \\
    &\approx (1-\gamma) \Bar{F}_{GP}(r-u_{\gamma}(q) \mid \tau(q),\xi(q)),
\end{align*}
where $\Bar{F}_{-}(\cdot) := 1 - F_{-}(\cdot)$ denotes the survivor function. The joint density of $(R,Q)$ in the region $\mathcal{U}_{\gamma} := \{ (r,q) \in (0,\infty) \times (-2,2] \mid r \geq u_{\gamma}(q)\}$ is then given by 
\begin{equation} \label{eqn:Joint_dens_GPD}
    f_{R,Q}(r,q) = f_Q(q)f_{R_{q}}(r \mid q) \approx (1-\gamma) f_Q(q) f_{GP}(r-u_{\gamma}(q) \mid \tau(q),\xi(q)),
\end{equation}
where $f_{GP}$ is the GP density function. Equation \eqref{eqn:Joint_dens_GPD} implies that the SPAR model is defined within the region $\mathcal{U}_{\gamma}$. 

To simplify the inference, we also assume that the functions $f_Q(q)$, 
$u_{\gamma}(q)$, $\tau(q)$ and $\xi(q)$ are finite and continuous over $q \in (-2,2]$ and satisfy the periodicity property $\lim_{q \to -2^+} f(q) = f(2)$. Such conditions are not guaranteed in general, and whether they are satisfied depends on the choice of margins, alongside the form of the dependence structure. \citet{Mackay2023} showed that the assumptions are valid for a wide range of copulas on Laplace margins, but using one-sided margins (e.g., exponential) or heavy-tailed margins can result in the assumptions not being satisfied for the same copulas. 

The SPAR model does not require specific marginal assumptions, and SPAR representations exist for variables with different marginal domains of attraction; however we consider these characteristics unlikely for phenomena in the Earth's environment. In applications, we typically assume either (i) a practical environmental setting, in which it is reasonable to assume that all variables are bounded (and then apply the model to standardised variables with zero mean and unit variance), or (ii) make marginal transformation to common scale. As discussed in \cite{Mackay2023}, there are theoretical reasons to prefer transformation to Laplace margins.

\section{Angular density estimation} \label{sec:angular_dens}
In this section, we consider the angular density $f_Q$ of equation \eqref{eqn:Joint_dens_GPD}, which we estimate using kernel density (KD) smoothing techniques. Such techniques offer many practical advantages: they are nonparametric, meaning no distributional assumptions for the underlying data are required, and they give smooth, continuous estimates of density functions. These features make KD techniques desirable for the estimation of $f_Q$. Note that other nonparametric smooth density estimation techniques are also available \citep[e.g.,][]{Gu1993,Randell2016}, but we do not consider these here. 

Unlike standard KD estimators \citep{Chen2017}, we require functional estimates that are periodic on the angular domain $(-2,2]$, motivating the use of circular density techniques \citep{Chaubey2022}. Given a sample $\{q_1,q_2,\dots,q_n\}$ from $Q$, the KD estimate of the density function is given by 
\begin{equation*}
    \hat{f}_{Q}(q; h) = \frac{1}{n} \sum_{i=1}^n K_h(q,q_i),
\end{equation*}
where $K_h$ denotes some circular kernel with bandwidth parameter $h$. The bandwidth controls the smoothness of the resulting density estimate, with the smoothness increasing as $h\to \infty$. The goal is typically to select $h$ as small as possible without overfitting. Within the literature, a wide range of circular kernels have been proposed; see \citet{Chaubey2022} for an overview. We restrict attention to one particular kernel since it is perhaps the most widely used in practice \citep{García–Portugués2013}. Specifically, we consider the von Mises kernel,  
\begin{equation} \label{eqn:von_mises_kernel}
    K_h(q,q_i) = \frac{1}{4 I_0(1/h)} \exp \left\{ \frac{1}{h} \cos \left((q-q_i)\frac{\pi}{2}\right)\right\},
\end{equation}
where $I_0$ is the modified Bessel function of order zero \citep{Taylor2008}. Here we have modified the kernel to have support on $(-2,2]$, rather than the usual support of $(-\pi,\pi]$. 

With a kernel selected, a critical issue when applying equation \eqref{eqn:von_mises_kernel} in practice is the choice of $h$. A variety of approaches have been proposed for automatically selecting the bandwidth parameter, including plug-in values \citep{Taylor2008}, cross-validation techniques \citep{Hall1987} and bootstrapping procedures \citep{DiMarzia2011}.

For our modelling approach, we opt not to use automatic selection techniques for the bandwidth parameter; instead, we select $h$ on a case-by-case basis, using the diagnostics proposed in Section \ref{subsec:diagnostics} to inform selection. In unreported results, we found many of the automatic selection methods to perform poorly in practice, and it has been shown that such techniques can fail for multi-modal densities \citep{Oliveira2012}. Multi-modality is often observed within the angular density \citep{Mackay2023}, suggesting it is better not to select $h$ using automatic techniques. 

\section{Conditional density estimation} \label{sec:cond_radial_dens}
We now consider the conditional density of equation \eqref{eqn:Joint_dens_GPD}. For simplicity, we assume that $\gamma \in (0,1)$ is fixed at some high level for each $q \in (-2,2]$. In practice, the choice of non-exceedance probability is non-trivial, and sensitivity analyses must be performed to ensure an appropriate value is selected; see Sections \ref{sec:practical_tools} and \ref{sec:case_study} for further details. Note that this is directly analogous to the threshold selection problem in univariate analyses; see \citet{Murphy2023} for a recent overview. 

To apply equation \eqref{eqn:POT_SPAR}, we require estimates of the threshold and GP parameter functions, denoted $u_{\gamma}(q)$, $\tau(q)$ and $\xi(q)$ respectively. As noted in Section \ref{subsec:mv_evt_modelling}, this is equivalent to performing a non-stationary peaks over threshold analysis on the conditional radial variable $R_q$, with $q$ viewed as a covariate. 

Throughout this article, we let $(\mathbf{r}, \mathbf{q}) := \{ (r_i,q_i) \mid i = 1, 2, \dots, n\}$ denote a sample of size $n \in \NN$ from $(R,Q)$. In this section we introduce two methods for inference. The first approach assumes the conditional radial distribution is locally stationary over a small angular range. In the second approach, spline-based modelling techniques are used to estimate the threshold and parameter functions as smoothly-varying functions of angle. 
The local stationary inference is used as a precursor to the spline-based inference, providing a useful comparison and `sense check' on results. 
%



\subsection{Local stationary inference} \label{subsec:local_inference}

We compute local stationary estimates at a fixed grid of values $\mathcal{Q}_{grid} := \{ -2 + 4i/M \mid i = 1, 2, \dots, M \} \subset (-2,2]$, where $M$ denotes some large positive integer, selected to ensure $\mathcal{Q}_{grid}$ has sufficient coverage on $(-2,2]$. For each $q \in \mathcal{Q}_{grid}$, we assume there exists a local neighbourhood $\mathcal{Q}_{q} = [q-\delta,q+\delta]$, $\delta>0$, such that the distribution of $R_{q^*}$ is stationary for $q^*\in\mathcal{Q}_q$. This is true in the limit as $\delta \to 0$, and a reasonable approximation for small $\delta$. 

In practice, rather than fixing the size of the interval, we select the $N$ nearest observations in terms of the angular distance from $q$, defined as $ d(q)_i := \min \{ \lvert q_i - q \rvert, 4 - \lvert q_i - q \rvert \}$, $i=1,...,n$, for some value $N\ll n$. Define $\mathcal{I}_q \subset \{1,2,\dots,n\}$ to be the index set of the $N$ smallest order statistics of $d(q)_i$. Local estimates of the threshold and parameter functions can be obtained from the corresponding radial set $\mathcal{R}_q := \{ r_i \mid i \in \mathcal{I}_q \}$. Specifically, we define $\hat{u}^l_{\gamma}(q)$ to be the $\gamma$ empirical quantile of $\mathcal{R}_q$, and $\hat{\sigma}^l(q)$ and $\hat{\xi}^l(q)$ to be maximum likelihood estimates of the GP distribution parameters obtained from the set $\{ r_i - \hat{u}^l_{\gamma}(q_i) \mid i \in \mathcal{I}_q, r_i > \hat{u}^l_{\gamma}(q_i) \}$. Choosing an appropriate value for $N$ involves a bias-variance trade-off; selecting too large (small) a value will increase the bias (variability) of the resulting pointwise threshold and parameter estimates. For our modelling procedure, this selection is not crucial, since local estimates are merely used as a means to inform the smooth estimation procedure presented in Section \ref{subsec:smooth_infer}. 

\subsection{Smooth inference for the SPAR model} \label{subsec:smooth_infer}
We now consider smooth estimation of the threshold and parameter functions. For this, we employ the approach of \citet{Youngman2019}, in which GAMs are used to capture covariate relationships; software for this approach is given in \citet{Youngman2020}. Our procedure is two-fold; we first estimate the threshold function $u_{\gamma}(q)$ for a given $\gamma$, then estimate the parameter functions $\tau(q)$ and $\xi(q)$ using the resulting threshold exceedances. 


This section is structured as follows. First, we provide a high-level overview of GAM-based modelling techniques. We then introduce procedures for estimating the threshold and parameter functions via the GAM framework. Finally, we discuss the selection of the basis dimensions required for the GAM formulations. 

\subsubsection{GAM-based procedures} \label{subsubsec:GAM_overview}
GAMs are a flexible class of regression models that allow for complex, non-linear relationships between response and predictor variables. They extend the traditional linear regression model by allowing the response to be modelled as a sum of smooth basis functions of the predictor variables. They are particularly useful when the relationship between the response and predictor variables is complex in nature and cannot be easily captured using standard parametric regression techniques.

Employing the GAM framework, the threshold and parameter functions can be represented through a sum of smooth basis functions, or splines. For an arbitrary function $g:(-2,2]\to\RR$, we write
\begin{equation} \label{eqn:GAM_form}
    g(q) = \beta_0 + \sum_{j=1}^k B_j(q) \beta_j,
\end{equation}
where $B_j$, $j \in \{1,2,\dots,k\}$ denote smooth basis functions, $\beta_j$, $j \in \{0,1,\dots,k\}$ denote coefficients and $k \in \NN$ denotes the basis dimension. To apply equation \eqref{eqn:GAM_form} in practice, one must first select a family of basis functions $B_j$, $j \in \{1,2,\dots,k\}$. A wide variety of bases have been proposed in the literature; see \citet{Perperoglou2019} for an overview. We restrict attention to one particular type of basis function known as a cubic spline. Cubic splines are widely used in practice to capture non-linear relationships, and exhibit many desirable properties, such as optimality in various respects, continuity and smoothness \citep{Wood2017}. Moreover, cubic splines can be modified to ensure periodicity by imposing conditions on the coefficients, resulting in a cyclic cubic spline. In the context of the SPAR framework, these properties are desirable to ensure the estimated threshold and parameters functions are smooth and continuous, and that they satisfy periodicity on the interval $(-2,2]$. 


With basis functions selected, an important consideration is the basis dimension size $k$; this corresponds to the number of knots of the spline function. This selection represents a trade-off, since selecting too many knots will result in higher computational burden and parameter variance, while selecting too few will not offer sufficient flexibility for capturing non-linear relationships. We consider this trade-off in detail in Section \ref{subsubsec:GAM_tune_paras}. 

Given $k$, the next step is to determine the knot locations; these are points where spline sections join. The knots should be more closely spaced in regions where more observations are available. In our case, we define knots at empirical quantiles of the angular variable $Q$ corresponding to a set of equally spaced probability levels; this is typical for spline based modelling procedures.

With a GAM formulated, the final step is to estimate the spline coefficients $\beta_j$, $j \in \{0,1,\dots,k\}$. Various methods have been proposed for this estimation \citep{Wood2017}. We have opted to use the restricted maximum likelihood (REML) approach of \citet{Wood2016}. For this technique, the log-likelihood function is penalised in a manner that avoids over-fitting, and cross-validation is used to automatically select the corresponding penalty parameters. Estimation via REML avoids the use of MCMC, which can be computationally expensive in practice; see \citet{Wood2017} for further details.

\subsubsection{Estimation of the threshold and GP parameter functions} \label{subsubsec:GP_funcs}
Estimation of the threshold function $u_{\gamma}(q)$ is equivalent to estimating quantiles of $R_q$ over $q \in (-2,2]$, motivating the use of quantile regression techniques. Employing the GAM framework with $g_{u_{\gamma}}$ defined as in equation \eqref{eqn:GAM_form}, we set $g_{u_{\gamma}}(q):= \log(u_{\gamma}(q))$, so that $u_{\gamma}(q) = \exp(g_{u_{\gamma}}(q))>0$. We then employ the approach of \citet{Youngman2019}, whereby a misspecified asymmetric Laplace model is assumed for $R_q$, and REML is used to estimate the coefficients associated with $g_{u_{\gamma}}$. By altering the pinball loss function typically used in quantile regression procedures \citep{Koenker2017}, this approach avoids computational issues that can often arise within such procedures; see the Supplementary Material for further details.

Similar to $u_\gamma$, we define $g_\tau(q) = \log(\tau(q))$ and $g_\xi(q) = \xi(q)$, with $g_\tau$, $g_\xi$ defined as in equation \eqref{eqn:GAM_form}. Again applying the approach of \citet{Youngman2019}, we estimate the coefficients associated with $g_\tau$ and $g_\xi$ using REML. Further details about this estimation procedure can be found in the Supplementary Material. 

\subsubsection{Selecting basis dimensions} \label{subsubsec:GAM_tune_paras}
An important consideration when specifying the GAM forms for both the threshold and parameter functions is the basis dimension. Selecting an appropriate dimension is essential for ensuring accuracy and flexibility in spline modelling procedures \citep{Wand2000,Perperoglou2019}. Generally speaking, selecting too few knots may result in functional estimates that do not capture the underlying covariate relationships, while parameter variance increases for larger dimensions. 

While some approaches have been proposed for automatic dimension selection \citep[e.g.,][]{Kauermann2011}, most available spline based modelling procedures select the dimension on a case-by-case basis using practical considerations. Moreover, as long as the basis dimension is sufficiently large enough, the resulting modelling procedure should be insensitive to the exact value, or the knot locations \citep{Wood2017}. This is due to the REML estimation framework, which penalises over-fitting, thus dampening the effect of adding additional knots to the spline formulation. As such, it is preferable in practice to select more knots than one believes is truly necessary to capture the covariate relationships. Therefore, we select reasonably large basis dimensions for the data sets considered in Sections \ref{sec:sim_study} and \ref{sec:case_study}. 

\section{Practical tools for SPAR model inference} \label{sec:practical_tools}
In this section, we introduce practical tools to aid with implementation of the inference frameworks presented in Sections \ref{sec:angular_dens} and \ref{sec:cond_radial_dens}. Specifically, we introduce a tool for quantifying uncertainty in the SPAR framework, alongside diagnostic tools for assessing goodness of fit. The latter tools can also be used to inform the selection of tuning parameters, such as the non-exceedance probability $\gamma$, the bandwidth parameter $h$, and the basis dimension. 

\subsection{Evaluating uncertainty} \label{subsec:SPAR_uncert}
When applying the SPAR modelling framework in practice, uncertainty will arise for each of the estimated components; namely, the angular density, threshold and parameter functions. In practice, this uncertainty is a result of sampling variability combined with model misspecification. The former arises due to finite sample sizes only partially representing the entire population, while the latter arises from modelling frameworks not fully capturing the complex features of the data. Quantifying this uncertainty is crucial for interpreting statistical results and making informed decisions based on the inherent modelling limitations.


To quantify uncertainty in SPAR model fits, we must consider each model component in turn. For this, we take a similar approach to \citet{Haselsteiner2021} and \citet{Murphy-Barltrop2023}, and consider a fixed angle $q \in \mathcal{Q}_{grid}$, with $\mathcal{Q}_{grid}$ defined as in Section \ref{subsec:local_inference}. We then quantify the estimation uncertainty for each model component while keeping the angle fixed. Specifically, we propose the following bootstrap procedure: for $b = 1, \dots, B$, where $B \in \NN$ denotes some large positive integer, do the following 
\begin{enumerate}
    \item Resample the original data set (with replacement) to produce a new sample of size $n$. 
    \item Compute the point estimate of the angular density at $q$, denoted $\hat{f}_{Q,b}(q)$, using the methodology described in Section \ref{sec:angular_dens}. 
    \item Compute the point estimates of the threshold, scale and shape parameters at $q$, denoted $\hat{u}_{\gamma,b}(q)$, $\hat{\tau}_b(q)$ and $\hat{\xi}_b(q)$ respectively, using the methodology described in Section \ref{subsec:smooth_infer}. 
\end{enumerate}
We remark that the choice of resampling procedure can be adapted to incorporate data sets exhibiting temporal dependence. In this case, rather than using a standard bootstrap, one can apply a block bootstrap \citep{Kunsch1989}. This sampling scheme retains temporal dependence in the resampled data set, ensuring the additional uncertainty that arises due to lower effective sample sizes is accounted for \citep{Politis1994}. See \citet{Keef2013a} and \citet{Murphy-Barltrop2023} for applications of block bootstrapping within the extremes literature. 

Given a significance level $\alpha \in (0,1)$, we use the outputs from the bootstrapping procedure to construct estimates of the median and $100(1 - \alpha)\%$ confidence interval for each model component at $q$. Considering the angular density, for example, these quantities are computed using the set $\{\hat{f}_{Q,b}(q) \mid b \leq k \leq B\}$. Assuming the estimation procedure is unbiased, one would expect $\Pr(\hat{f}^{\alpha/2}_{Q,b}(q) \leq f_{Q}(q) \leq \hat{f}^{1-\alpha/2}_{Q,b}(q)) \approx (1-\alpha)$, where $\hat{f}^{\alpha/2}_{Q,b}(q)$ and $\hat{f}^{1-\alpha/2}_{Q,b}(q)$ denote the empirical $\alpha/2$ and $1 - \alpha/2$ quantile estimates from $\{\hat{f}_{Q,b}(q) \mid 1 \leq b \leq B\}$, respectively. 

Repeating this procedure for all $q \in \mathcal{Q}_{grid}$ allows one to evaluate uncertainty over the angular domain for each model component, thus quantifying the SPAR model uncertainty. This in turn allows us to evaluate uncertainty in quantities computed from the SPAR model, such as isodensity contours or return level sets; see Sections \ref{sec:sim_study} and \ref{sec:case_study}. 



\subsection{Evaluating goodness of fit} \label{subsec:diagnostics}

We present a localised diagnostic to assess the relative performance of the SPAR model fits in different regions, similar to that used in \cite{mackay2024}. Consider a partition of the angular domain around $q \in \{-1.5,-1,-0.5,0,0.5,1,1.5,2\}$, corresponding to a variety of regions in $\RR^2$. For each $q$, take the local radial window $\mathcal{R}_q$ as defined in Section \ref{subsec:local_inference}. Treating $\mathcal{R}_q$ as a sample from $R_q$, we compare the observed quantiles with the fitted SPAR model quantiles, resulting in a localised QQ plot. Similar to before, uncertainty can be quantified via non-parametric bootstrapping. Comparing the resulting QQ plots over different angles, and different values of $\gamma$ and $k$, provides another means to assess model performance.  

Finally, we propose comparing the estimated angular density, obtained using the methodology of Section \ref{sec:angular_dens}, with the corresponding density function computed from the histogram. This comparison allows one to assess whether the choice of bandwidth parameter, $h$, is appropriate for a given data structure. 

\section{Simulation study} \label{sec:sim_study}
\subsection{Study set up}
In this section, we evaluate the performance of the smooth inference framework introduced in Section \ref{subsec:smooth_infer} via simulation. We do not consider the local estimation approach of Section \ref{subsec:local_inference} here, since our proposed local estimates are only meant as a means of assessing smooth SPAR estimates when the true values are unknown.

\begin{figure}[!ht]
    \centering
    \includegraphics[width=\textwidth]{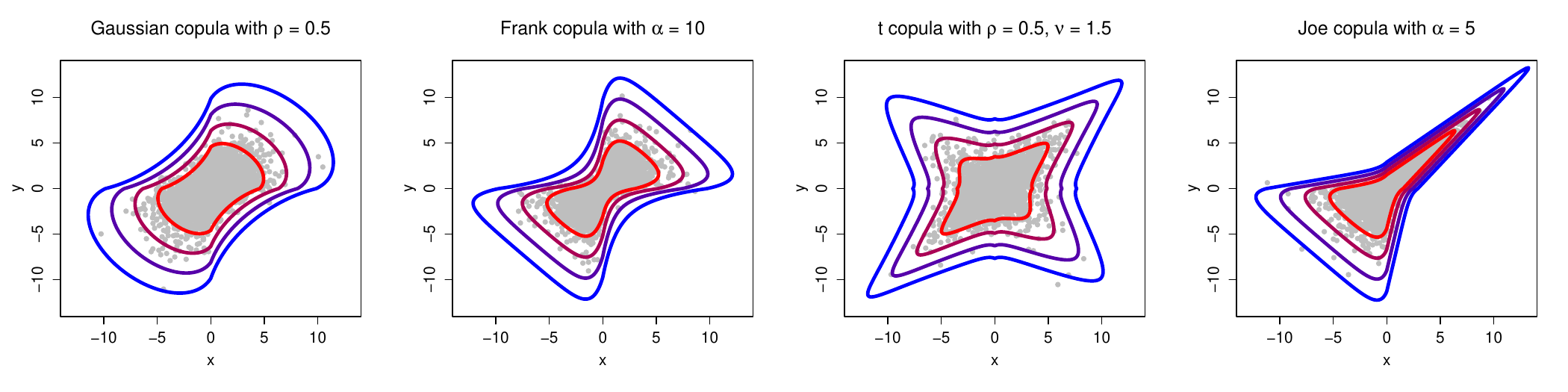}
    \caption{Example data sets of size $n=10,000$ and isodensity contours from each copula on standard Laplace margins. The red to blue lines in each plot represent the joint density levels $p \in \{10^{-3},10^{-4},10^{-5},10^{-6} \}$.}
    \label{fig:copula_examples}
\end{figure}

We consider four copulas on standard Laplace margins; Gaussian, Frank, t and Joe, as defined in the Supplementary Material. These distributions represent a range of dependence structures. Note that analogous dependence coefficients to $\chi$ can be defined to quantify the strength of extremal dependence in other regions of the plane \citep[see][]{Mackay2023}. In the following, we denote the four quadrants of $\RR^2$ as $\mathbb{Q}_1, ..., \mathbb{Q}_4$. For $\rho>0$, the Gaussian copula has intermediate dependence in $\mathbb{Q}_1$ and $\mathbb{Q}_3$ \citep{Hua2011}, and negative dependence in $\mathbb{Q}_2$ and $\mathbb{Q}_4$. The Frank copula is AI in all quadrants, whereas the t copula is AD in all quadrants. Finally, the Joe copula is AD in $\mathbb{Q}_1$, negatively dependent in $\mathbb{Q}_2$ and $\mathbb{Q}_4$, and AI in $\mathbb{Q}_3$. Samples from each copula are shown in Figure \ref{fig:copula_examples}, together with the corresponding values of the copula parameters used in the simulation studies. In each case, the sample size is $n=10,000$. One can observe the variety in dependence structures, as evidenced by the shape of data clouds. For the distributions considered here, the asymptotic shape parameter function is $\xi(q)=0$, $q\in(-2,2]$, and the asymptotic scale parameter functions can be derived analytically \citep[see][]{Mackay2023}. The true values of the threshold functions $u_{\gamma}(q)$ and angular density functions $f_Q(q)$ can be calculated using numerical integration. Hence, in all cases, the target values for the SPAR model parameters are known.

To evaluate performance, we simulate $500$ samples from each distribution and for every sample, apply the methods outlined in Section \ref{sec:angular_dens} and \ref{subsec:smooth_infer} to estimate all SPAR model components. Using these estimates, we compute isodensity contours, defined as $\{ (x,y) \in \RR^2 \mid f_{X,Y}(x,y) = p \}$ for some $p$. In particular, we consider $p \in \{10^{-3}, 10^{-4}, 10^{-5}, 10^{-6}\}$; the corresponding true contours for each distribution are given in Figure \ref{fig:copula_examples}. These density values represent regions of low probability mass, corresponding to joint extremal behaviour. Moreover, estimates of the joint density are appropriate for evaluating the overall performance of the SPAR model, since in practice, capturing the joint density is crucial for ensuring one can accurately extrapolate into the joint tail. 


Alongside isodensity contours, we also compare the estimated GP scale parameter functions and angular density functions to their corresponding target values. For the former, we remark that for each copula, the conditional radial distribution $f_{R_q}(r|q)$ only converges to a GP distribution in limit as $r\to\infty$, implying we are unlikely to accurately estimate the asymptotic GP parameter functions for finite samples; see \citet{Mackay2023}. We remark that even with this caveat, we still obtain high quality estimates for the isodensity contours at extreme levels. 

Uncertainty in the estimation procedure is quantified by adapting the procedure of Section \ref{subsec:SPAR_uncert} across the $500$ simulated samples. This allows us to compute median estimates and confidence intervals for isodensity contours, scale functions and angular density functions. 

Although the choice of coordinate system does not affect whether the SPAR model assumptions are satisfied, the asymptotic SPAR parameter functions do depend on the coordinate system. Since smooth, continuous splines are used to represent the GP parameter functions, the choice of coordinate system may affect the quality of model fits. The simulation study is therefore conducted using both $L1$ and $L2$ polar coordinates. 



\subsection{Tuning parameters} \label{subsec:simstudy_tuneparas}
We first consider the tuning parameters required for the smooth inference procedure, as outlined in Section \ref{subsubsec:GAM_tune_paras}. For each copula, the threshold and scale parameter functions appeared to vary in a similar manner over angle. Furthermore, we fix a constant value of the shape parameter with angle, i.e., $\xi(q) = \xi \in \RR$ for all $q \in (-2,2]$, since for each copula, the tail behaviour remains constant over angles \citep{Mackay2023}. As discussed in Section \ref{sec:case_study}, this is not true in the general case, so fixing $\xi(q)$ to be constant is an additional constraint imposed on the model.

As noted previously, it is better to select a basis dimension $k$ that is larger than one would expect to be necessary. We considered a range of dimensions in the interval $[5,50]$, and compared the resulting model fits across each of the four copulas. From this analysis, we found that setting $k=25$ was sufficiently flexible for capturing the angular dependence for both the threshold and scale functions. 


We are also required to select a non-exceedance probability $\gamma \in (0,1)$. As observed in \citet{Mackay2023}, each of the four copulas exhibits a different rate of convergence to the asymptotic form. Therefore, different values of $\gamma$ may be appropriate for these different dependence structures. However, we instead opt to keep $\gamma$ fixed across all copulas. This is for consistency in the estimation framework, as well as to show that even in the case when the model is misspecified, the corresponding inference framework is still robust enough to approximate the true model. We considered a range of $\gamma$ values, restricting our attention to the interval $[0.5,0.95]$, and compared the resulting model fits. As expected, the performance for each copula varied non-homogeneously across $\gamma$ values. Ultimately, we found that setting $\gamma = 0.8$ appeared sufficient for approximating the conditional radial tails for each dependence structure. In particular, this value appeared high enough to give approximate convergence to a GP distribution model, without giving a large degree of variability in model estimates.  

Finally, for estimation of the angular density, we fix the bandwidth parameter at $h = 1/50$ for each copula. Our results show that for these copula examples, this bandwidth is sufficiently flexible to approximate the true angular density functions across the majority of angles. 

\subsection{Results} \label{subsec:simstudy_results}
\begin{figure}[htp]
    \centering
    \begin{subfigure}[b]{0.21\textwidth}
        \centering
        \includegraphics[width=\textwidth]{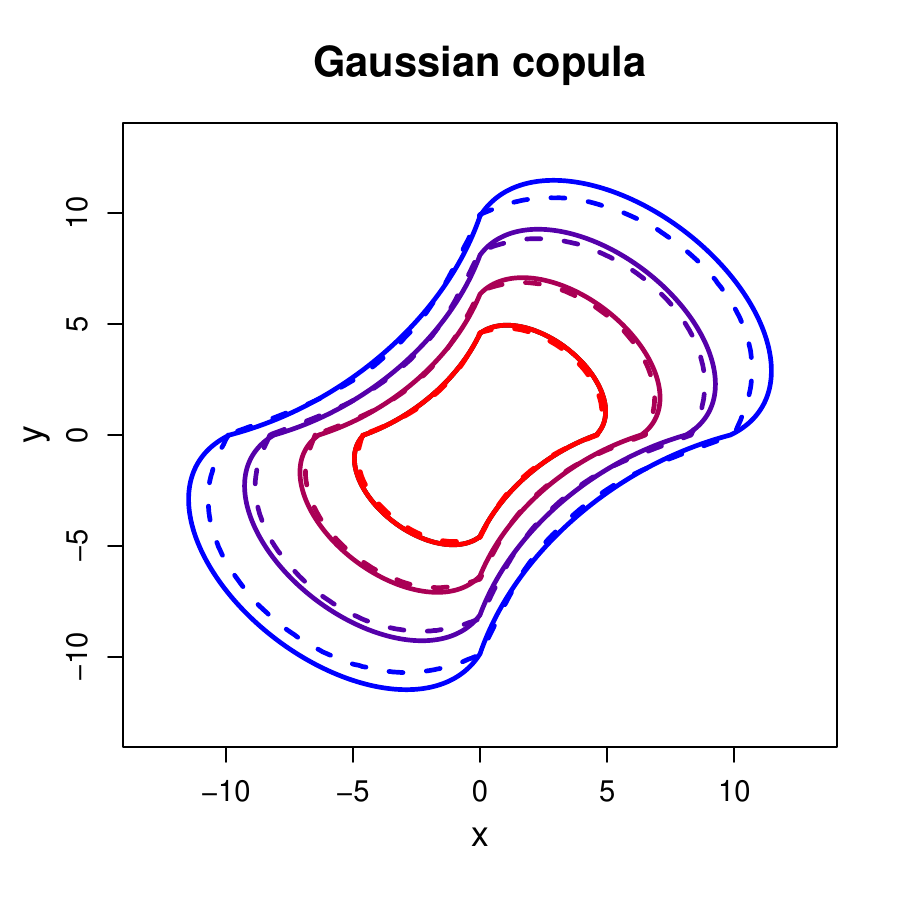}
    \end{subfigure}
    \quad
    \begin{subfigure}[b]{0.21\textwidth}  
        \centering 
        \includegraphics[width=\textwidth]{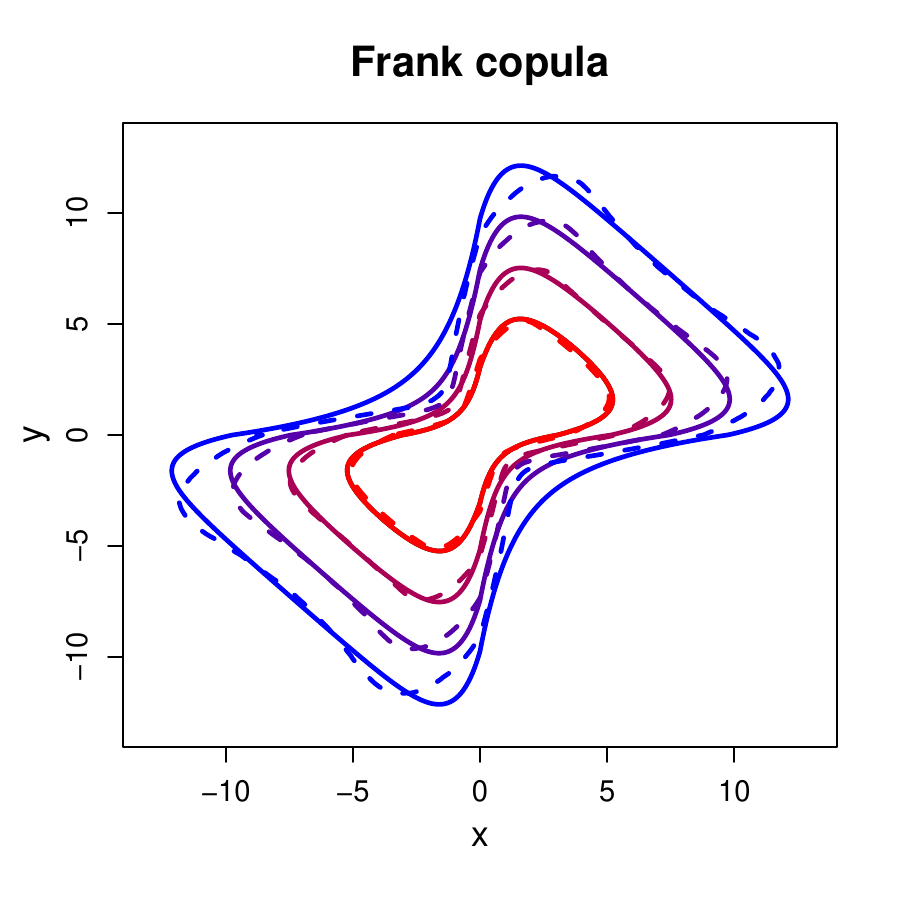}
    \end{subfigure}
    \quad
    \begin{subfigure}[b]{0.21\textwidth}   
        \centering 
        \includegraphics[width=\textwidth]{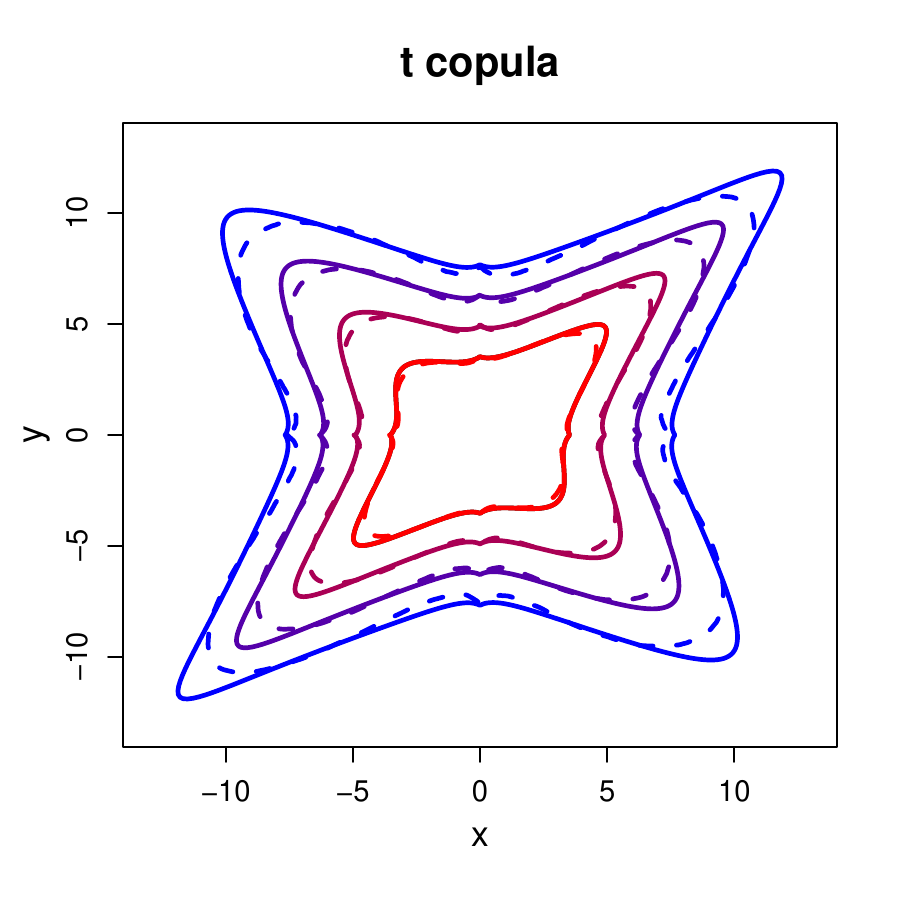}
    \end{subfigure}
    \quad
    \begin{subfigure}[b]{0.21\textwidth}   
        \centering 
        \includegraphics[width=\textwidth]{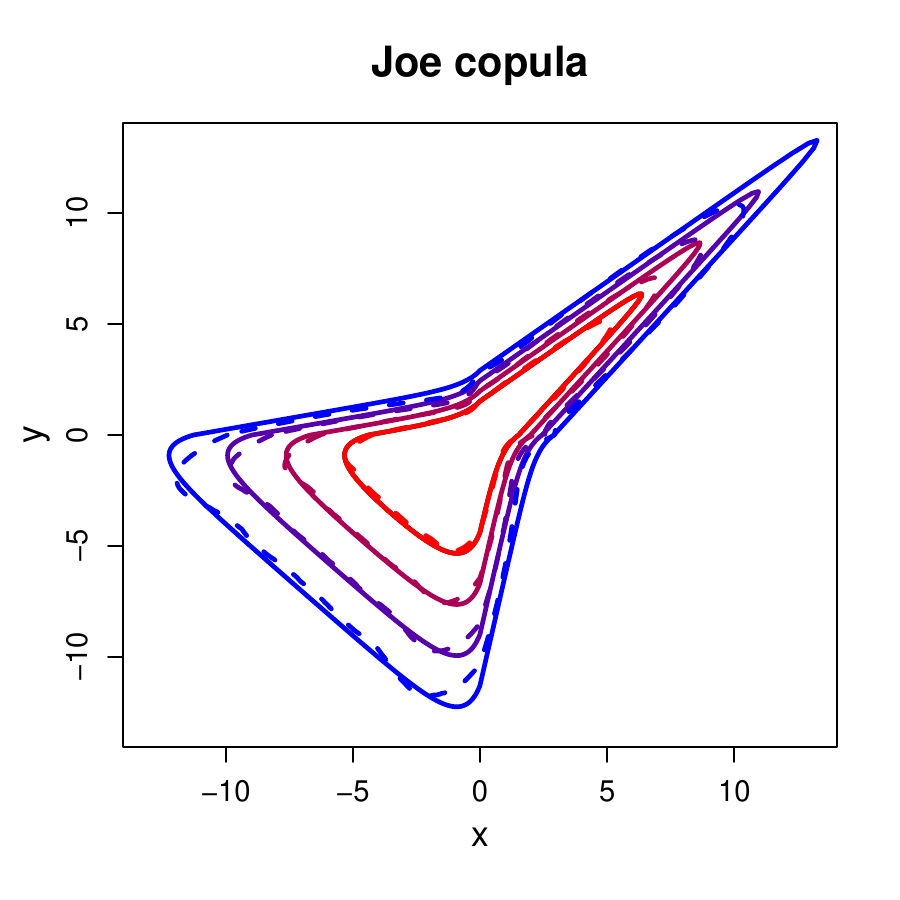}
    \end{subfigure}
    \caption{Comparison of true (thick lines) and median estimated (dashed lines) isodensity contours under the $L1$ coordinate system. In each plot, the red to blue lines represent the joint density levels $p \in \{10^{-3},10^{-4},10^{-5},10^{-6} \}$.}    \label{fig:equidensity_L1}
\end{figure}

Figure \ref{fig:equidensity_L1} compares the median estimates of isodensity contours, obtained using the $L1$ coordinate system, to the true contours at a range of low density levels; the corresponding plots for the $L2$ coordinate system are given in the Supplementary Material. For both coordinate systems, one can observe generally good agreement between the sets of contours, suggesting the modelling framework is, on average, capturing the dependence structure of each copula. Furthermore, plots comparing the median estimates from both coordinate systems can also be found in the Supplementary Material. These plots show a similar overall performance for both systems, with perhaps a slight preference for the $L1$ estimates. 



Plots comparing the estimated GP scale parameter functions, and associated confidence intervals, to the known asymptotic functions are given in the Supplementary Material. In some angular regions, the estimated isodensity contours and scale functions from the SPAR model do not agree with the known values; for example, in the region around $q = 0.5$ for the Joe copula. In this case, there is a sharp cusp in the asymptotic GP scale parameter function. Similarly, there is a sharp cusp in the true GP scale parameter for the t copula at $q=\pm0.5, \pm1.5$. As the inference framework assumes that the scale is a smooth function of angle, this behaviour is not properly captured. Despite the GAM representation not being able to capture these cusps, the overall performance of the estimated SPAR model is still reasonable in these regions. Furthermore, there is poor agreement between the estimated and asymptotic scale functions for the Frank copula. This is likely due to the relatively slow convergence of this distribution to its asymptotic form, and hence the poorer approximation by the GP distribution.



\begin{figure}[ht]
    \centering
    \begin{subfigure}[b]{0.21\textwidth}
        \centering
        \includegraphics[width=\textwidth]{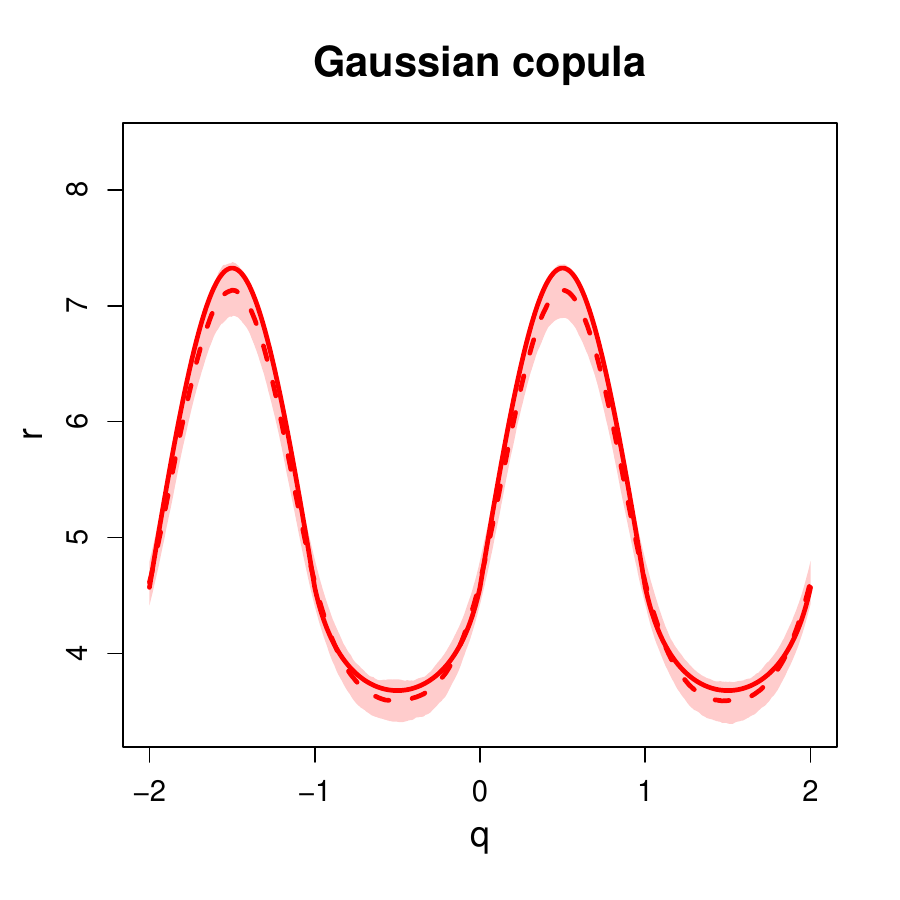}
    \end{subfigure}
    \quad
    \begin{subfigure}[b]{0.21\textwidth}  
        \centering 
        \includegraphics[width=\textwidth]{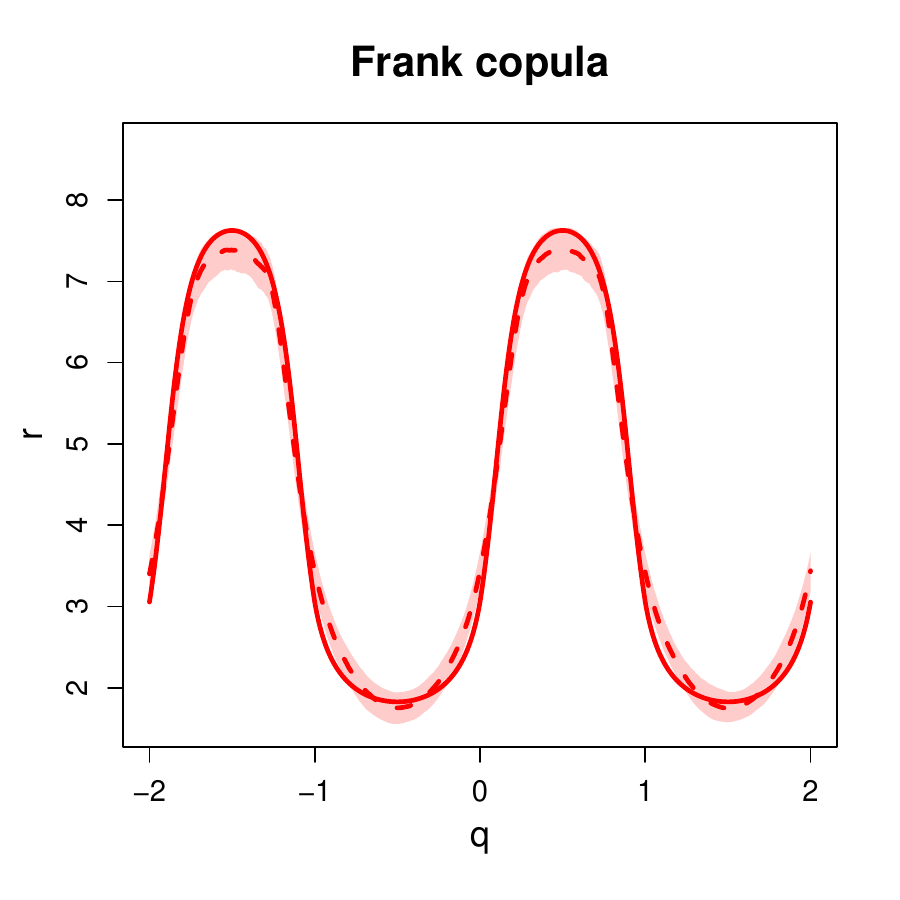}
    \end{subfigure}
    \quad
    \begin{subfigure}[b]{0.21\textwidth}   
        \centering 
        \includegraphics[width=\textwidth]{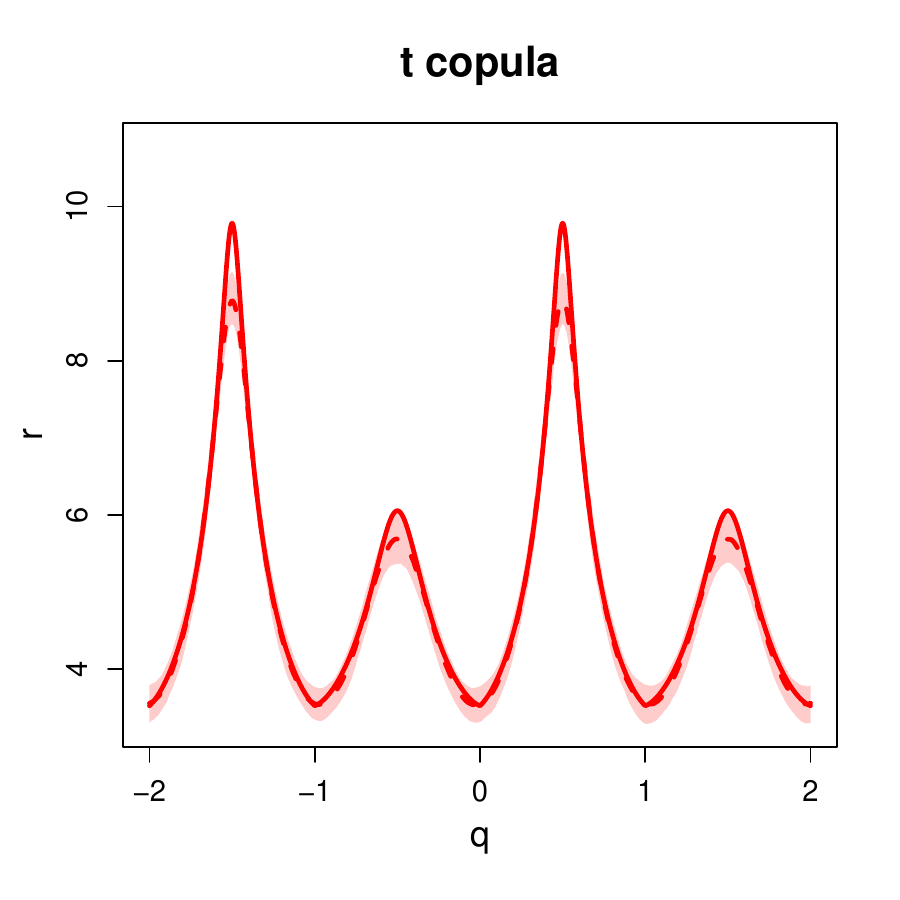}
    \end{subfigure}
    \quad
    \begin{subfigure}[b]{0.21\textwidth}   
        \centering 
        \includegraphics[width=\textwidth]{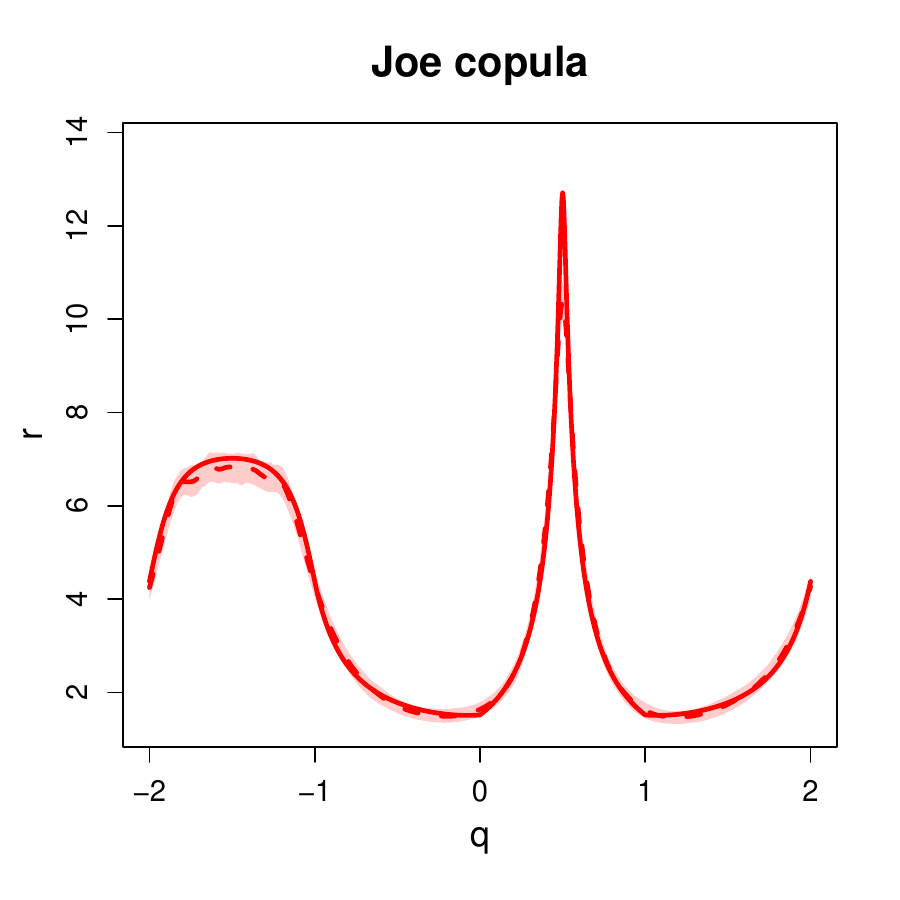}
    \end{subfigure}
    \vskip\baselineskip
       \begin{subfigure}[b]{0.21\textwidth}
        \centering
        \includegraphics[width=\textwidth]{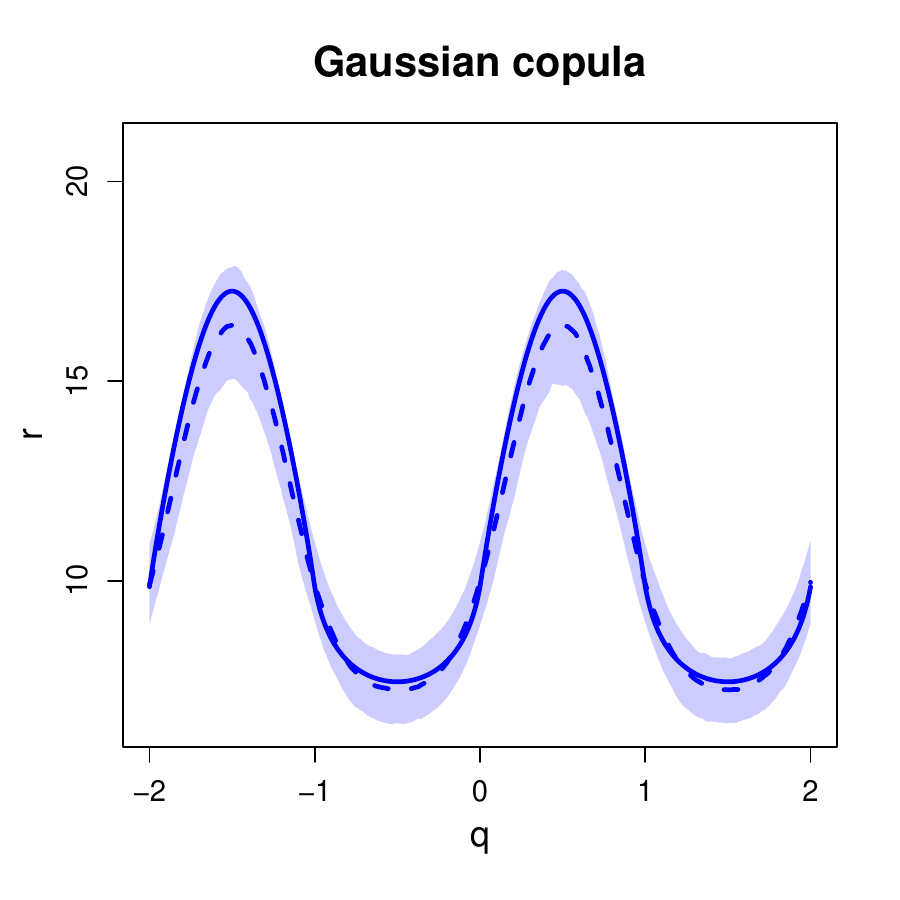}
    \end{subfigure}
    \quad
    \begin{subfigure}[b]{0.21\textwidth}  
        \centering 
        \includegraphics[width=\textwidth]{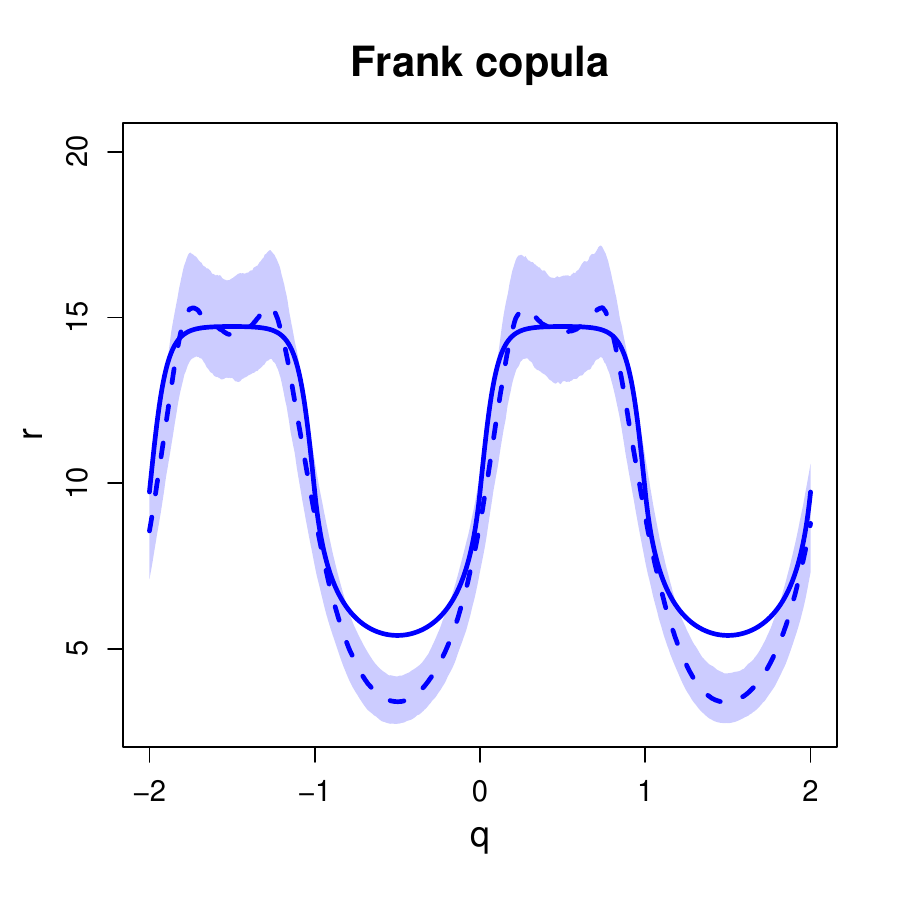}
    \end{subfigure}
    \quad
    \begin{subfigure}[b]{0.21\textwidth}   
        \centering 
        \includegraphics[width=\textwidth]{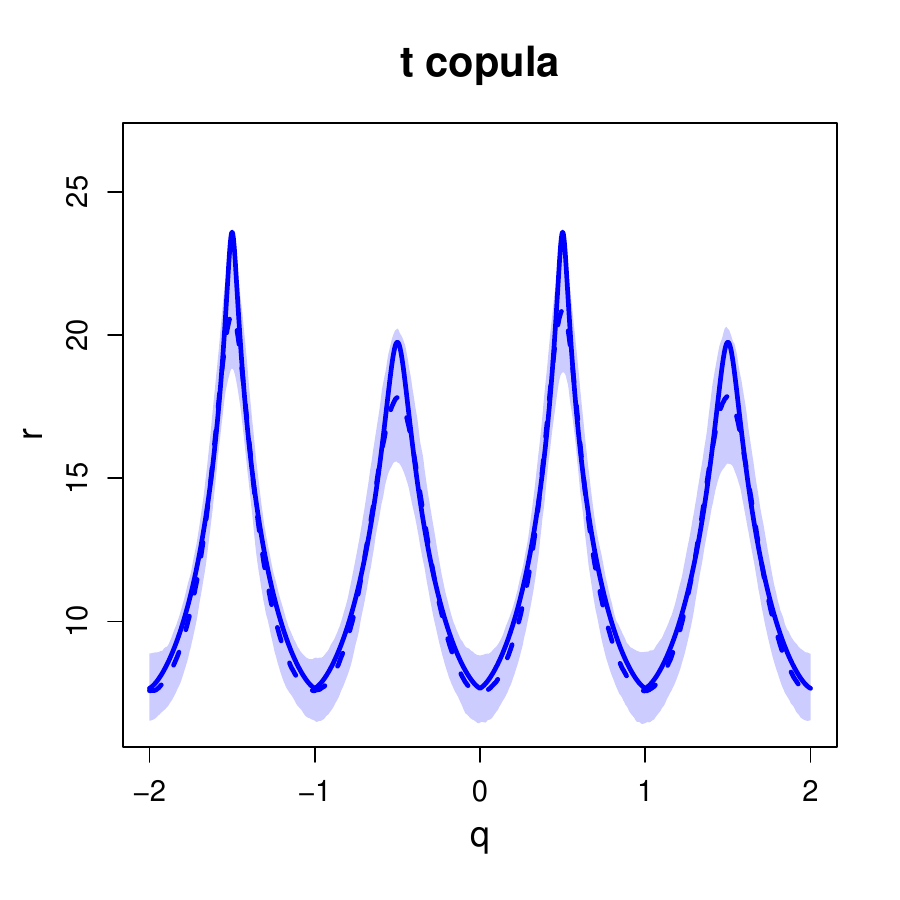}
    \end{subfigure}
    \quad
    \begin{subfigure}[b]{0.21\textwidth}   
        \centering 
        \includegraphics[width=\textwidth]{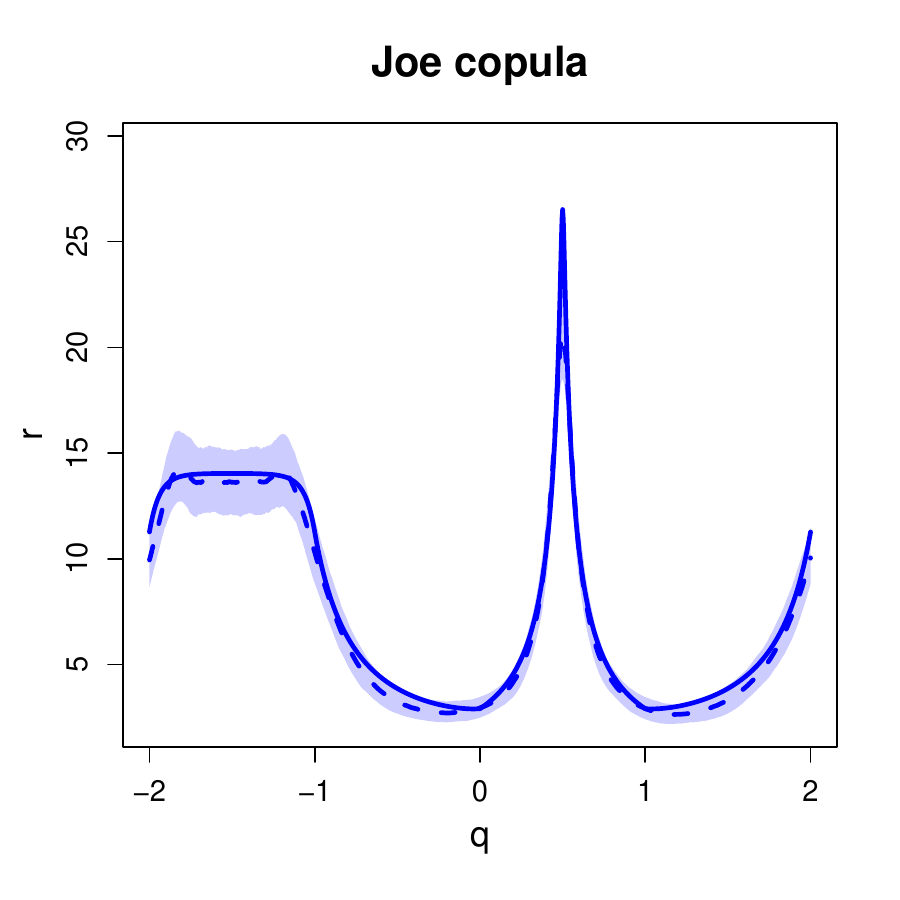}
    \end{subfigure}
    
    \caption{Comparison of median estimated isodensity contours (dashed lines), with 95\% confidence intervals (shaded region), to true contours (solid lines) for joint density level $p=10^{-3}$ (top row) and $p=10^{-6}$ (bottom row), with estimates obtained using $L1$ polar coordinates.}
    \label{fig:uncertainty_equidensity_L1}
\end{figure}

Next, we consider the uncertainty of the isodensity contours for $p \in \{10^{-3},10^{-6} \}$ in the radial-angular space. Figure \ref{fig:uncertainty_equidensity_L1} shows the median contour estimates, along with estimated 95\% confidence intervals, obtained under the $L1$ coordinate system; the corresponding plots for $L2$ coordinates are given in the Supplementary Material. One can observe that the true contours are generally well captured within the estimated uncertainty regions. The exception in the $10^{-6}$ density contour for the Frank copula in $\mathbb{Q}_2$ and $\mathbb{Q}_4$, owing to the aforementioned slow rate of convergence to the asymptotic form for this copula.

Finally, we compare median estimates of the angular density functions, alongside estimated confidence regions, to the corresponding true density functions for each copula. These results are given in the Supplementary Material. Overall, we observe close agreement between the estimated and true functions at the majority of angles. However, we note that the KD estimation framework appeared unable to fully capture the modal regions for the Frank, t and Joe copulas; see Section \ref{sec:discussion} for further discussion. 

Overall, when compared to the truth, the SPAR model estimates perform well for each copula. This observation suggests that our proposed inference framework, with appropriate tuning parameters, can capture the extremal dependence structure across a range of copulas with differing dependence classes. This illustrates both the flexibility and robustness of the SPAR approach, and its advantages over many alternative multivariate models.

\section{Case Study} \label{sec:case_study}

In this section, we apply the techniques introduced in Sections \ref{sec:angular_dens} and \ref{sec:cond_radial_dens} to the data sets A, B and C introduced in Section \ref{subsec:motiv_examples}. We show that the resulting model fits are physically plausible and capture the complex dependence features of each data set. We also apply the tools introduced in Section \ref{sec:practical_tools} to quantify uncertainty and assess goodness of fit; the resulting diagnostics indicate generally good performance.  

\subsection{Pre-processing}
The simulation study in Section \ref{sec:sim_study} considered data on standard Laplace margins. This is because the SPAR model assumptions are satisfied for a wide range of copulas on Laplace margins, and the resulting asymptotic representations are relatively simple \citep{Mackay2023}. However, the SPAR framework does not pre-suppose any particular choice of margins, and \citet{Mackay2023} also showed that SPAR representations arise for random vectors with bounded and heavy tailed margins. 

For the case of the metocean data sets, the margins are unknown. In practice, estimation of the marginal distributions, as is common in many extreme value analyses, introduces a high degree of additional modelling uncertainty, and poor marginal estimates affect the quality of the resulting multivariate inference \citep{Towe2023}. Therefore, we opt not to transform the margins of the metocean time series, and to instead fit the SPAR model on the original scale of the data. With suitable selections of tuning parameters, we demonstrate below that our inference framework is flexible enough to capture the observed extremal dependence structures for the metocean data sets without the need for marginal transformation.

When modelling phenomena (such as ocean waves) in the Earth's environment, physical constraints (e.g. limited capacity for energy transfer from surface wind caused by atmospheric low pressure systems, wave steepness limits) typically support the assumption that tails of marginal distributions of random variables are bounded. When variables are presented on different physical scales (e.g. mm vs. km), we are careful to standardise them to zero mean and unit standard deviation prior to analysis. For variables which are believed to have unbounded tails, we would transform to common margins prior to analysis; in this case, as outlined in \cite{Mackay2023}, we favour transformation to common standard Laplace margins.


An important consideration for the SPAR model is where to place the origin of the polar coordinate system. When using Laplace margins, a natural choice is to locate the origin at $(x,y)=(0,0)$. When working on the original scale of the data, the choice is less clear. One option would be to place the polar origin at $(x,y)=(0,0)$. However, this would restrict the range of angles for which the SPAR model offers a useful representation, since both variables we are considering here take only positive values. To account for this, we normalise the data to have zero mean and unit variance, and select the polar origin at $(x,y)=(0,0)$ in the normalised variable space. Define normalised variables $(\Tilde{T}_z,\Tilde{H}_s) := ((T_z - \mu_{T_z})/\sigma_{T_z},(H_s - \mu_{H_s})/\sigma_{H_s})$, where $(\mu_{T_z}, \mu_{H_s})$ and $(\sigma_{T_z}, \sigma_{H_s})$ denote the estimated means and standard deviations of $(T_z,H_s)$, respectively. 

We henceforth assume that for each data set, the normalised joint density function, $f_{\Tilde{T}_z,\Tilde{H}_s}$, satisfies the assumptions of the SPAR model, namely that the conditional radial variable converges to a generalised Pareto distribution, in the sense of Equation~2.2, and that the functions $f_Q(q)$, $u_{\gamma}(q)$, $\tau(q)$ and $\xi(q)$ satisfy the finiteness and continuity assumptions of Section~2.2. We then apply the statistical techniques introduced in Sections \ref{sec:angular_dens}, \ref{sec:cond_radial_dens} and \ref{sec:practical_tools}; these results are presented in Section \ref{subsec:case_study_val}. Throughout this section, we present all results for the $L1$ coordinate system; the corresponding results for $L2$ coordinates are given in the Supplementary Material, with both systems resulting in similar model fits.

We remark that each metocean time series exhibits non-negligible temporal dependence. Following Section \ref{subsec:SPAR_uncert}, we apply block bootstrapping throughout this section whenever quantifying uncertainty, with the block size set to $4$ days. This block size appeared appropriate to account for the observed dependence in each time series. Note that temporal dependence could alternatively be accounted for by `declustering' the data and only modelling peak values. However, in multivariate applications, what constitutes a `peak value' is ambiguous since the extremes of each variable do not necessarily occur simultaneously; see \citet{mackay2021correlation} and \citet{mackay2023-DIFORM} for further discussion.


\subsection{Tuning parameters}
Prior to inference, we must first select each of the relevant tuning parameters for the methodologies discussed in Sections \ref{sec:angular_dens} and \ref{sec:cond_radial_dens}. Since the true dependence features are unknown, we use local estimates of the SPAR model, obtained using the framework of Section \ref{subsec:local_inference}, to inform the choice of basis dimensions for the smooth inference procedure of Section \ref{subsec:smooth_infer}. 


To obtain local estimates, we are required to specify the number of reference angles $M$, the number of order statistics $N$, and the non-exceedance probability $\gamma$. For the first two values, we set $M=200$ and $N=500$; we found these values to be adequate to ensure a high degree of coverage over the angular interval $(-2,2]$, and to give angular windows that appeared approximately stationary. For selecting $\gamma$, we tested a range of probabilities in the interval $[0.5,0.95]$. Through this testing, we set $\gamma = 0.7$, since this value appeared to give approximate convergence to a GP tail across the majority of local windows. The same non-exceedance probability is also used for the smooth SPAR model estimates. 



With local estimates obtained, we then consider smooth estimation of the SPAR components. Notably, for each of the time series, we observe clear trends in the locally estimated shape parameter function; it would therefore not be appropriate to specify this parameter as constant. This makes sense when one considers the shapes of the data clouds illustrated in Figure \ref{fig:metocean_dataset}; the radial behaviour varies significantly over angles. Furthermore, both metocean variables are bounded below by $0$; therefore, we would expect shorter tails in angular directions that intersect the axes. A range of basis dimensions were tested, and from this analysis, we fixed $k=35$ for the threshold and scale functions, and $k=12$ for the shape function. These values appeared to offer adequate flexibility for capturing the trends observed over the angular variable.


Finally, for angular density estimation, we follow Section \ref{subsec:simstudy_tuneparas} and set the bandwidth $h = 1/50$. This value offered sufficient flexibility to capture the observed angular distributions. 



\subsection{Results} \label{subsec:case_study_val}
\begin{figure}[htp]
    \centering
    \includegraphics[width=\textwidth]{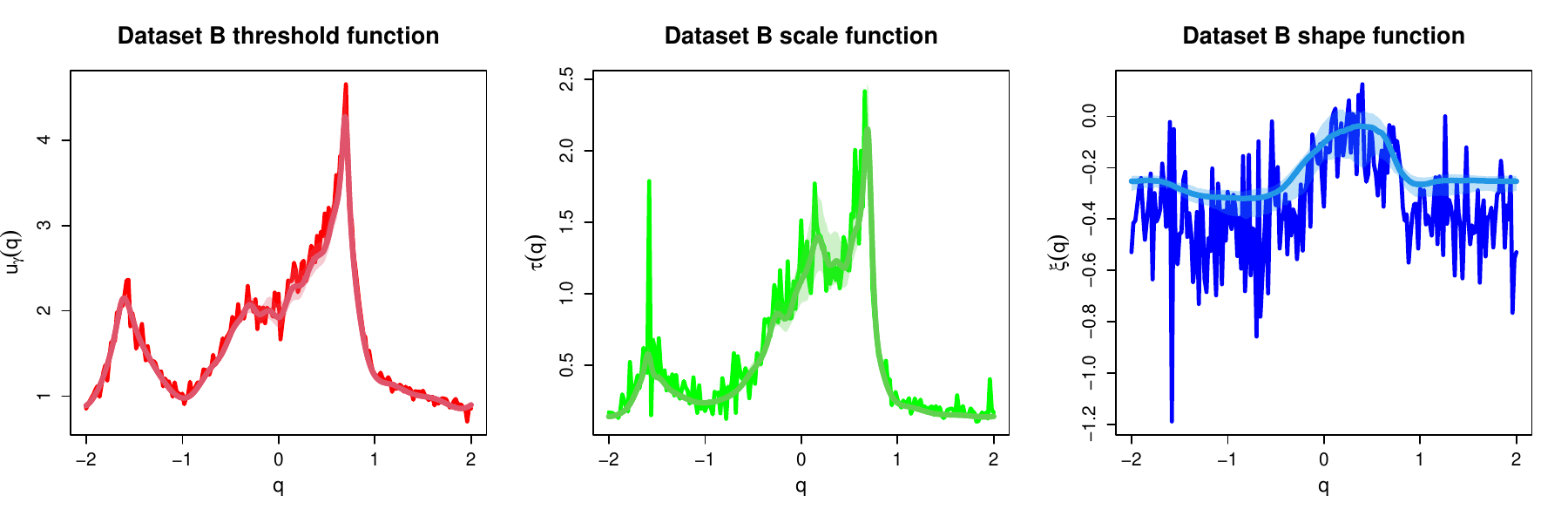}
    \caption{Comparison of estimated local (rough lines) and smooth threshold (red, left), scale (green, middle) and shape (blue, right) functions for data set B with the $L1$ coordinate system, with shaded regions denoting 95\% confidence intervals.}
    \label{fig:local_smooth_B}
\end{figure}

Figure \ref{fig:local_smooth_B} compares the threshold and parameter function estimates for data set B from the local and smooth inference procedures; the corresponding plots for data sets A and C are given in the Supplementary Material. The shaded regions in this figure denote the 95\% bootstrapped confidence intervals for the smooth model fits. One can observe generally good agreement for each component of the SPAR model. We remark that the local estimates appear unstable, and hence unreliable, for certain angles; this is not surprising, given the small sample size of the angular window. However, the general overall agreement suggests the smooth SPAR estimates are accurately capturing the observed dependence features for each data set, providing evidence that the chosen tuning parameters are appropriate. 

\begin{figure}[h]
    \centering
    \begin{subfigure}[b]{0.3\textwidth}
        \centering
        \includegraphics[width=\textwidth]{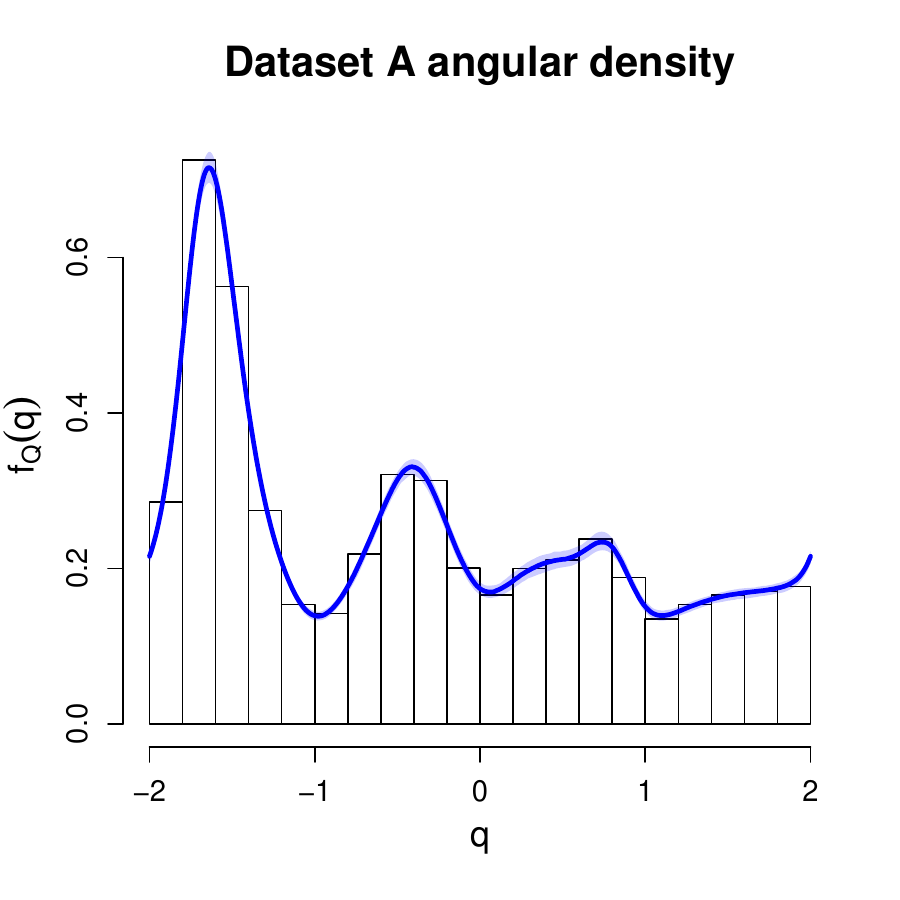}
    \end{subfigure}
    \quad
    \begin{subfigure}[b]{0.3\textwidth}  
        \centering 
        \includegraphics[width=\textwidth]{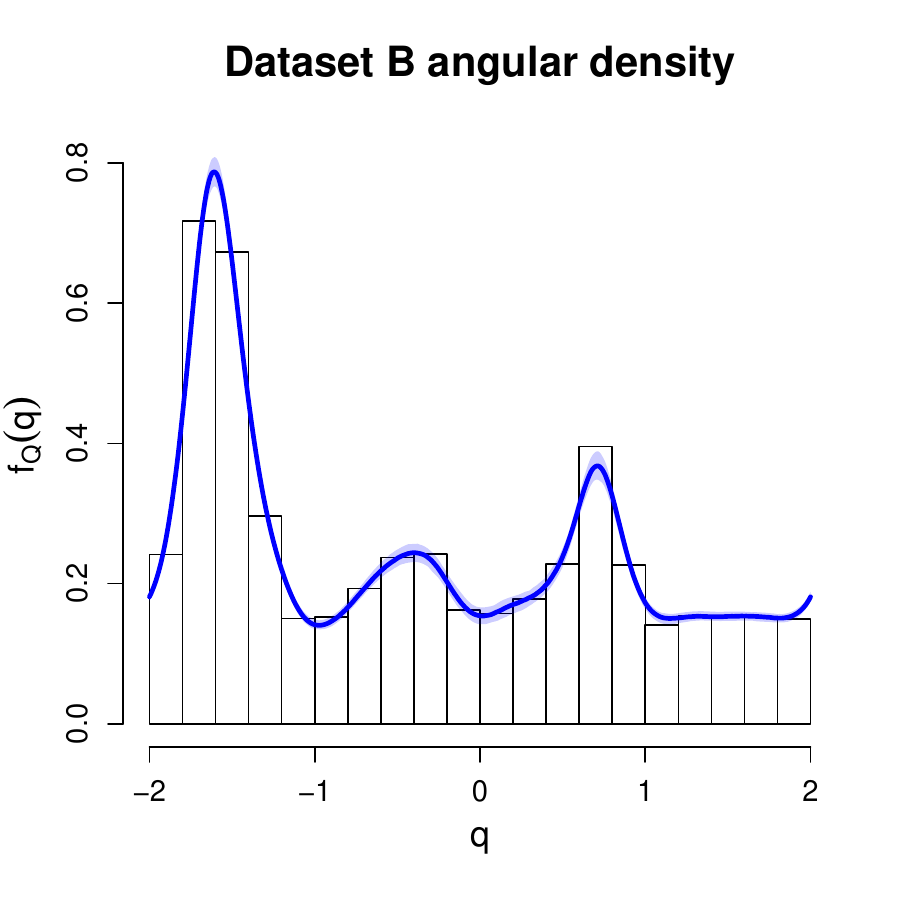}
    \end{subfigure}
    \quad
    \begin{subfigure}[b]{0.3\textwidth}  
        \centering 
        \includegraphics[width=\textwidth]{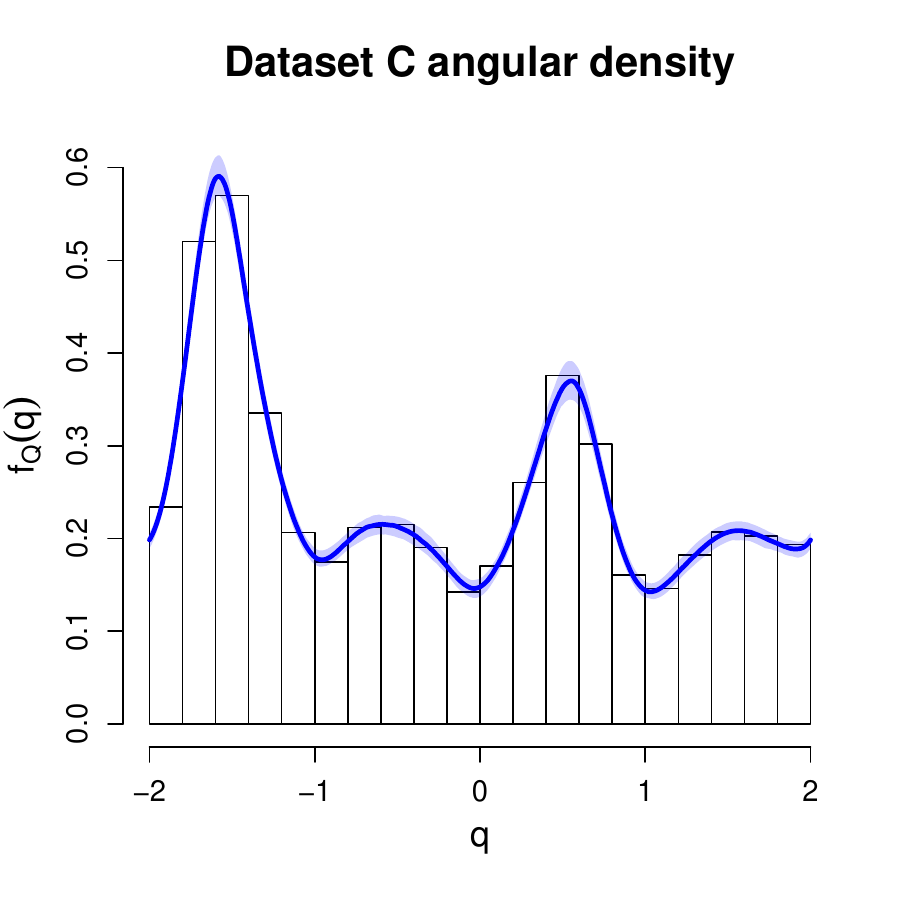}
    \end{subfigure}
    \caption{Comparison of estimated median (blue lines) angular density functions to histograms for data sets A (left), B (centre), and C (right) with the $L1$ coordinate system. The shaded regions in each plot denote the estimated 95\% confidence intervals.}
    \label{fig:angular_density}
\end{figure}

Following Section \ref{subsec:diagnostics}, we compare the median angular density functions from the KD estimation technique with empirical histograms. These comparisons are given in Figure \ref{fig:angular_density} for each data set, and one can observe good agreement between the estimated quantities.

\begin{figure}[htp]
    \centering
    \begin{subfigure}[b]{0.3\textwidth}
        \centering
        \includegraphics[width=\textwidth]{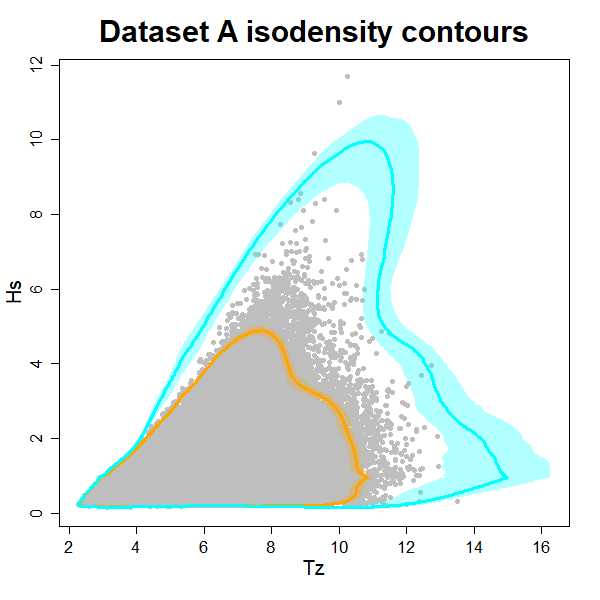}
    \end{subfigure}
    \quad
    \begin{subfigure}[b]{0.3\textwidth}  
        \centering 
        \includegraphics[width=\textwidth]{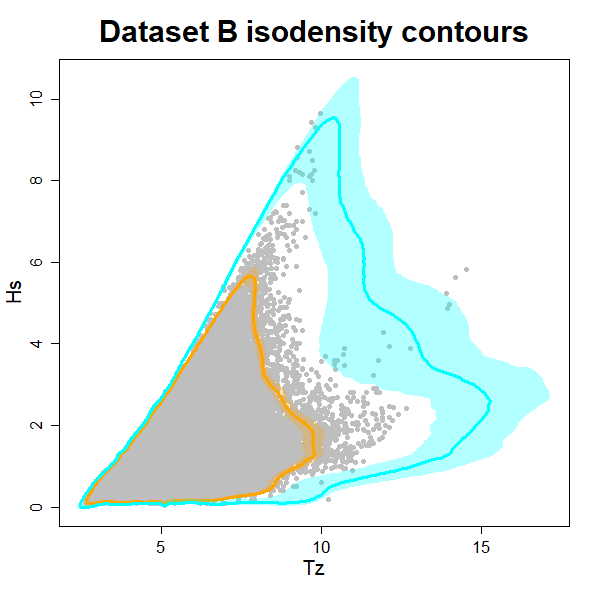}
    \end{subfigure}
    \quad
    \begin{subfigure}[b]{0.3\textwidth}  
        \centering 
        \includegraphics[width=\textwidth]{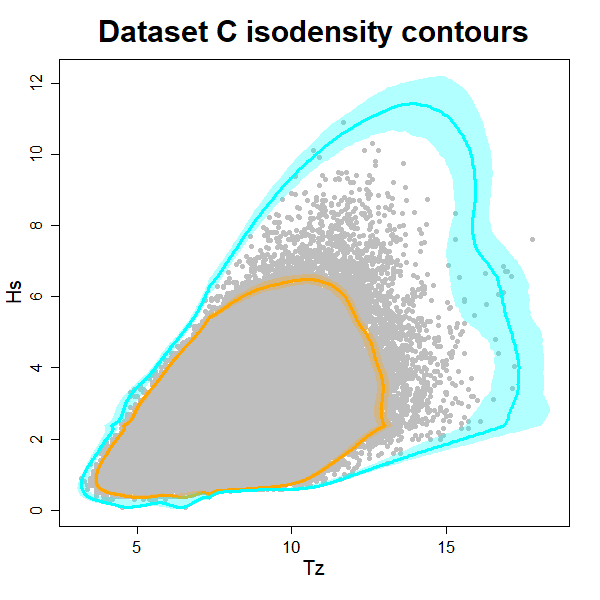}
    \end{subfigure}
    \caption{Estimated median isodensity contours at $p = 10^{-3}$ (orange lines) and $p =10^{-6}$ (cyan lines) for data sets A (left), B (centre), and C (right) with the $L1$ coordinate system. The shaded region for each contour denote the 95\% bootstrapped confidence intervals.}
    \label{fig:equidensity_contours}
\end{figure}

Estimates of isodensity contours are shown in Figure \ref{fig:equidensity_contours}. These joint density contours are given on the original scale of the data, rather than on the normalised scale, and we consider the joint density levels $p \in \{ 10^{-3},10^{-6}\}$, corresponding to regions of low probability mass. The estimated isodensity contours appear to capture the shape and structure of each data cloud well. Furthermore, we note that the SPAR model appears able to capture the observed asymmetric dependence structures, illustrating the flexibility and robustness of this modelling framework. 


To further demonstrate the utility of the SPAR framework, we also use the fitted model to obtain return level sets for each of the data sets. Return level sets are commonly used in ocean engineering for the design of offshore structures. They can be defined in various ways, but are generally defined in terms of marginal probabilities under various rotations of the coordinate axes, or in terms of the probability of an observation falling anywhere outside the set (the so-called `total exceedance probability' of a contour); see \citet{Mackay2021}. \citet{Papastathopoulos2024} noted that SPAR-type models offer a natural way for total exceedance probability contours to be constructed. For exceedance probability $a \in[0,1]$ with $a < 1 - \gamma$, the radius of the contour at angle $q$ is the $(1-a - \gamma)/(1-\gamma)$ quantile of the GP distribution with parameter vector $(u_{\gamma}(q),\xi(q),\tau(q))$. For any angle $q \in (-2,2]$, the probability of an observation exceeding this radius is equal to $a$; consequently, the probability of observing data outside of the resulting contour set is equal to $a$. When observations are independent and the distribution is stationary, we can define such sets in terms of return periods; given a number of years $K \in \NN$, the $K$-year return level set is the set corresponding to the probability $a:=1/n_yK$, where $n_y$ denotes the number of observations per year. One would expect to observe data points outside of the return level set once, on average, every $K$ years. Given the temporal dependence observed within the metocean data sets, we note that such an interpretation is not possible due to clustering of extreme events, and as such these estimates are conservative \citep{mackay2021correlation}. However, the resulting return level sets can still provide a useful summary of joint extreme behaviour. 

\begin{figure}[htp]
    \centering
    \begin{subfigure}[b]{0.3\textwidth}
        \centering
        \includegraphics[width=\textwidth]{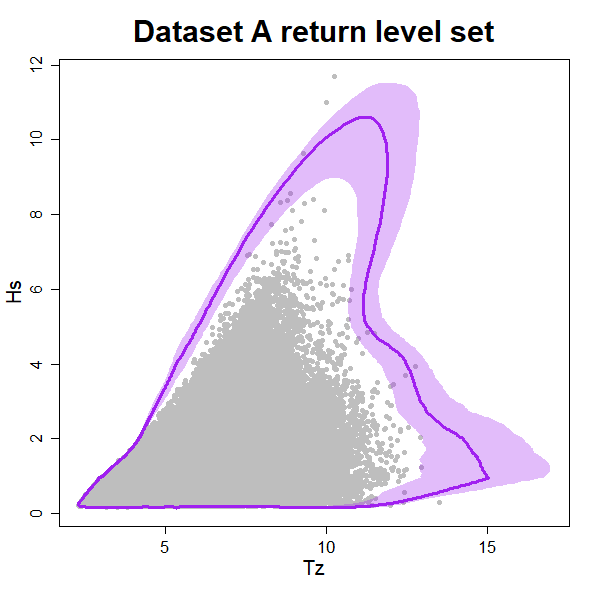}
    \end{subfigure}
    \quad
    \begin{subfigure}[b]{0.3\textwidth}  
        \centering 
        \includegraphics[width=\textwidth]{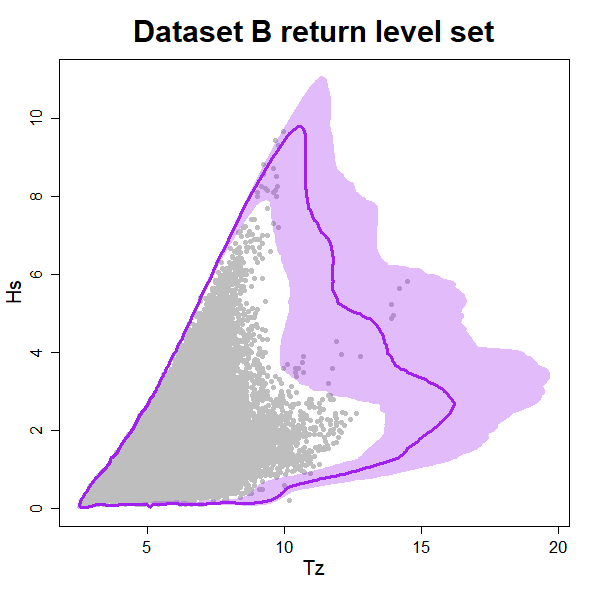}
    \end{subfigure}
    \quad
    \begin{subfigure}[b]{0.3\textwidth}  
        \centering 
        \includegraphics[width=\textwidth]{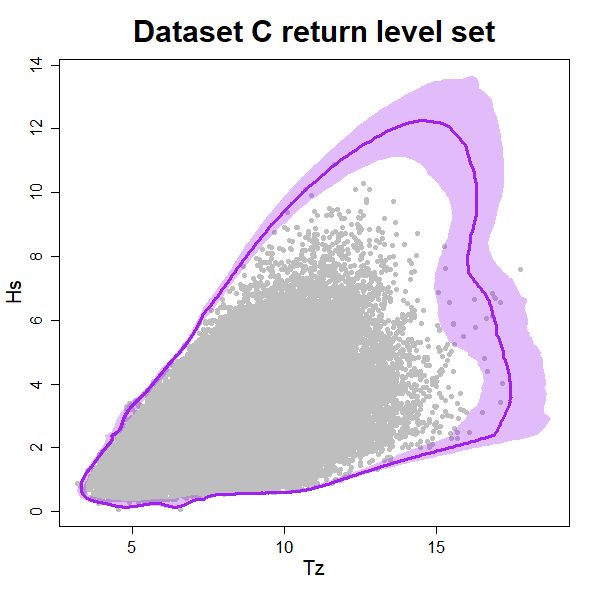}
    \end{subfigure}
    \caption{Estimated median $10$ year return level sets (purple lines) for data sets A (left), B (centre), and C (right) with the $L1$ coordinate system. The shaded region for each return level set denotes the 95\% bootstrapped confidence region.}
    \label{fig:return_level_sets}
\end{figure}

Plots of estimated median $10$ year return level sets for each data set are given in Figure \ref{fig:return_level_sets}, along with 95\% bootstrapped confidence intervals. These sets, obtained by computing GP distribution quantiles from the fitted model, appear sensible in shape and structure when compared to the data cloud. Moreover, given the lengths of observation windows of each data set, we would not expect to observe many datapoints outside of the return level set; this is clearly true in every case. Furthermore, a comparison of return level sets from the two coordinate systems is given in the Supplementary Material, where one can observe generally good agreement between the estimated sets


We note that simulation from the SPAR model is straightforward; a simulation scheme is given in the Supplementary Material, alongside examples for each of the metocean data sets.  





Inspection of the local (angular) QQ plots introduced in Section~Section \ref{subsec:diagnostics} suggests that the performance of the fitted SPAR model varies across different angular regions. Although we observe generally good agreement between quantiles, there is clearly better agreement for certain angles, suggesting rates of convergence to the GP tail model may vary over angle for these data sets.

\section{Discussion} \label{sec:discussion}
In this paper, we have introduced a novel inference framework for the SPAR model of \citet{Mackay2023}. We have explored the properties of this framework, and introduced practical tools for quantifying uncertainty and assessing goodness of fit. Furthermore, we have applied this framework to simulated and real data sets in Sections \ref{sec:sim_study} and \ref{sec:case_study}, with results indicating that the proposed framework captures joint tail behaviour across a wide range of data structures. Moreover, this framework has been recently applied in \citet{mackay2024}, where the authors show the SPAR model can accurately capture joint extremes of wind speeds and wave heights, and extreme response distributions for a variety of metocean data sets. Our proposed modelling framework is one of the first multivariate extreme value modelling techniques that can be applied without marginal transformation, offering an advantage over competing approaches and removing a significant degree of model variability.

Noting that the SPAR model has only been developed recently, this work is the first attempt to apply this modelling framework in practice, and it is likely that other inference approaches will follow. While our proposed framework performs well in general, we acknowledge there exist some shortcomings that could provide the motivation for future work. 

The results from Section \ref{sec:sim_study} indicate that the proposed angular density estimation framework from Section \ref{sec:angular_dens} performs poorly for some copulas in regions around the angular mode(s). However, we note that even with this caveat, the KD estimation framework appeared adequate for capturing the angular distribution for the observed data sets in Section \ref{sec:case_study}. Future work could explore whether using alternative angular density estimation approaches \citep[e.g.,][]{Gu1993,Randell2016} could further improve performance.  

Observe that for the AD copulas considered in Section \ref{sec:sim_study}, the true isodensity functions exhibit clear cusps, where the underlying GP scale function is non-differentiable. Such sections cannot be captured under the current framework, since the use of cyclic cubic splines for smooth estimation imposes differentiability at all angles. Future work could explore how such behaviour could be captured in the inference framework. For example, one could use a spline representation that allows for superimposed knots. Combined with a more general spline inference procedure, this alternative representation could allow for optimal estimation of both the number and locations of knots, while simultaneously giving cusps in the estimated SPAR functions \citep{hastie2009elements}.

From Section \ref{sec:case_study}, one can observe that for the estimated isodensity contours and return levels obtained using the $L1$ coordinate system, there exist distinct cusps at certain angles; these arise due to the square shape of $\mathcal{U}_1$. We acknowledge that such cusps are not realistic for practical applications, and consequently, estimates from the $L2$ coordinate system may be preferable in such settings.

When estimating the shape parameter functions in Section \ref{sec:case_study}, we did not impose any functional constraints, even though the variables considered must have finite lower and upper bounds, and hence cannot be in the domain of attraction of a GP distribution with a non-negative shape parameter. However, even without bounding the shape parameter, we note that across all of our model fits, the estimated shape functions were almost always homogeneously negative, indicating the proposed framework is flexible enough to detect the form of tail behaviour directly from the data. Future work could explore whether imposing physical constraints on the shape function improves the quality of model fits.

Following on from Section \ref{subsec:case_study_val}, it appears that having one non-exceedance probability for all angles may not be optimal for fitting the SPAR model in practice due to different rates of convergence at different angles. Exploring techniques for selecting and estimating threshold functions with varying rates of exceedance \citep[e.g.,][]{Northrop2011} remains an open area for future work. More generally, the results in Section~\ref{sec:sim_study} demonstrate that SPAR inference performs well for extremes from the known bivariate distributions considered. In Section~\ref{sec:case_study}, we demonstrate that SPAR inference generates physically reasonable estimates of extremes from real metocean data sets, and have provided diagnostic evidence that SPAR model fits are reasonable. It would be useful in future to compare the characteristics of SPAR estimates against those from competitor schemes, particularly those which require a two-stage inference of first estimating marginal extreme value models and transformation to some standard marginal scale of choice, followed by estimation of a dependence model on standard scale. Of course, these comparisons will only be possible for the intervals of the angular domain where the two-stage model is valid. In contrast, SPAR inference is useful on the full angular domain, using variables standardised to zero mean and unit variance.

In the current work, we have chosen to use particular approaches to estimate each of the angular and conditional radial models. The non-stationary extreme value literature provides a range of alternative representations, used routinely in environmental applications and in ocean engineering in particular (see e.g. \citealt{JnsEA15}, \citealt{Randell2016}, \citealt{Zanini2020}). Various software tools are also available for the task, including \cite{texmex} and \cite{TowEA24}.

As discussed in Sections~\ref{sec:angular_dens} and \ref{sec:sim_study}, SPAR inference involves the specification of various tuning parameters, regulating the characteristics of the angular and radial models estimated. In fact, a SPAR inference is computationally the same as a non-stationary extreme value inference. Indeed, studies (e.g. \citealt{JnsEA15}, \citealt{TndEA24}) have already been conducted to evaluate the relative performance of different representations for the tail of the conditional radial component, and its sensitivity to tuning parameter setting. Nevertheless, future studies are recommended to assess the sensitivity of SPAR inference to choice of tuning parameter.

We have restricted attention to the bivariate setting throughout this work. This decision was made for simplicity, as well as the fact many of the examples given in \citet{Mackay2023} are for bivariate vectors. Using the proposed inference techniques as a starting point, a natural avenue for future work would therefore be expanding the modelling framework to the general $d$-dimensional setting. 


A non-stationary SPAR model for the joint behaviour of extremes of variables $X,Y$ which are non-stationary with respect to angular covariate $\Theta$ can be constructed relatively straightforwardly. For example, we might adopt a SPAR representation ($(R,Q)|\Theta$, say) for $(X,Y)|\Theta$, and a 2-D basis representation for smooth functions on the $(Q,\Theta)$ angular domain (using e.g splines, see \citealt{Wood2003}, \citealt{Randell2016}, \citealt{Youngman2019}) and a generalised Pareto conditional tail for $R|(Q,\Theta)$. In an environmental context, this would be an appealing model for significant wave height and period, non-stationary with respect to wave direction.

Finally, we note that in \citet{Mackay2023}, the authors also derive a link between the SPAR model and the limit set representation for multivariate extremes. Specifically, the radius of the limit set at a fixed angle is given by the asymptotic shape parameter of the SPAR representation. We believe the inference approach we have proposed could be adapted for the estimation of limit sets, though additional care will be required given estimates obtained from finite sample sizes seldom equal limiting asymptotic quantities in practice.



\section*{Supplementary Material}
\begin{itemize}
    \item \textbf{Supplementary Material for ``Inference for multivariate extremes via a
    semi-parametric angular-radial model"}: File containing supporting figures and additional information about the REML procedures and simulation study. (.pdf file)
\end{itemize}

\section*{Declarations}
\subsection*{Ethical Approval}
Not Applicable
\subsection*{Availability of supporting data}
The data sets analysed in Section \ref{sec:case_study} are freely available online at \url{https://github.com/EC-BENCHMARK-ORGANIZERS/EC-BENCHMARK}. 
\subsection*{Competing interests}
The authors have no relevant financial or non-financial interests to disclose.
\subsection*{Funding}
This work was supported by EPSRC grant numbers EP/L015692/1 and EP/Y016297/1. 
\subsection*{Authors' contributions}
All authors contributed to the study conception and design. Model development was performed by Ed Mackay and Phil Jonathan. Material preparation and initial analysis were performed by Callum Murphy-Barltrop. The first draft of the manuscript was written by Callum Murphy-Barltrop and all authors commented on previous versions of the manuscript. All authors read and approved the final manuscript.


\subsection*{Acknowledgments}
This paper is based on work partly completed while Callum Murphy-Barltrop was part of the EPSRC funded STOR-i centre for doctoral training (EP/L015692/1). Ed Mackay was funded by the EPSRC Supergen Offshore Renewable Energy Hub, United Kingdom (EP/Y016297/1). We would like to thank Ben Youngman for his assistance with the \verb|evgam| package in the \verb|R| computing language.

\clearpage

\begin{center}
\textbf{\large Supplementary Material to `Inference for bivariate extremes via a semi-parametric angular-radial model'}
\end{center}
\setcounter{equation}{0}
\setcounter{figure}{0}
\setcounter{table}{0}
\setcounter{section}{0}
\makeatletter
\renewcommand{\theequation}{S\arabic{equation}}
\renewcommand{\thefigure}{S\arabic{figure}}
\renewcommand{\thesection}{S\arabic{section}}

\section{Additional information for Section 4.4.2}
In this section, we provide further details about the REML schemes used to estimate the GAM parameters associated with the threshold and parameter functions. We first consider the GAM formulation of the threshold function $\log(u_{\gamma}(q)) = g_{u_{\gamma}}(q)$, which we estimate via quantile regression techniques. For finite sample sizes, conditional quantile regression requires us to compute 
\begin{equation} \label{eqn:quant_reg_opt}
    \hat{\pmb{\beta}}_{u_{\gamma}} := \underset{\pmb{\beta}_{u_{\gamma}} \in \mathbb{R}^{k+1}}{\arg \min } \sum_{i=1}^n\left(\rho_\gamma\left( r_i-\exp(g_{u_{\gamma}}(q_i))\right)\right),
\end{equation}
where $\pmb{\beta}_{u_{\gamma}}$ denotes the spline coefficients associated with $g_{u_{\gamma}}(q)$, and $\rho_\gamma(x) = (\gamma - 1)x$ for $x < 0$ and $\gamma x$ for $x \geq 0$ \citep{Koenker2017}. However, optimisation of \eqref{eqn:quant_reg_opt} is non-trivial and computational issues often result due to the non-differentiability of $\rho_\gamma$ at zero.


To overcome these computational issues, \citet{Youngman2019} proposed the following mis-specified model 
\begin{equation*} 
    (R \mid \exp(g_{u_{\gamma}}(q)), \sigma(q),\gamma,c,Q=q) \sim \text{ALD}(\exp(g_{u_{\gamma}}(q)), \sigma(q),\gamma,c),
\end{equation*}
with corresponding density function 
\begin{equation*}
    f_{ALD}(r) = \frac{\gamma(1-\gamma)}{\sigma(q)}\exp \left\{ -\rho_{\gamma,c}\left( \frac{r - \exp(g_{u_{\gamma}}(q))}{\sigma(q)} \right) \right\}, \; r \in \RR,
\end{equation*}
where $\sigma(q)>0$ and $\rho_{\gamma,c}$ denotes the modified check function of \citet{Oh2011}. This check function is defined as
\begin{equation*}
    \rho_{\gamma, c}(x)= \begin{cases}(\gamma-1)(2 x+c) & \text { for } x<c \\ (1-\gamma) x^2 / c & \text { for }-c \leq x<0 \\ \gamma x^2 / c & \text { for } 0 \leq x<c \\ \gamma(2 x-c) & \text { for } c \leq x\end{cases}
\end{equation*}
for $c>0$, where $c$ is chosen close to zero. Unlike the standard loss function $\rho_{\gamma}$, $\rho_{\gamma, c}$ is differentiable at zero, offering significant advantages for inference in terms of speed and computational efficiency; see \citet{Oh2011} for further details. For our framework, we set $c = 0.5$; this is the default value suggested in \citet{Youngman2020}. 

Setting $\log (\sigma(q)) = g_{\sigma}(q)$, where $g_{\sigma}(q)$ denotes a GAM formulation from equation (4.5) of the main article, and letting $\pmb{\beta}_\sigma$ denote the associated coefficient vector, the log-likelihood associated with the mis-specified model given of equation \eqref{eqn:quant_reg_opt} is given by 
\begin{equation} \label{eqn:log_like_ALD}
    \ell(\pmb{\beta}_{u_{\gamma}}, \pmb{\beta}_\sigma ; \mathbf{r},\mathbf{q}, \gamma,c) \propto - \sum_{i=1}^n g_{\sigma}(q_i) - \sum_{i=1}^n  \rho_{\gamma,c}\left(\frac{r_i-\exp(g_{u_{\gamma}}(q_i))}{\exp(g_{\sigma}(q_i))}\right).
\end{equation}
Treating $\sigma(q)$ as a nuisance parameter function, maximisation of equation \eqref{eqn:log_like_ALD} with respect to $\pmb{\beta}_{u_{\gamma}}$ is equivalent to minimisation of equation \eqref{eqn:quant_reg_opt}. Consequently, the resulting estimate for $u_{\gamma}(q)$ gives an estimate of the conditional quantile function for $R_{q}$. For further details, along with detailed examples, see \citet{Yu2001} and \citet{Geraci2007}. Note that the same spline basis dimensions are used for both $u_{\gamma}$ and $\sigma$.  




Given an estimate of $u_{\gamma}(q)$, let $\mathcal{I}^o_{\gamma} := \{ i \in \{1,2,\dots,n\} \mid r_i \geq u_\gamma(q_i) \}$ denote the observed indices of threshold exceedances. Recalling the GAM formulations for the GP scale and shape, $\log(\tau(q)) = g_\tau(q)$ and $\xi(q) = g_\xi(q)$, the log-likelihood function of the GP tail model is given by
\begin{equation} \label{eqn:gp_ll}
    \ell(\pmb{\beta}_\tau, \pmb{\beta}_\xi; \mathbf{r}_{\mathcal{I}^o_{\gamma}},\mathbf{q}_{\mathcal{I}^o_{\gamma}}) = - \sum_{i \in \mathcal{I}^o_{\gamma}} g_\tau(q_i) + \sum_{i \in \mathcal{I}^o_{\gamma}} \left(-\frac{1}{g_\xi(q_i)} - 1\right) \log \left(1 + \frac{g_\xi(q_i) (r_i - u_{\gamma}(q_i)) }{\exp(g_\tau(q_i))} \right),
\end{equation}
where $\mathbf{r}_{\mathcal{I}^o_{\gamma}} := \{ r_i \mid i \in \mathcal{I}^o_{\gamma}\}$, $\mathbf{q}_{\mathcal{I}^o_{\gamma}} := \{ q_i \mid i \in \mathcal{I}^o_{\gamma}\}$, and $\pmb{\beta}_\tau, \pmb{\beta}_\xi$ denote the coefficient vectors associated with $g_\tau$ and $g_\xi$, respectively. Minimising equation \eqref{eqn:gp_ll} with respect to $\pmb{\beta}_\tau$ and $\pmb{\beta}_\xi$ results in estimates of the parameter functions. 

Minimisation of equations \eqref{eqn:log_like_ALD} and \eqref{eqn:gp_ll} is achieved via REML, with the corresponding penalty parameters estimated via cross validation. We refer to \citet{Wood2016} and \citet{Wood2017} for further details. 

\section{Additional information and plots for Section 6} \label{sec:sec6_additional}

In Section 6 of the main article, we consider four copula examples. The first is the Gaussian copula; given $\rho \in [-1,1]$, termed the Pearson correlation coefficient, this is given by 
\[
C(u_1, u_2; \rho) = \boldsymbol{\Phi} \left(\Phi^{-1}(u_1), \Phi^{-1}(u_2); \Sigma\right), \; \Sigma := \left(\begin{matrix}
    1 & \rho \\
    \rho & 1
\end{matrix} \right)
\]
where $\boldsymbol{\Phi}$ is the bivariate Gaussian cumulative distribution function with correlation matrix $\Sigma$ and $\Phi^{-1}$ is the inverse of the standard univariate Gaussian cumulative distribution function. The parameter $\rho$ controls the form and strength of dependence. 

Secondly, for any $\alpha \neq 0$, the Frank copula is defined as 
\[
C(u_1, u_2; \alpha) = -\frac{1}{\alpha} \log\left(1 + \frac{\prod_{i=1}^{2} (e^{-\alpha u_i} - 1)}{e^{-\alpha} - 1}\right),
\]
with the parameter $\alpha$ controlling the strength and form of dependence.

Next, for any $\rho \in [-1,1]$ and $\nu>0$, the t copula is given by
\[
C(u_1, u_2; \rho,\nu) = T\left(t^{-1}_\nu(u_1), t^{-1}_\nu(u_2); \Sigma,\nu\right)
\]
where $T$ is the bivariate t cumulative distribution function with correlation matrix $\Sigma$ and $\nu$ degrees of freedom, and $t^{-1}_\nu$ is the inverse of the univariate t cumulative distribution function with $\nu$ degrees of freedom. The strength of dependence in the tails is controlled by both $\nu$ and $\rho$ \citep{chan2008tail}.

Finally, given any $\alpha>0$, the Joe copula is given by 
\[
C(u_1, u_2; \alpha) = 1 - \left(1 - \sum_{i=1}^{2} u_i^\alpha\right)^\frac{1}{\alpha}
\]
where $\alpha > 0$ controlling the strength of dependence.

For each copula, we simulate data on standard Laplace margins, for which the marginal distribution is given by 
\begin{equation} \label{eqn:stan_lap}
 F_L(x) := \frac{1}{2} \left(1 + \text{sgn}(x) \cdot \left(1 - e^{-|x|}\right)\right), \; x \in \RR.   
\end{equation}
To achieve this, we first simulate data from each copula on standard uniform margins using the \verb|copula| package in the \verb|R| computing language. We then transform this data to standard Laplace using the inverse of equation \eqref{eqn:stan_lap}. 

\begin{figure}[htp]
    \centering
    \begin{subfigure}[b]{0.21\textwidth}
        \centering
        \includegraphics[width=\textwidth]{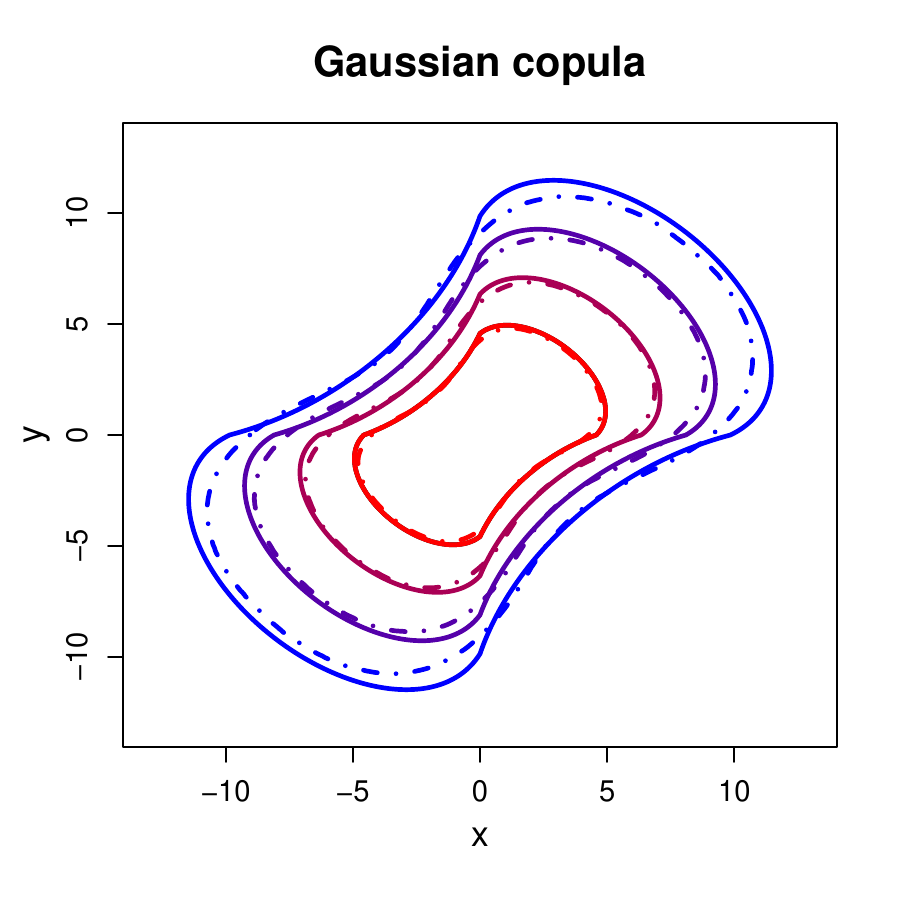}
    \end{subfigure}
    \quad
    \begin{subfigure}[b]{0.21\textwidth}  
        \centering 
        \includegraphics[width=\textwidth]{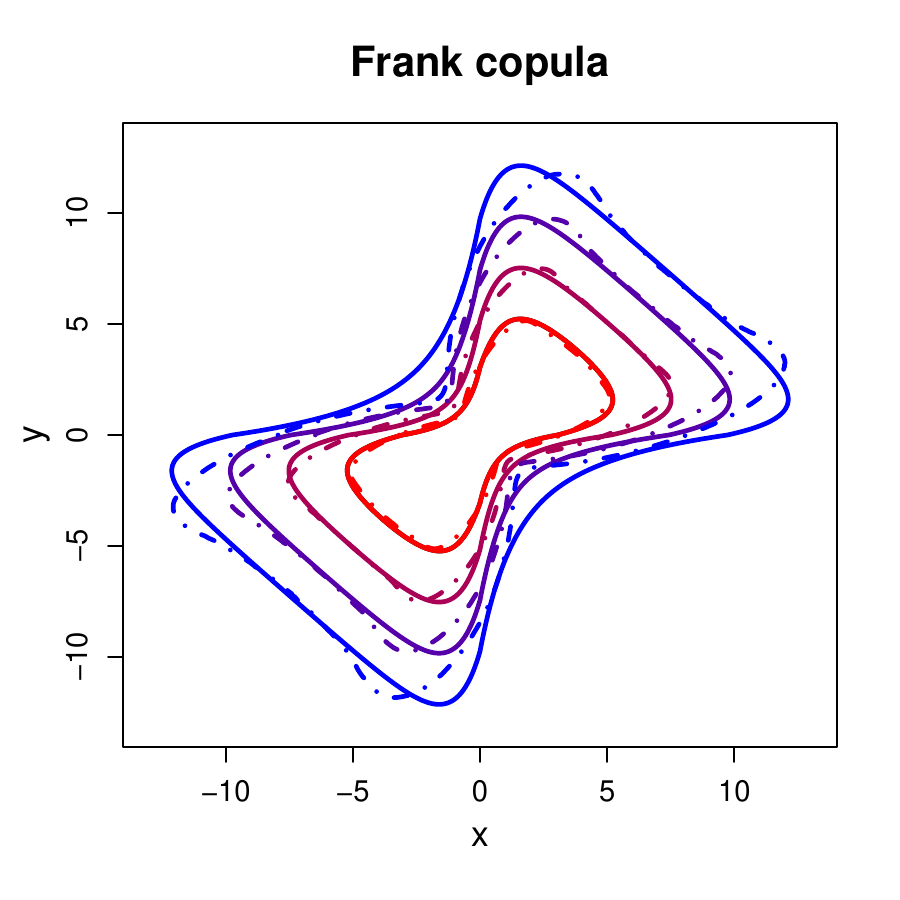}
    \end{subfigure}
    \quad
    \begin{subfigure}[b]{0.21\textwidth}   
        \centering 
        \includegraphics[width=\textwidth]{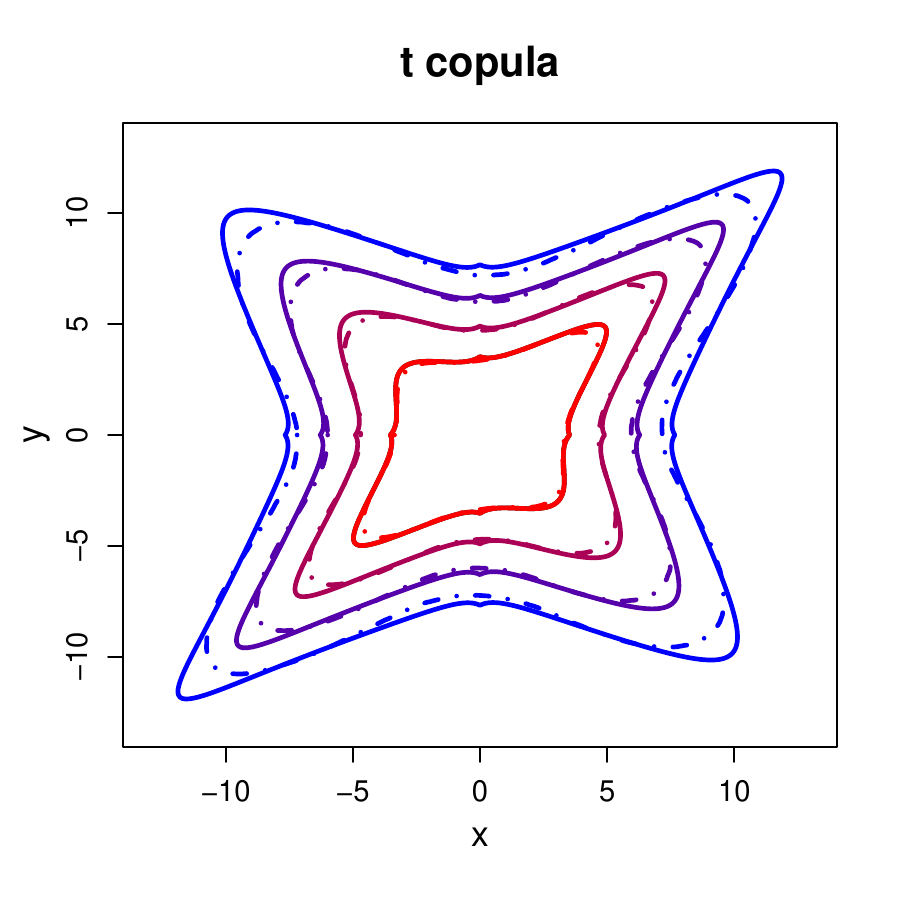}
    \end{subfigure}
    \quad
    \begin{subfigure}[b]{0.21\textwidth}   
        \centering 
        \includegraphics[width=\textwidth]{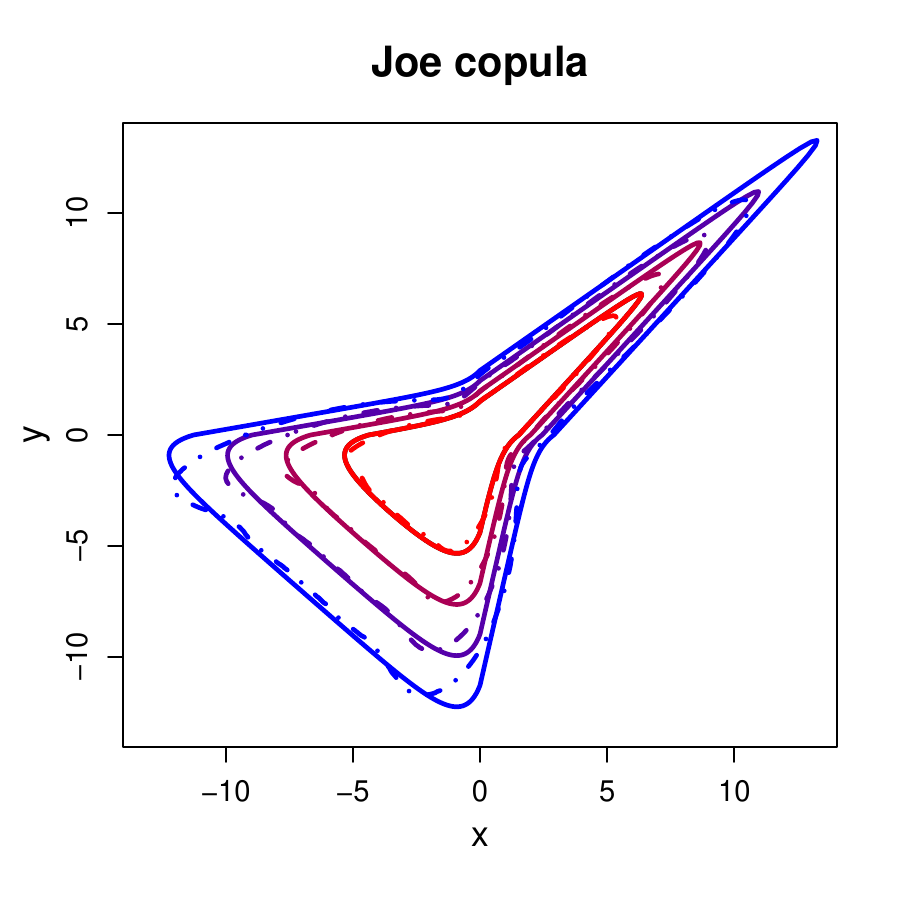}
    \end{subfigure}
    \caption{Comparison of true (thick lines) and median estimated (dot-dashed lines) isodensity contours under the $L2$ coordinate system. In each plot, the red to blue lines represent the joint density levels $p \in \{10^{-3},10^{-4},10^{-5},10^{-6} \}$.}
    \label{fig:equidensity_L2}
\end{figure}

Figure \ref{fig:equidensity_L2} compares the median estimates of isodensity contours, obtained using the $L2$ coordinate system, to the true contours at a range of low density levels.

\begin{figure}[H]
    \centering
    \begin{subfigure}[b]{0.21\textwidth}
        \centering
        \includegraphics[width=\textwidth]{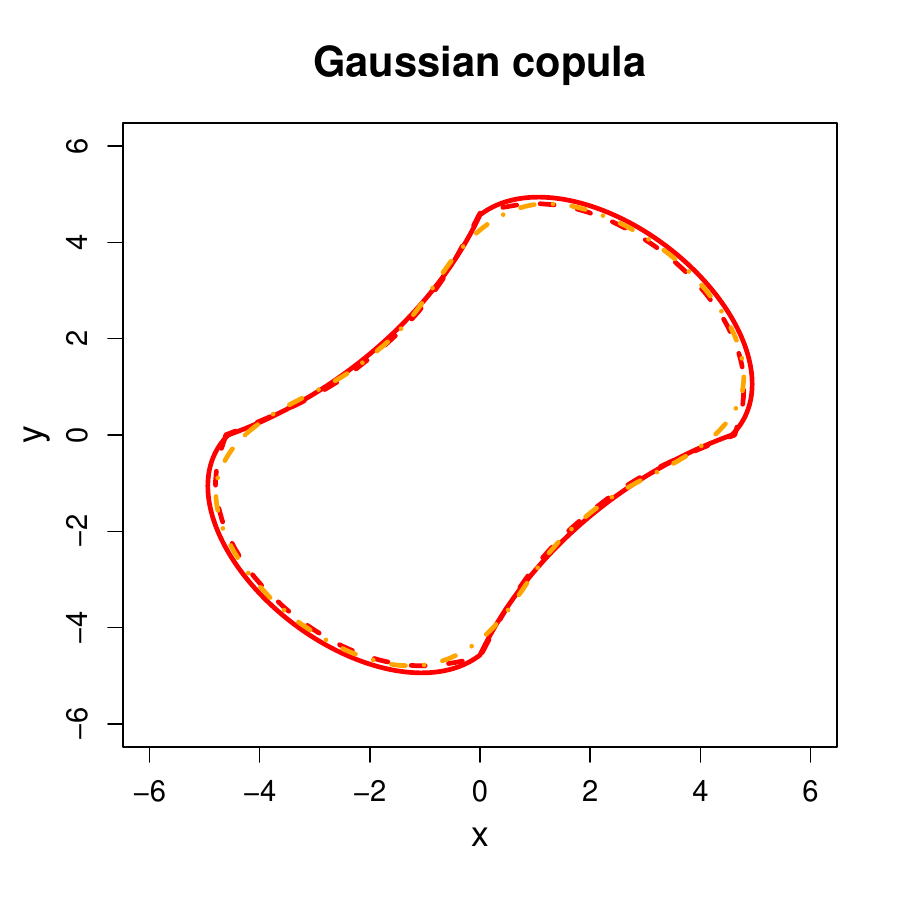}
    \end{subfigure}
    \quad
    \begin{subfigure}[b]{0.21\textwidth}  
        \centering 
        \includegraphics[width=\textwidth]{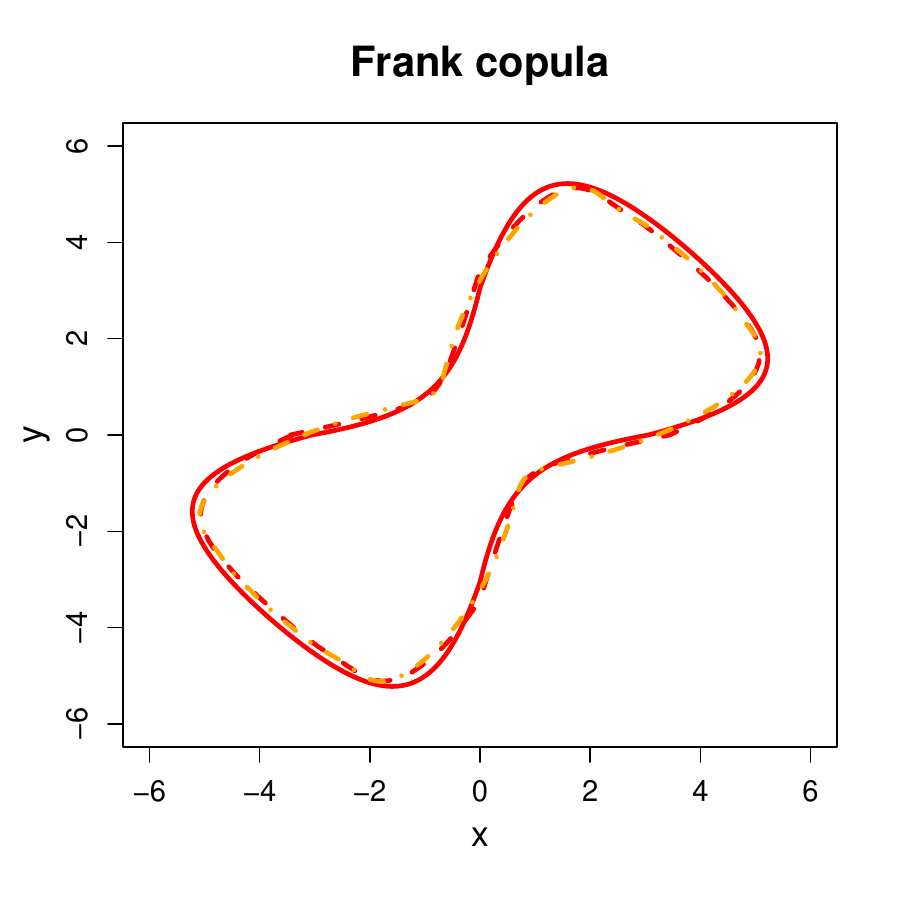}
    \end{subfigure}
    \quad
    \begin{subfigure}[b]{0.21\textwidth}   
        \centering 
        \includegraphics[width=\textwidth]{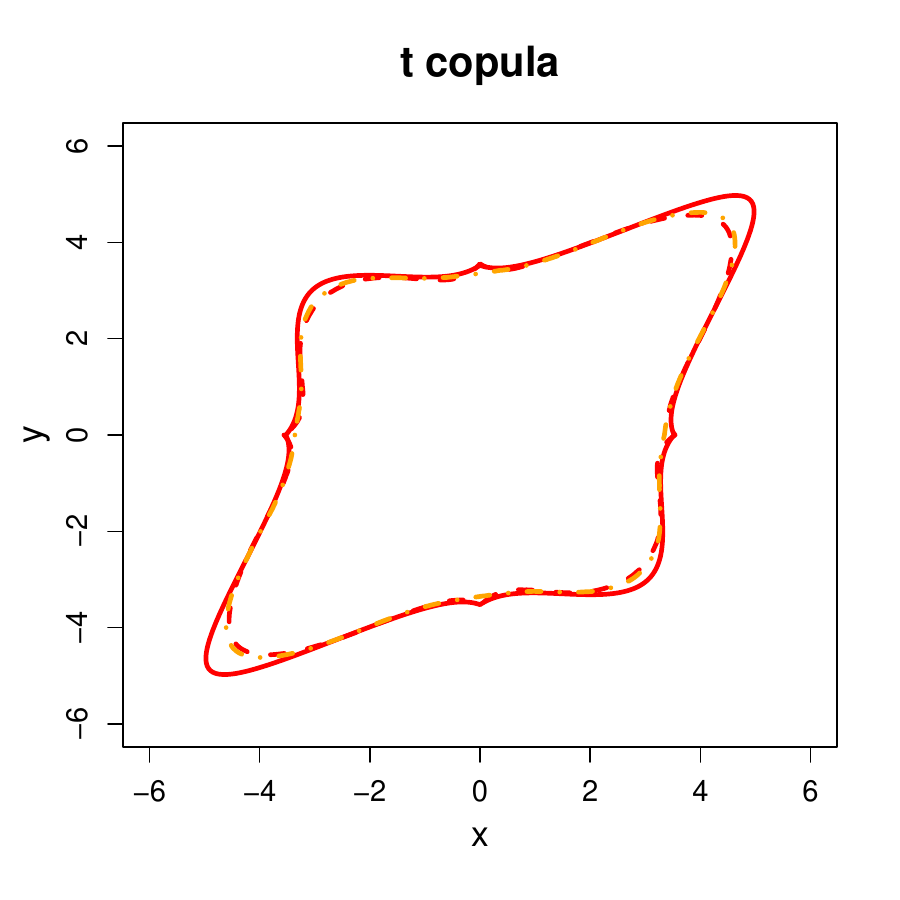}
    \end{subfigure}
    \quad
    \begin{subfigure}[b]{0.21\textwidth}   
        \centering 
        \includegraphics[width=\textwidth]{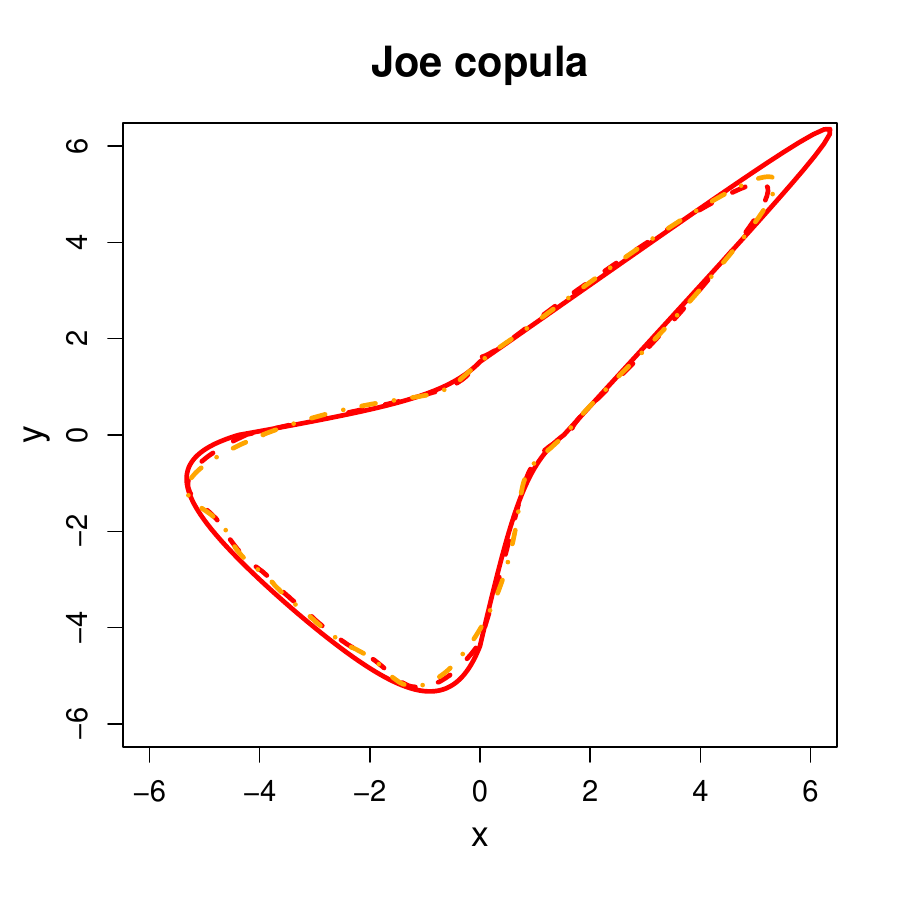}
    \end{subfigure}
    \vskip\baselineskip
    \begin{subfigure}[b]{0.21\textwidth}
        \centering
        \includegraphics[width=\textwidth]{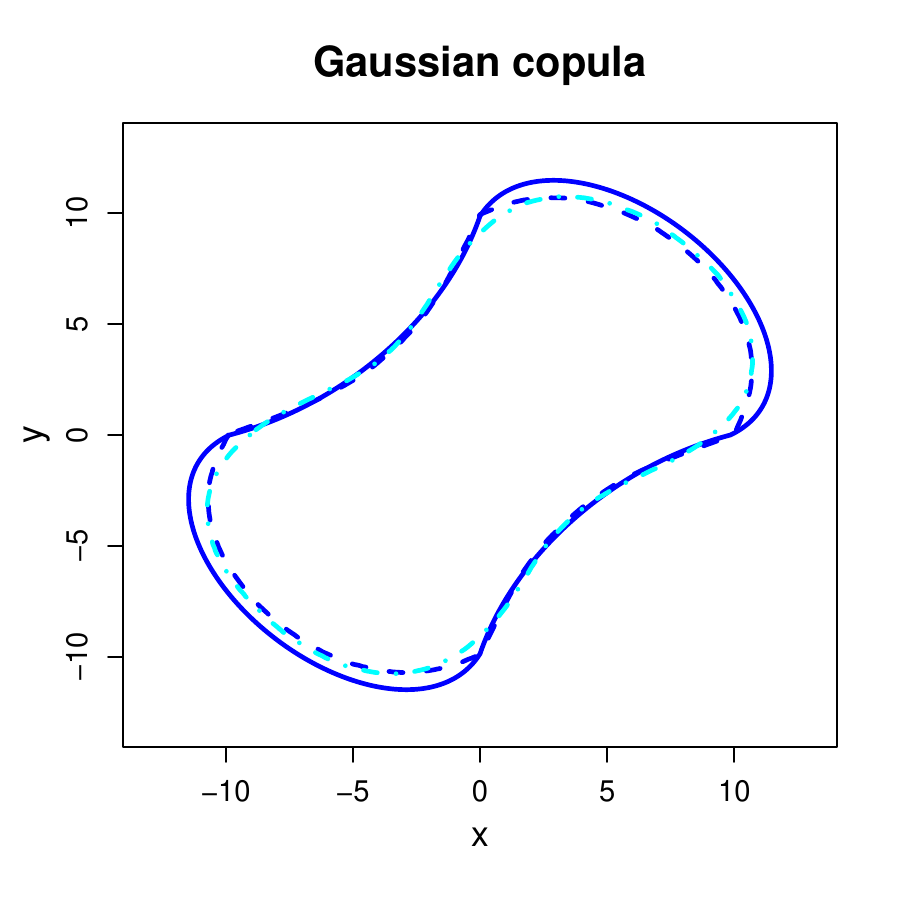}
    \end{subfigure}
    \quad
    \begin{subfigure}[b]{0.21\textwidth}  
        \centering 
        \includegraphics[width=\textwidth]{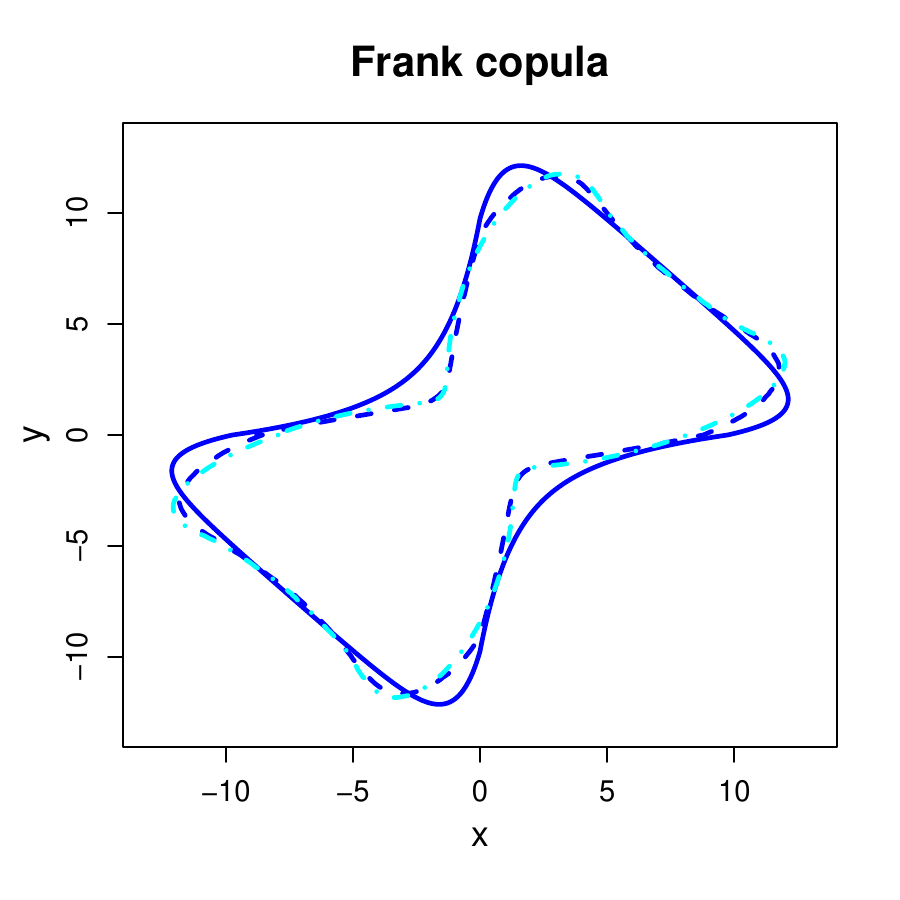}
    \end{subfigure}
    \quad
    \begin{subfigure}[b]{0.21\textwidth}   
        \centering 
        \includegraphics[width=\textwidth]{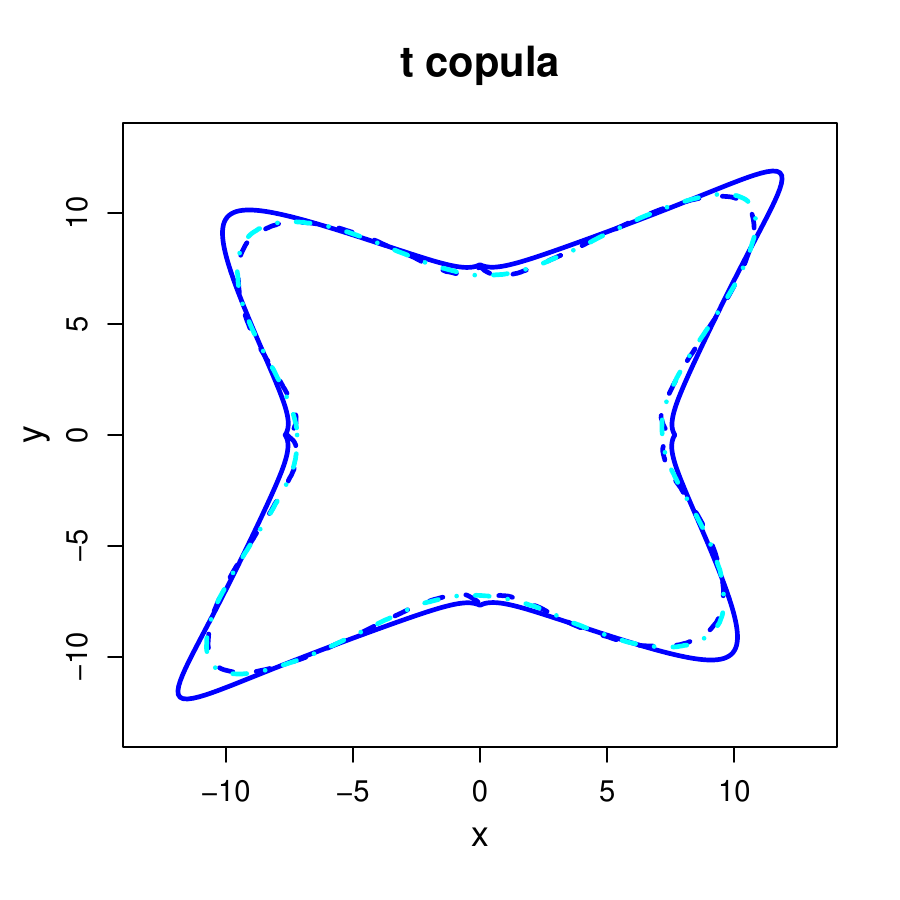}
    \end{subfigure}
    \quad
    \begin{subfigure}[b]{0.21\textwidth}   
        \centering 
        \includegraphics[width=\textwidth]{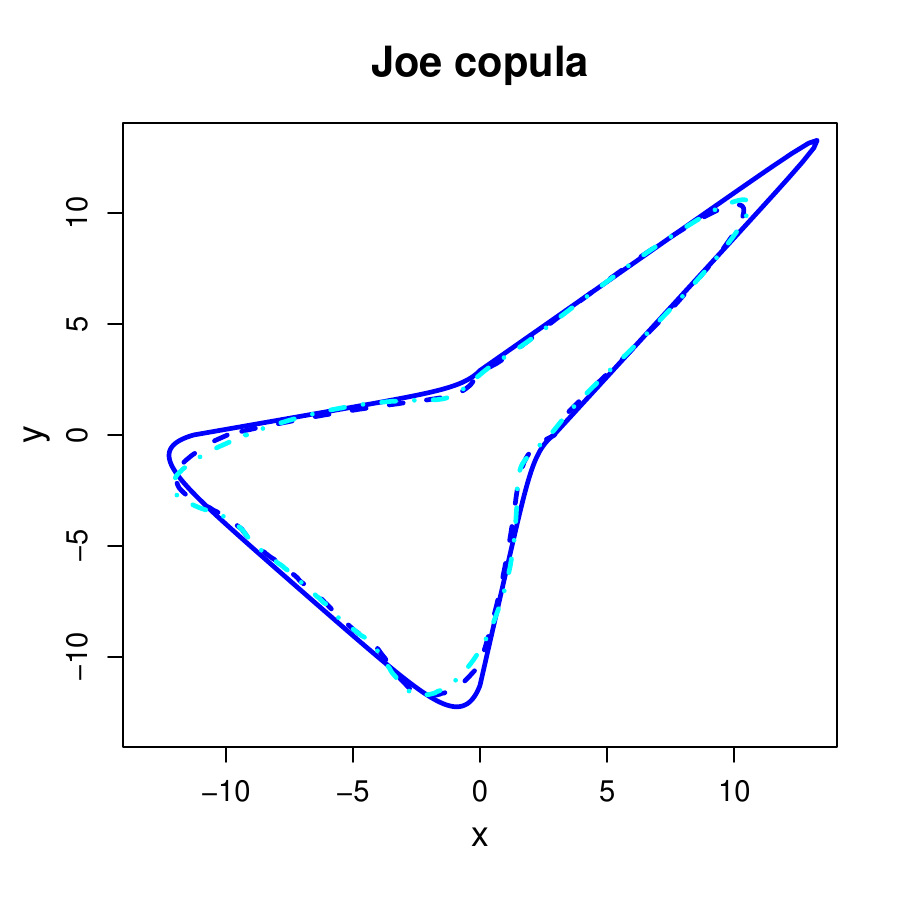}
    \end{subfigure}
    
    \caption{Comparison of estimated median isodensity contours to truth (solid lines) for the $L1$ (dashed lines) and $L2$ (dot-dashed lines) coordinate systems. The top and bottom rows give the comparisons for $p=10^{-3}$ and $p=10^{-6}$, respectively.}
    \label{fig:comparisons_angular_systems}
\end{figure}

Figure \ref{fig:comparisons_angular_systems} compares the median isodensity contours estimates from the two coordinate systems for two density levels $p \in \{ 10^{-3}, 10^{-6} \}$. One can observe that the differences between the sets of median estimates are negligible. 

\begin{figure}[H]
    \centering
    \begin{subfigure}[b]{0.21\textwidth}
        \centering
        \includegraphics[width=\textwidth]{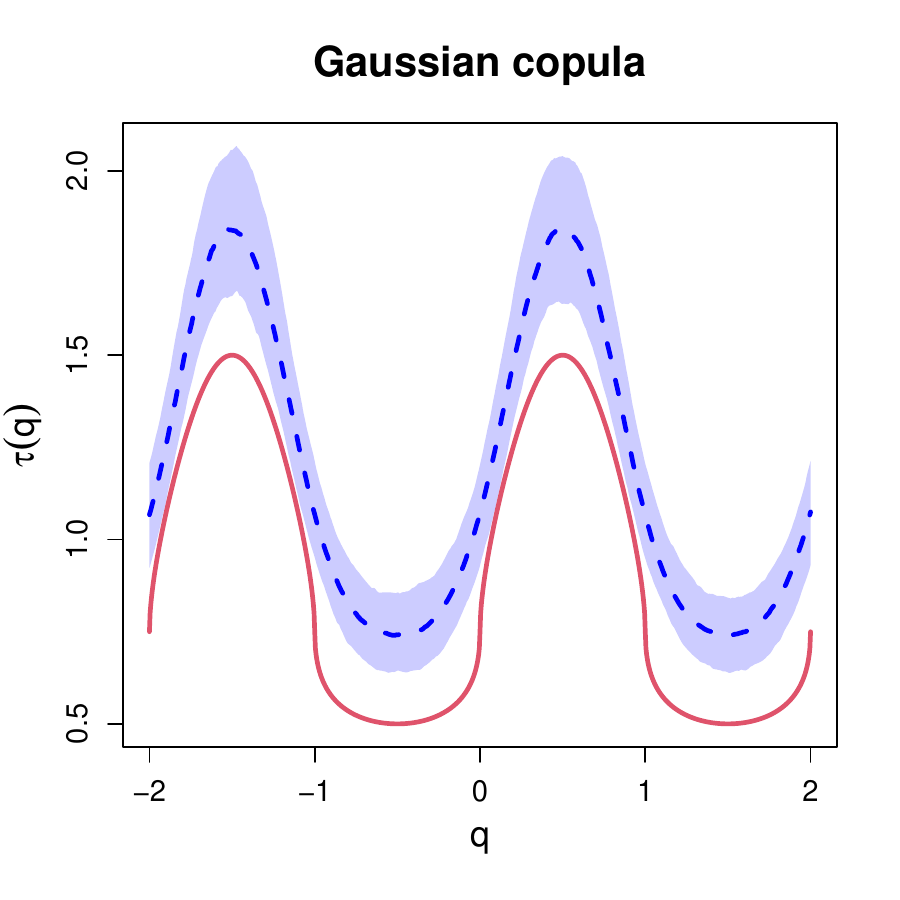}
    \end{subfigure}
    \quad
    \begin{subfigure}[b]{0.21\textwidth}  
        \centering 
        \includegraphics[width=\textwidth]{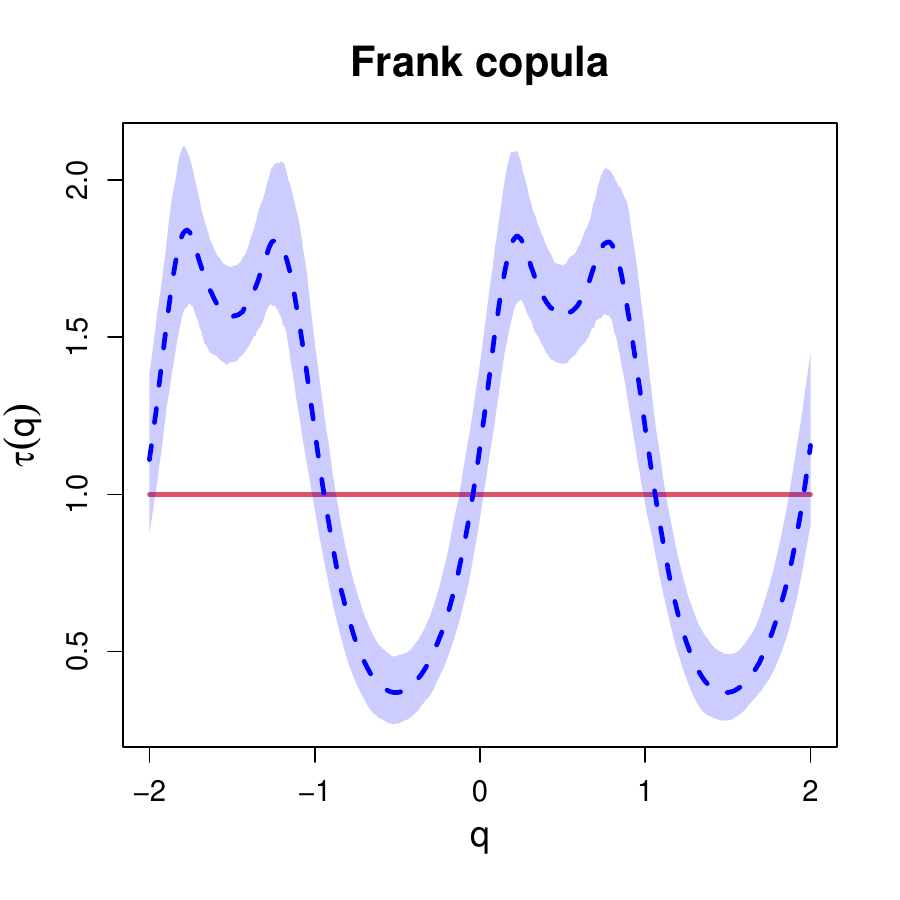}
    \end{subfigure}
    \quad
    \begin{subfigure}[b]{0.21\textwidth}   
        \centering 
        \includegraphics[width=\textwidth]{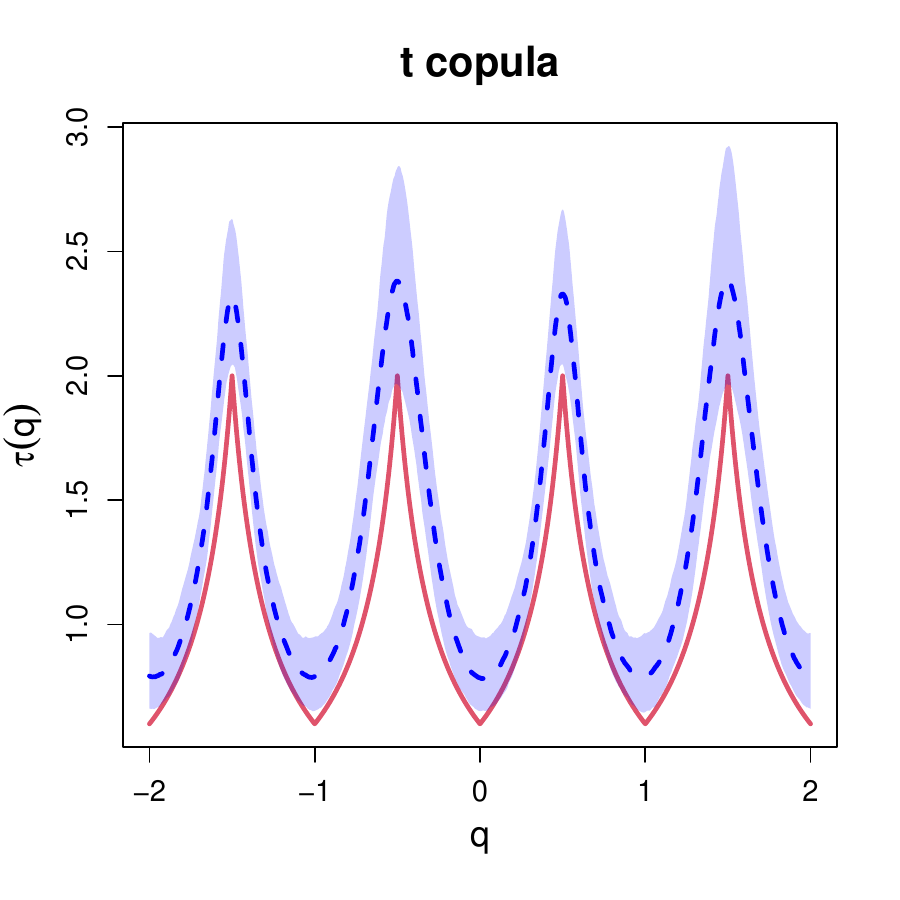}
    \end{subfigure}
    \quad
    \begin{subfigure}[b]{0.21\textwidth}   
        \centering 
        \includegraphics[width=\textwidth]{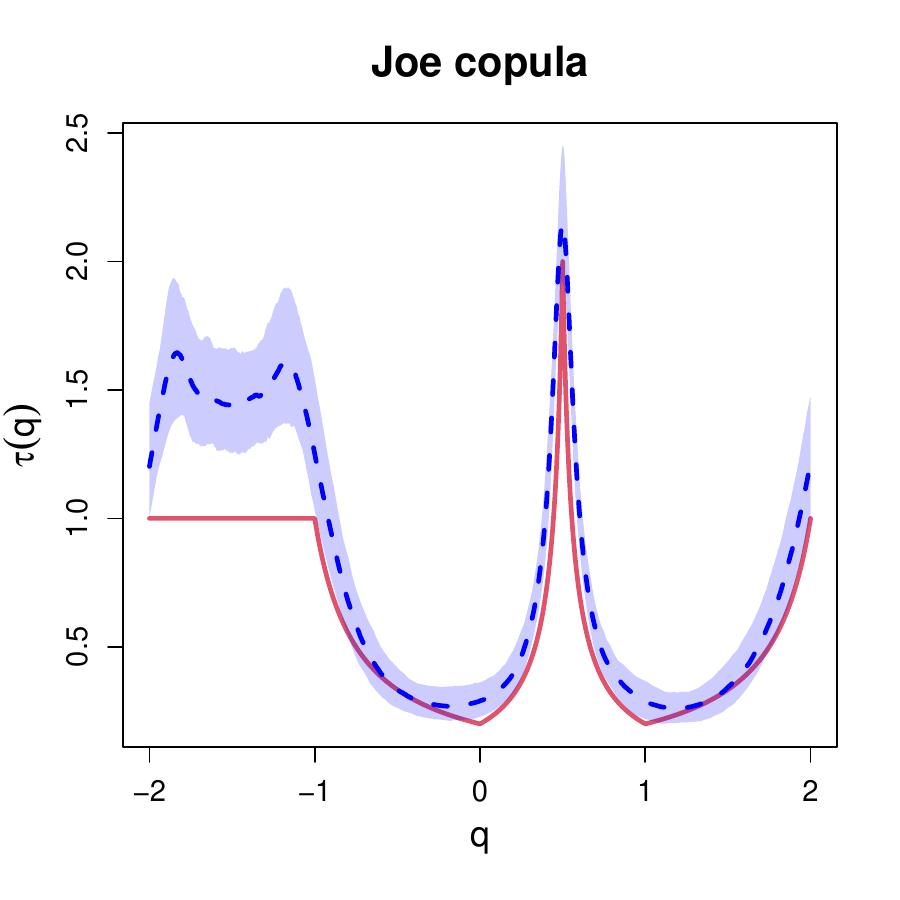}
    \end{subfigure}
    
    \caption{Comparison of true (solid red lines) and median estimated (dashed blue lines) scale parameter functions under the $L1$ coordinate system. In each plot, the shaded regions give the estimated 95\% confidence intervals.}
    \label{fig:sigma_ests_L1}
\end{figure}

\begin{figure}[H]
    \centering
    \begin{subfigure}[b]{0.21\textwidth}
        \centering
        \includegraphics[width=\textwidth]{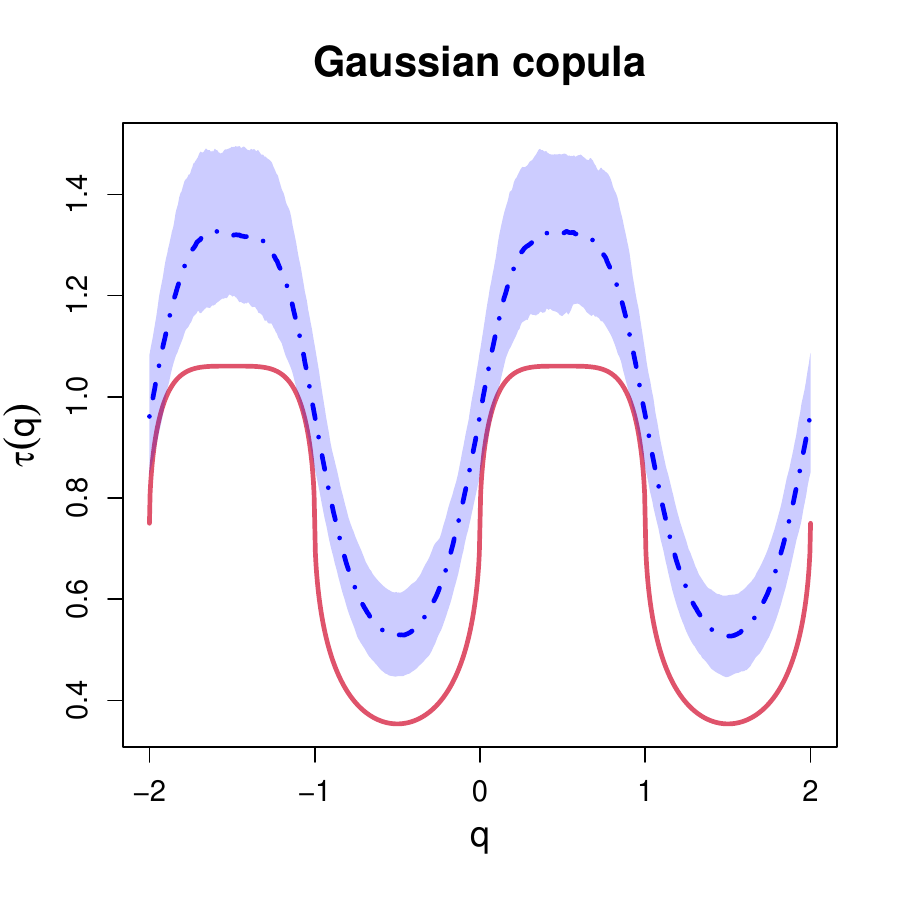}
    \end{subfigure}
    \quad
    \begin{subfigure}[b]{0.21\textwidth}  
        \centering 
        \includegraphics[width=\textwidth]{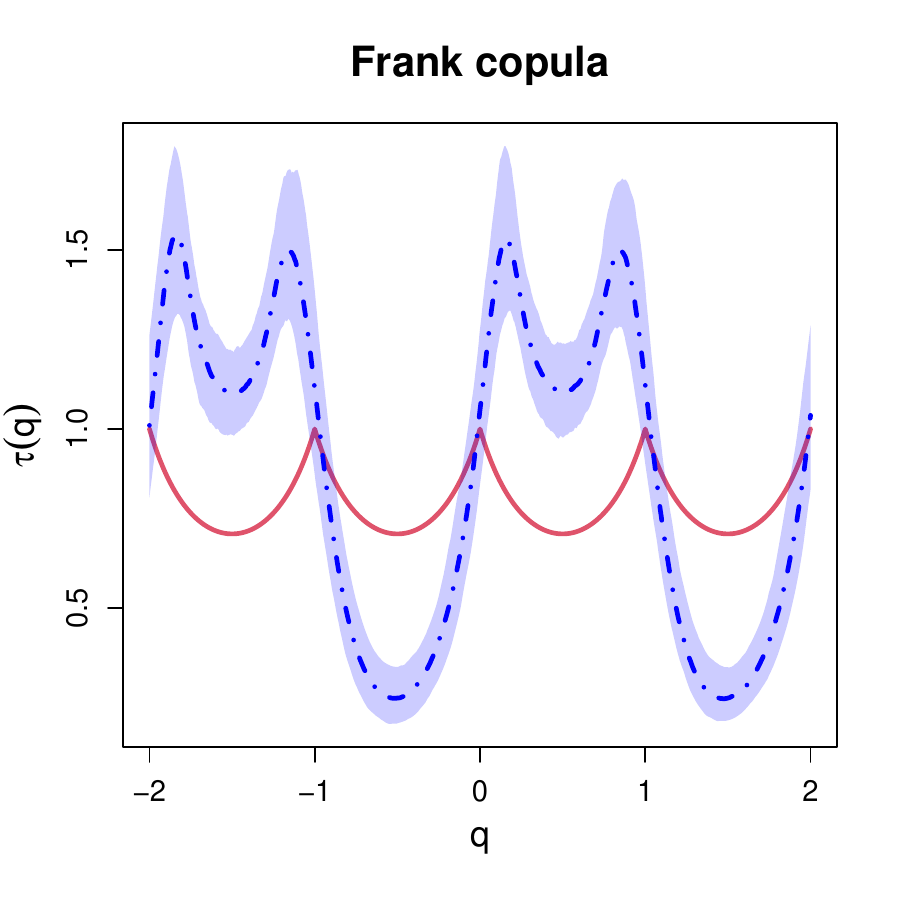}
    \end{subfigure}
    \quad
    \begin{subfigure}[b]{0.21\textwidth}   
        \centering 
        \includegraphics[width=\textwidth]{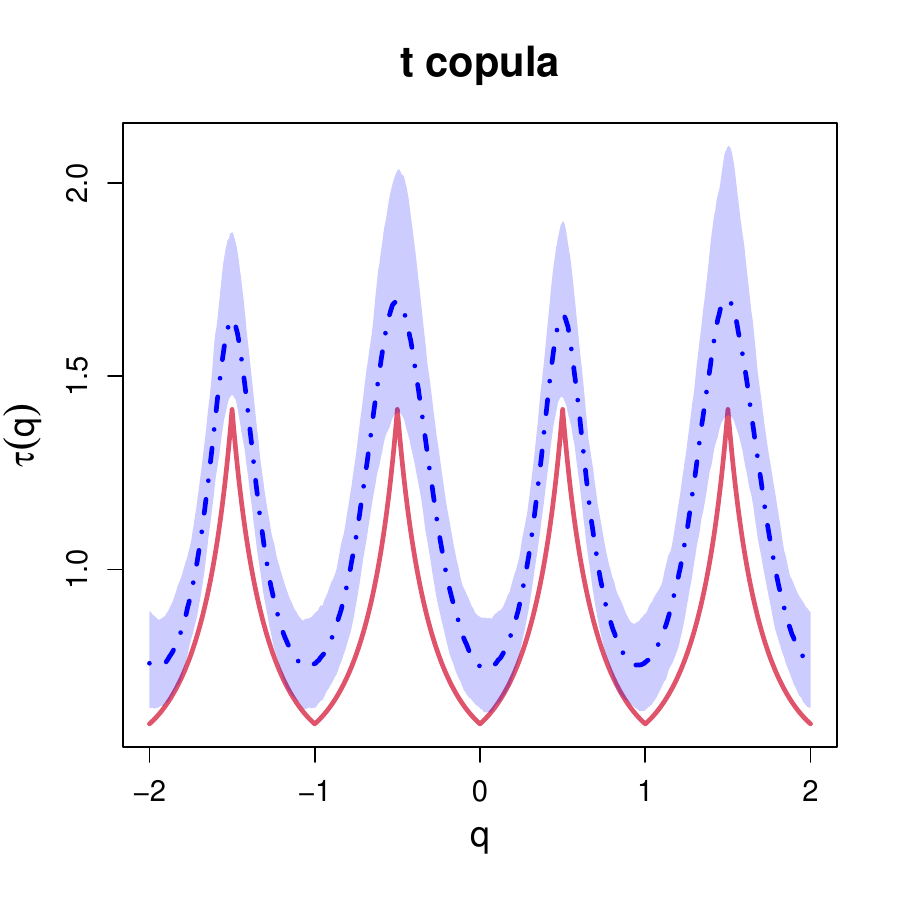}
    \end{subfigure}
    \quad
    \begin{subfigure}[b]{0.21\textwidth}   
        \centering 
        \includegraphics[width=\textwidth]{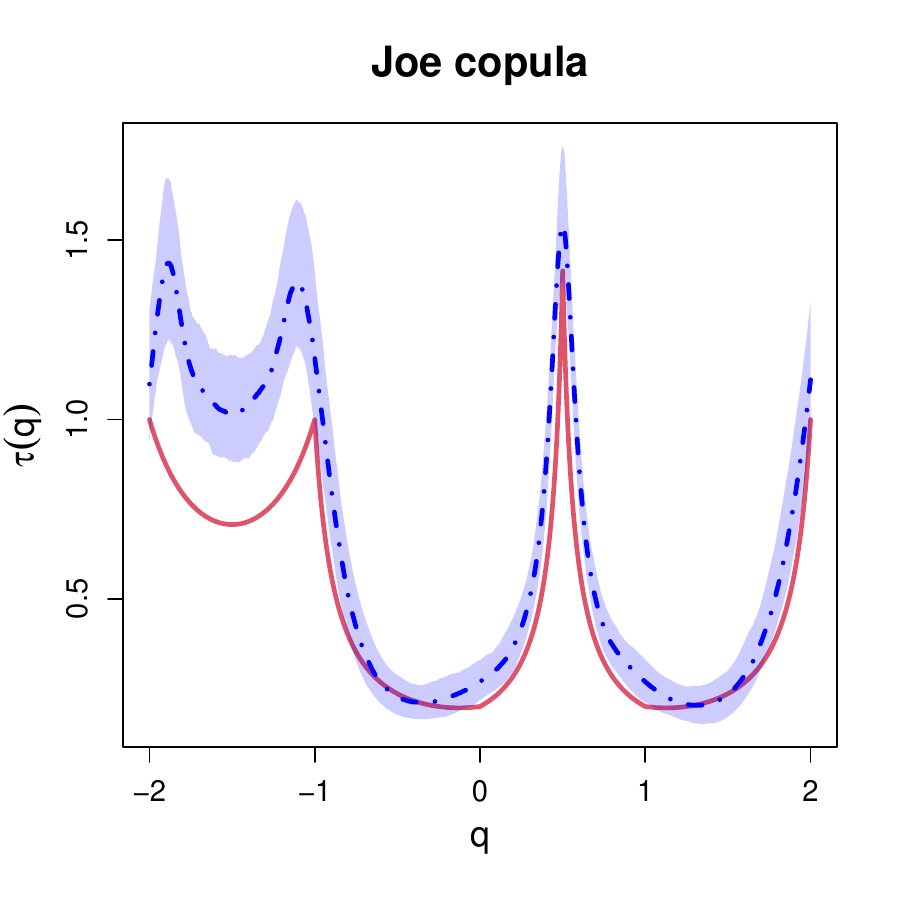}
    \end{subfigure}
    
    \caption{Comparison of true (solid red lines) and median estimated (dot-dashed blue lines) scale parameter functions under the $L2$ coordinate system. In each plot, the shaded regions give the estimated 95\% confidence intervals.}
    \label{fig:sigma_ests_L2}
\end{figure}

Figures \ref{fig:sigma_ests_L1} and \ref{fig:sigma_ests_L2} illustrate the median scale parameter estimates, with confidence intervals, for the $L1$ and $L2$ coordinate systems, respectively. Considering the fact estimates were computed using finite sample sizes, these plots illustrate reasonable agreement between the estimated and asymptotic scale parameter functions in most cases. One notable exception is the Frank copula, for which the estimated scale functions perform poorly. This is likely due to the relatively slow convergence of this distribution to its asymptotic form, as discussed in \citet{Mackay2023}. Furthermore, the majority of the $\xi$ estimates obtained over the simulated data sets are slightly negative. This discrepancy from the asymptotic value ($\xi = 0$) partly explains why the estimated scale functions estimates are biased high in most cases, since the scale and shape parameters are negatively correlated under the maximum likelihood framework \citep{hosking1987}. 
%


\begin{figure}[H]
    \centering
    \begin{subfigure}[b]{0.21\textwidth}
        \centering
        \includegraphics[width=\textwidth]{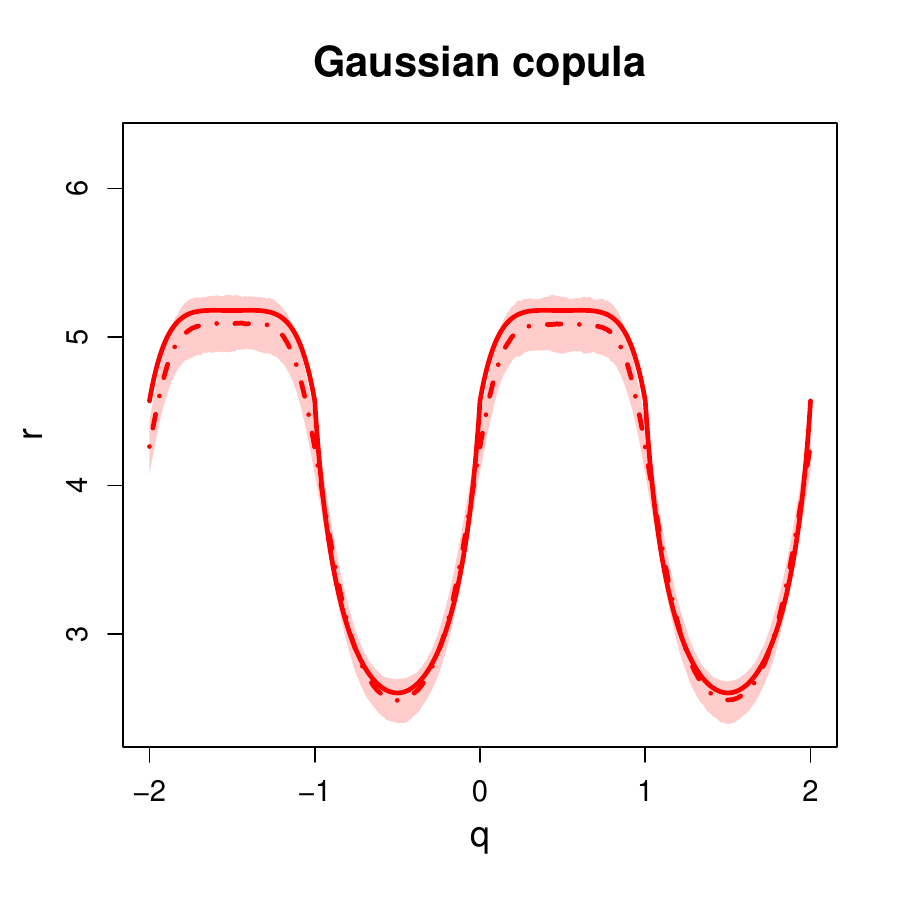}
    \end{subfigure}
    \quad
    \begin{subfigure}[b]{0.21\textwidth}  
        \centering 
        \includegraphics[width=\textwidth]{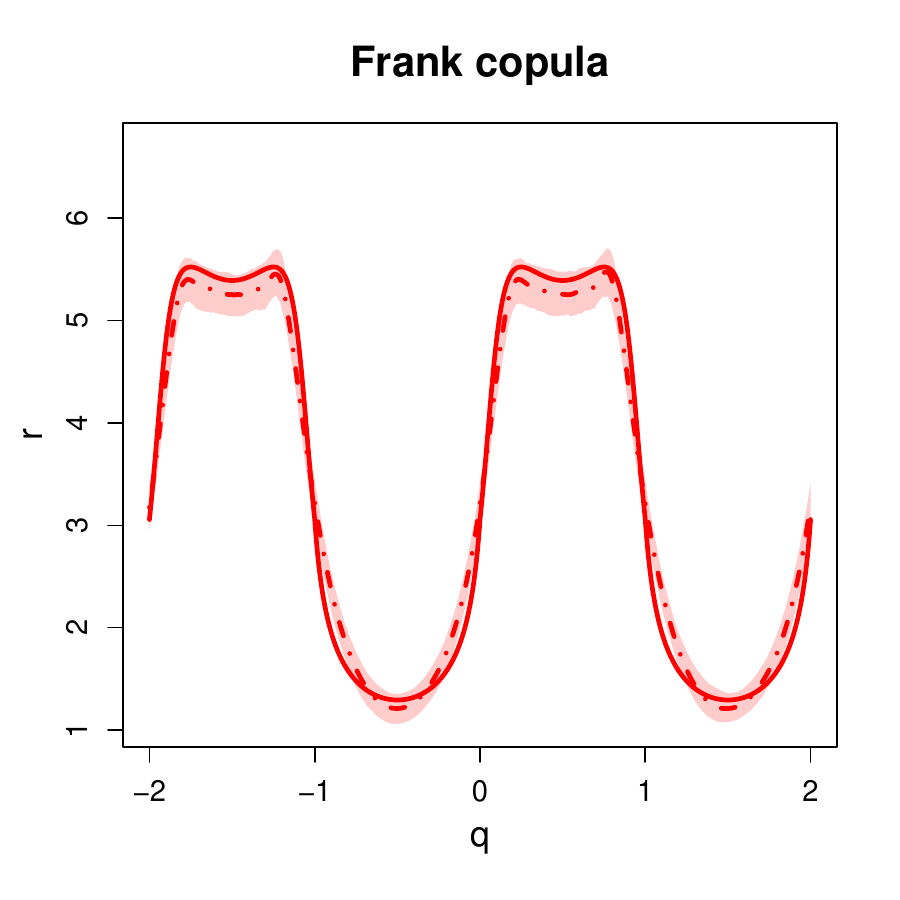}
    \end{subfigure}
    \quad
    \begin{subfigure}[b]{0.21\textwidth}   
        \centering 
        \includegraphics[width=\textwidth]{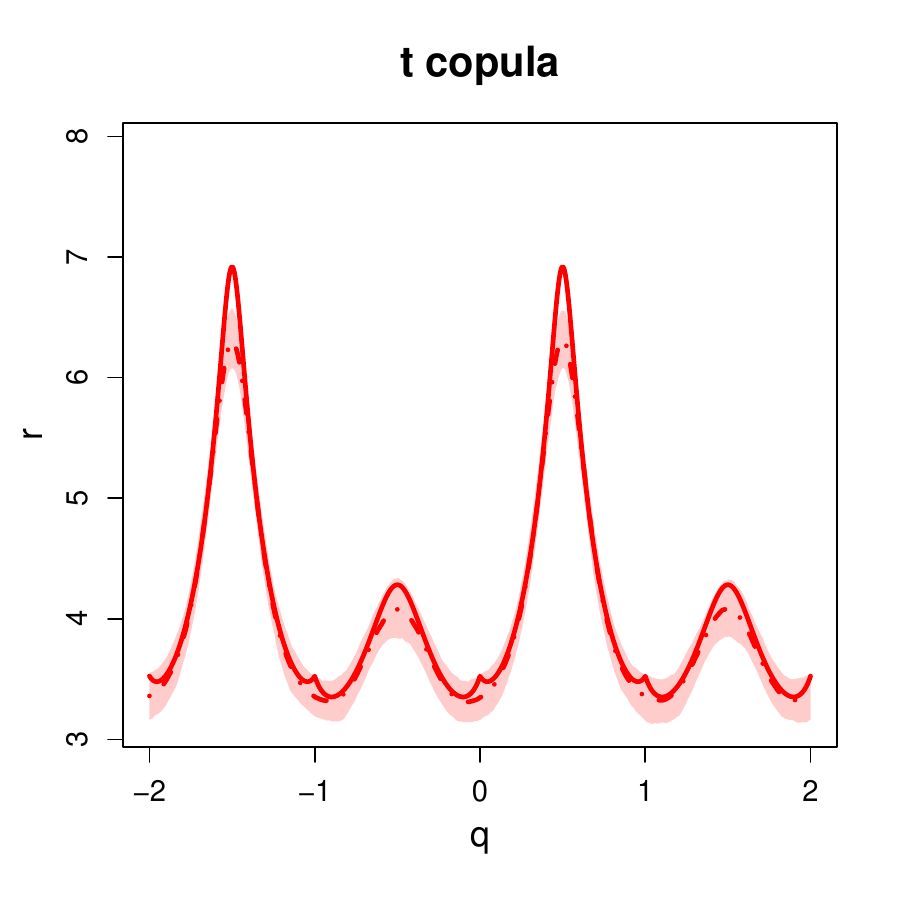}
    \end{subfigure}
    \quad
    \begin{subfigure}[b]{0.21\textwidth}   
        \centering 
        \includegraphics[width=\textwidth]{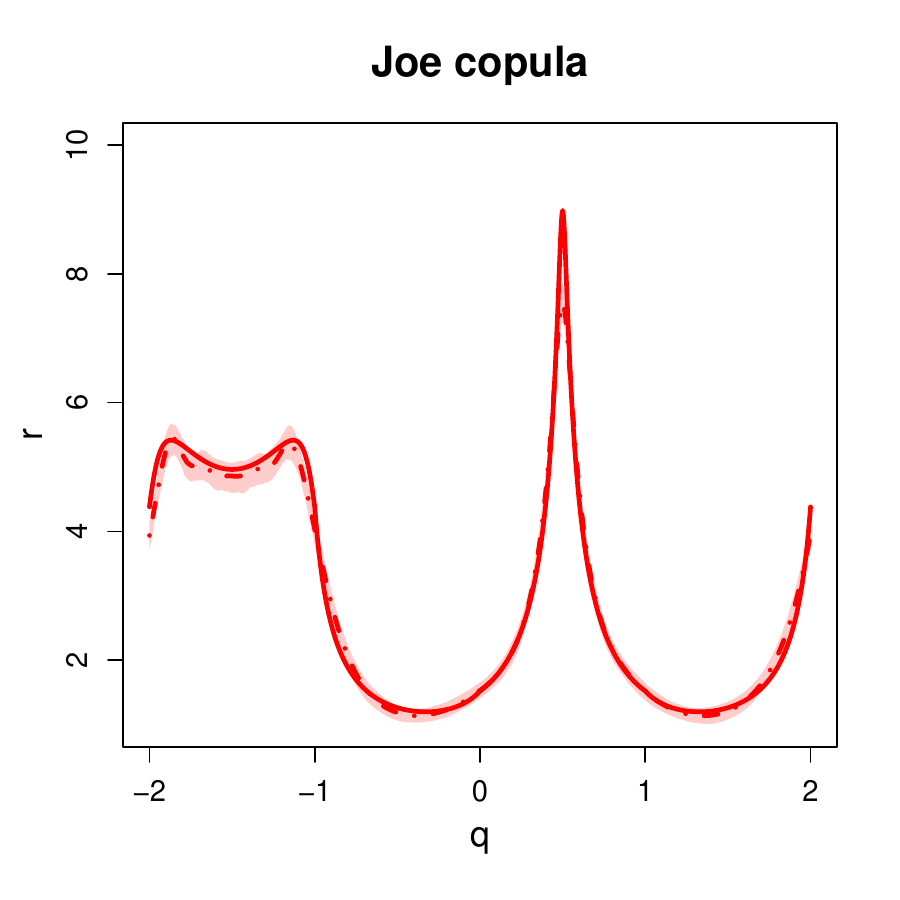}
    \end{subfigure}
    \vskip\baselineskip
       \begin{subfigure}[b]{0.21\textwidth}
        \centering
        \includegraphics[width=\textwidth]{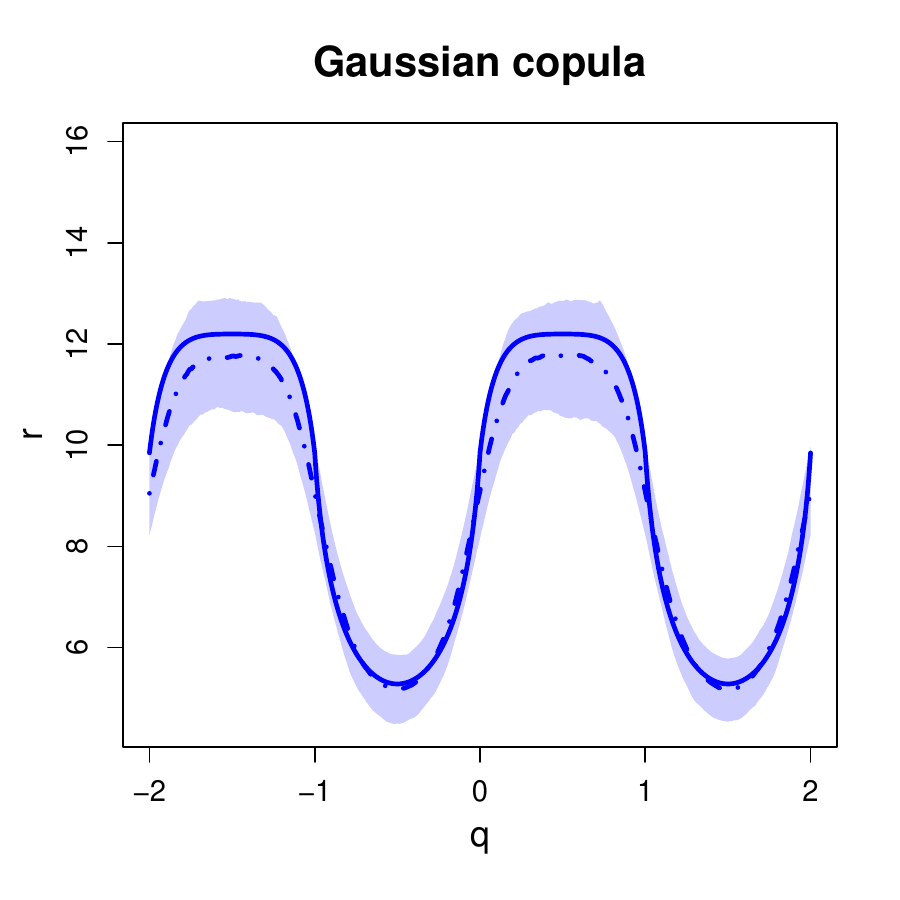}
    \end{subfigure}
    \quad
    \begin{subfigure}[b]{0.21\textwidth}  
        \centering 
        \includegraphics[width=\textwidth]{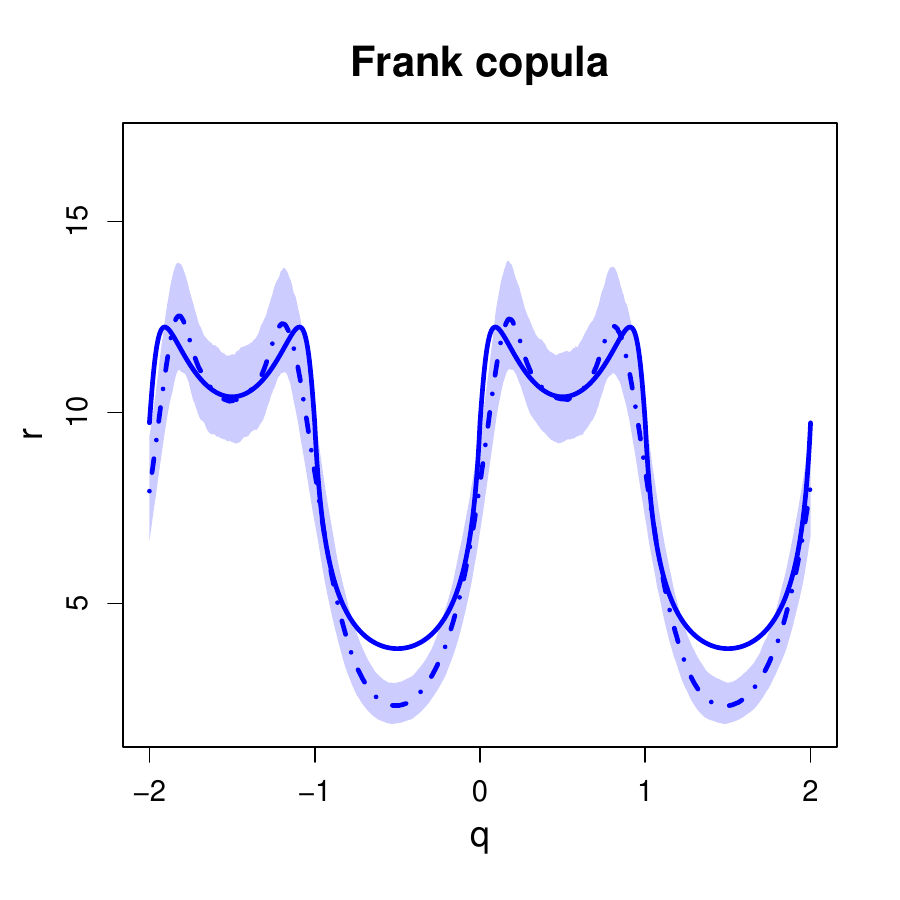}
    \end{subfigure}
    \quad
    \begin{subfigure}[b]{0.21\textwidth}   
        \centering 
        \includegraphics[width=\textwidth]{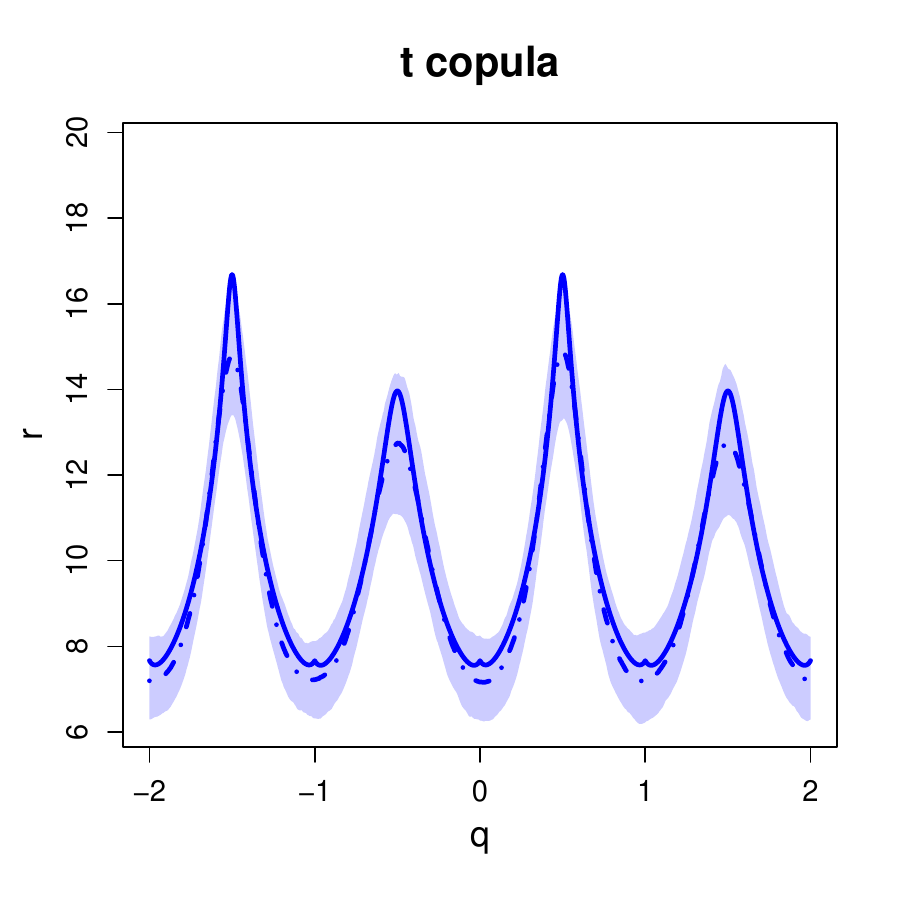}
    \end{subfigure}
    \quad
    \begin{subfigure}[b]{0.21\textwidth}   
        \centering 
        \includegraphics[width=\textwidth]{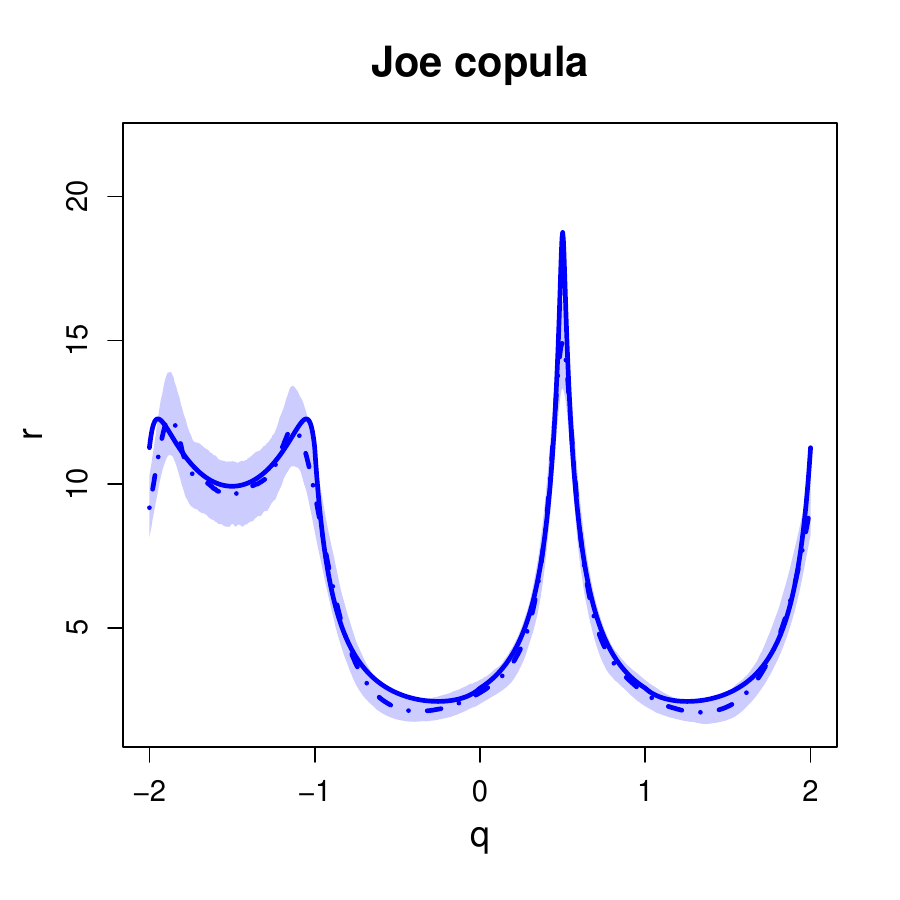}
    \end{subfigure}
    
    \caption{Comparison of median estimated isodensity contours (dashed lines), with 95\% confidence intervals (shaded region), to true contours (solid lines) for joint density level $p=10^{-3}$ (top row) and $p=10^{-6}$ (bottom row), with estimates obtained using $L2$ polar coordinates.}
    \label{fig:uncertainty_equidensity_L2}
\end{figure}

Figure \ref{fig:uncertainty_equidensity_L2} illustrates the median contour estimates for $p \in \{10^{-3},10^{-6} \}$ on the radial-angular scale for the $L2$ coordinate system, along with estimated 95\% confidence intervals. As with the $L1$ coordinates, the estimated confidence intervals capture the true contours in most cases. 

\begin{figure}[H]
    \centering
    \begin{subfigure}[b]{0.21\textwidth}
        \centering
        \includegraphics[width=\textwidth]{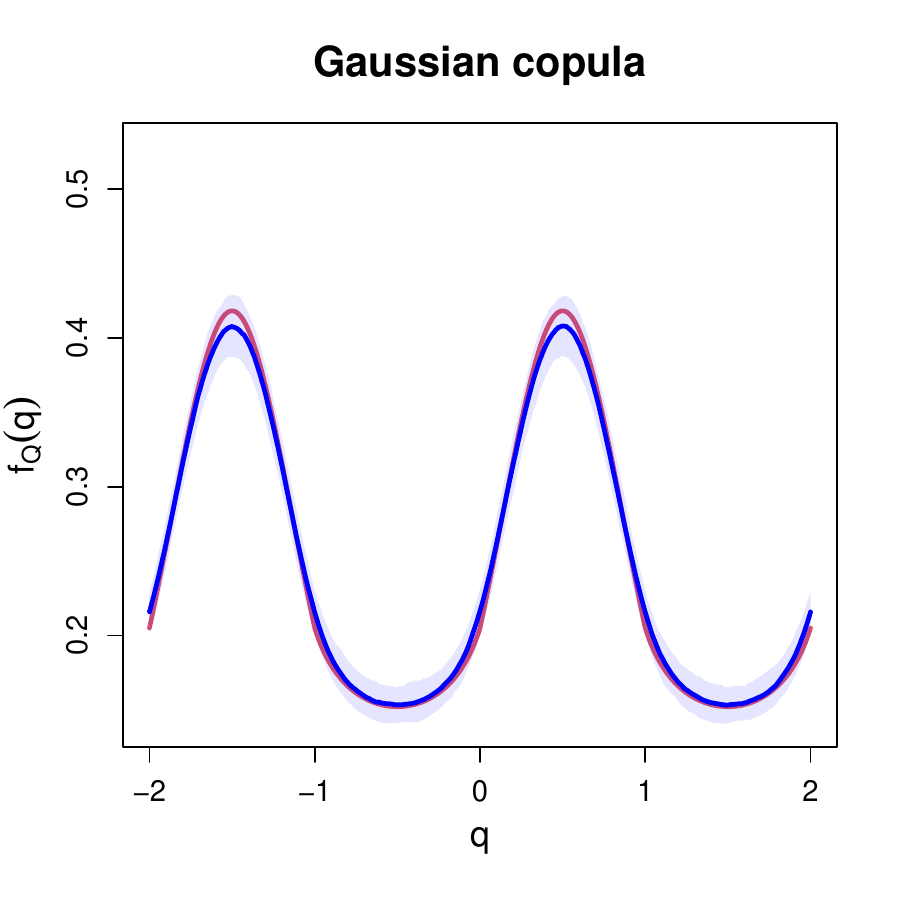}
    \end{subfigure}
    \quad
    \begin{subfigure}[b]{0.21\textwidth}  
        \centering 
        \includegraphics[width=\textwidth]{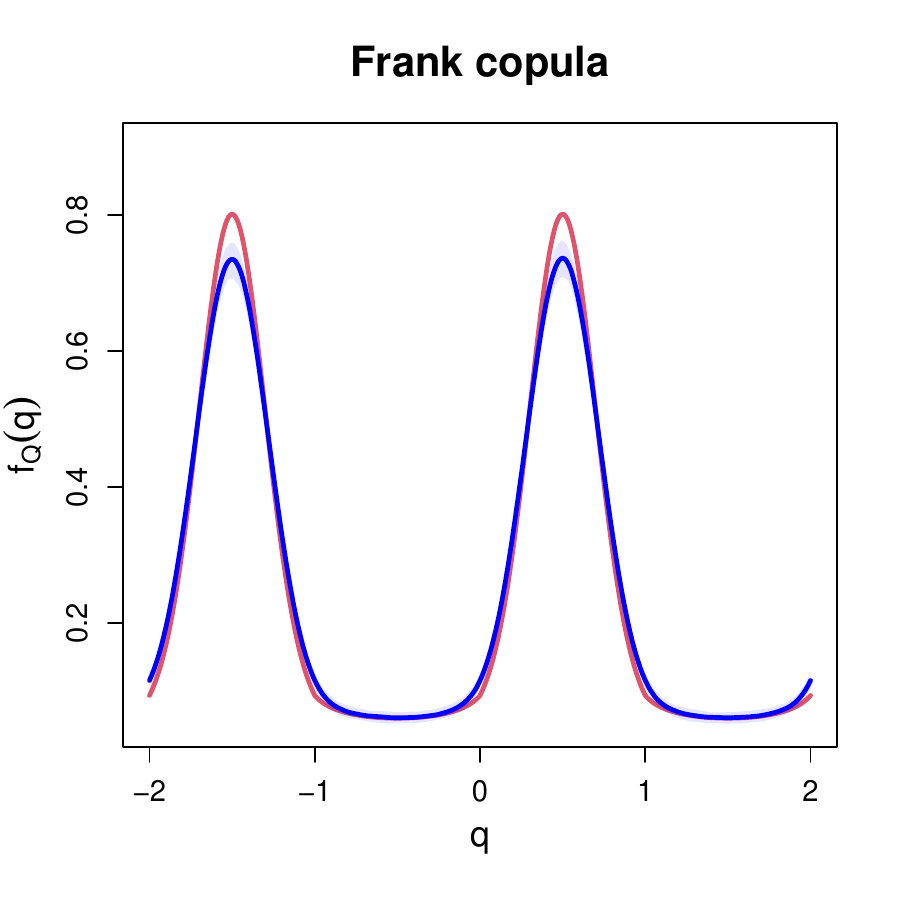}
    \end{subfigure}
    \quad
    \begin{subfigure}[b]{0.21\textwidth}   
        \centering 
        \includegraphics[width=\textwidth]{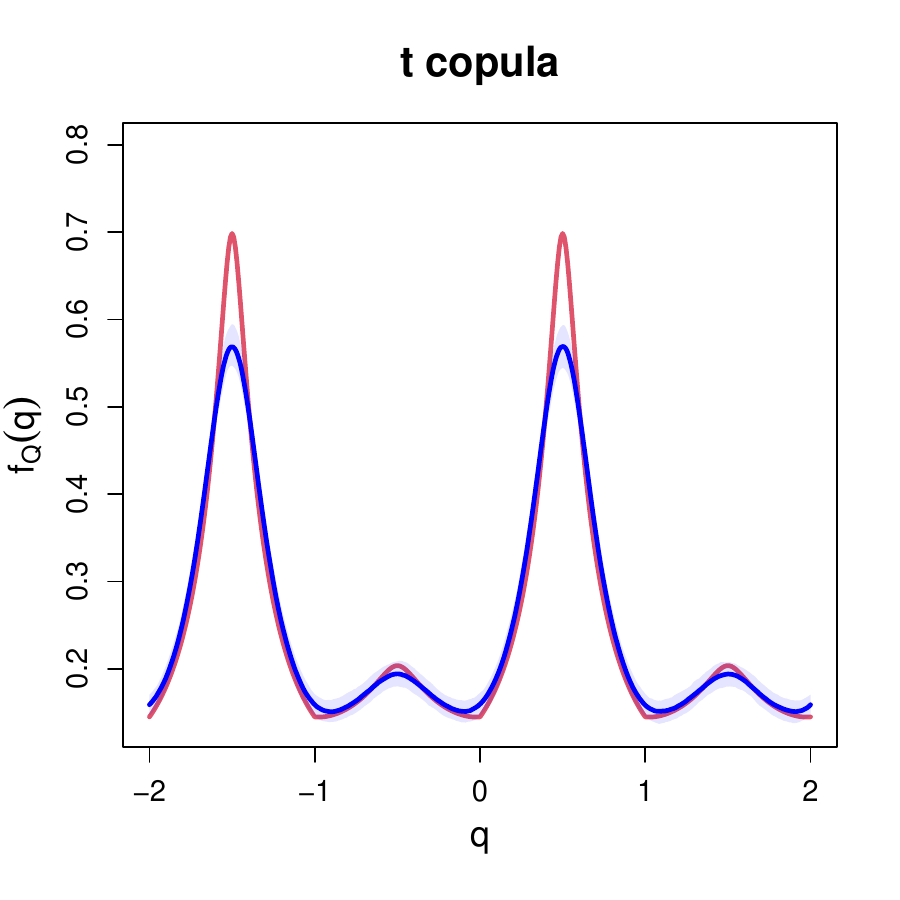}
    \end{subfigure}
    \quad
    \begin{subfigure}[b]{0.21\textwidth}   
        \centering 
        \includegraphics[width=\textwidth]{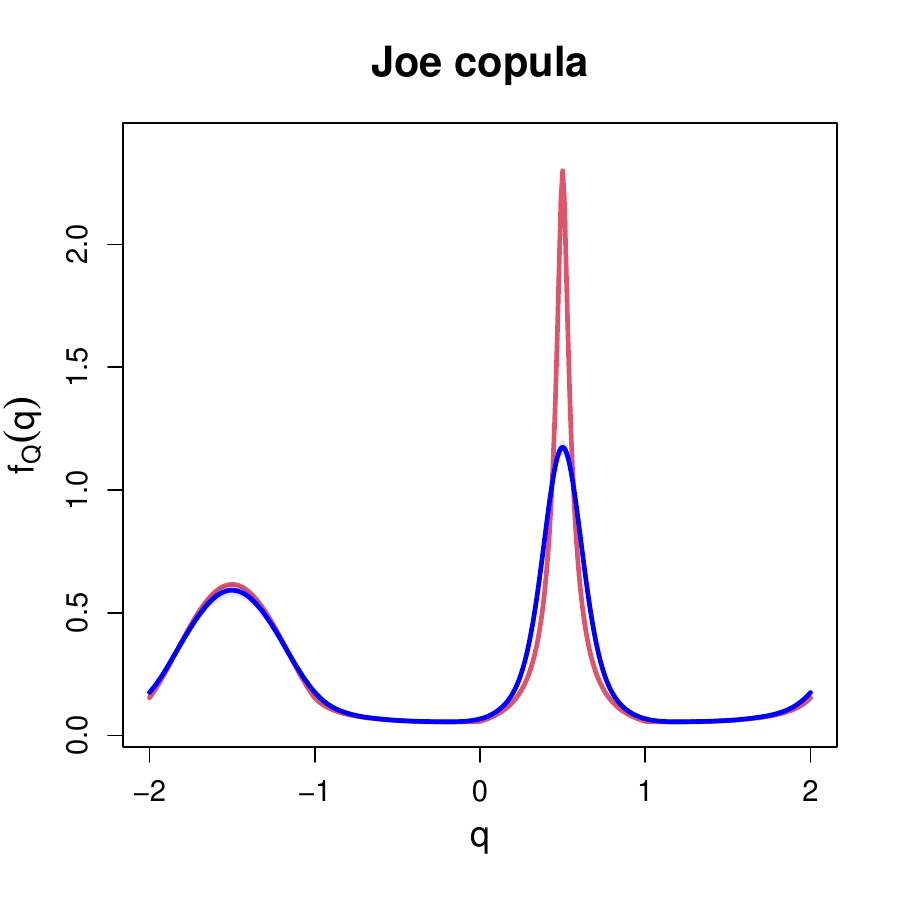}
    \end{subfigure}
    \caption{Comparison of true (red) and estimated median (blue) angular density functions, alongside 95\% confidence intervals, for the $L1$ coordinate system. In each plot, the shaded regions illustrate the estimated confidence intervals.}
    \label{fig:angdensity_L1}
\end{figure}

\begin{figure}[H]
    \centering
    \begin{subfigure}[b]{0.21\textwidth}
        \centering
        \includegraphics[width=\textwidth]{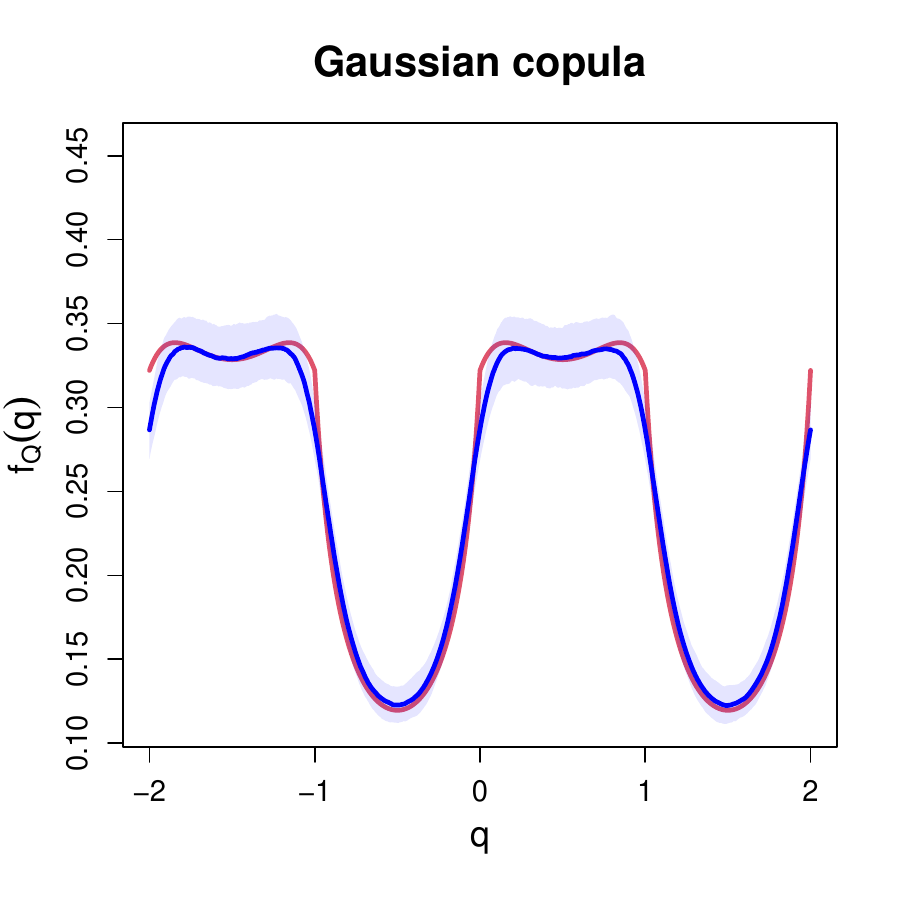}
    \end{subfigure}
    \quad
    \begin{subfigure}[b]{0.21\textwidth}  
        \centering 
        \includegraphics[width=\textwidth]{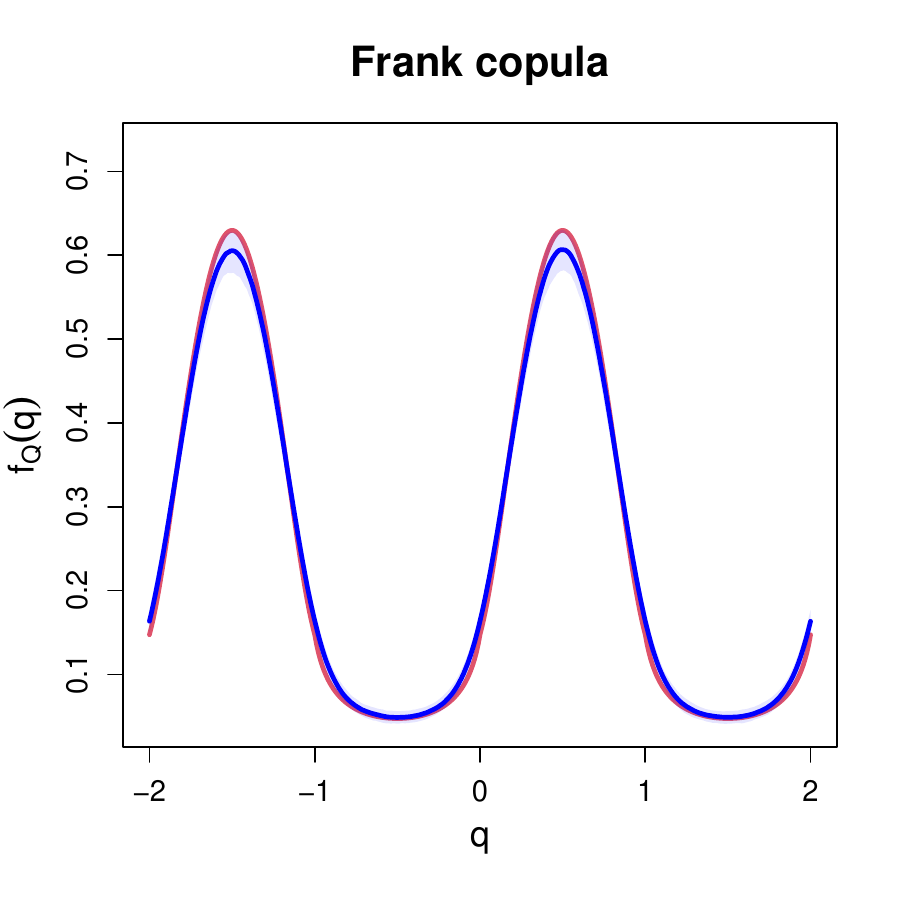}
    \end{subfigure}
    \quad
    \begin{subfigure}[b]{0.21\textwidth}   
        \centering 
        \includegraphics[width=\textwidth]{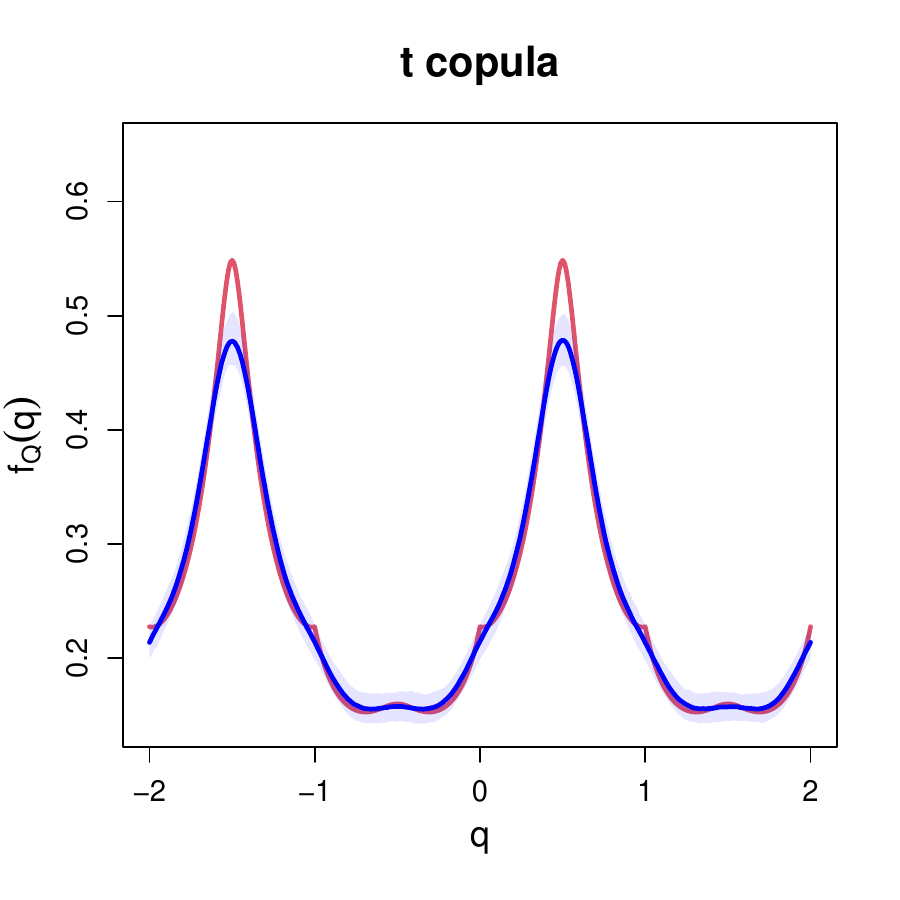}
    \end{subfigure}
    \quad
    \begin{subfigure}[b]{0.21\textwidth}   
        \centering 
        \includegraphics[width=\textwidth]{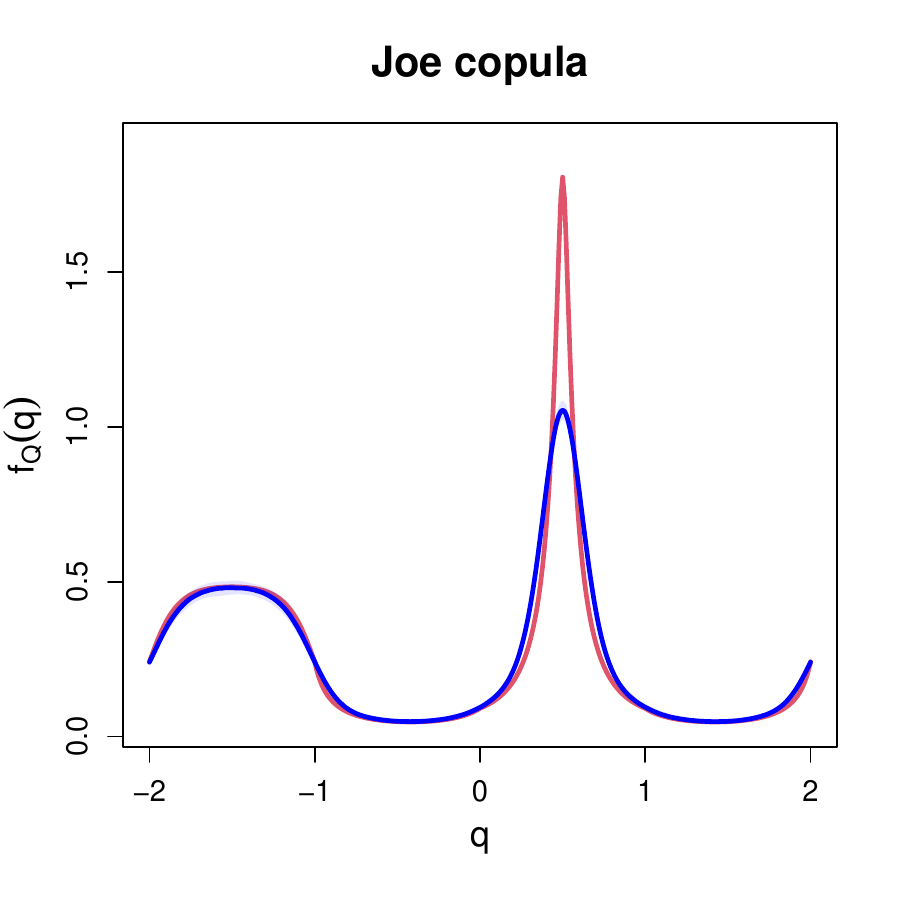}
    \end{subfigure}
    \caption{Comparison of true (red) and estimated median (blue) angular density functions, alongside 95\% confidence intervals, for the $L2$ coordinate system. In each plot, the shaded regions illustrate the estimated confidence intervals.}
    \label{fig:angdensity_L2}
\end{figure}

Figures \ref{fig:angdensity_L1} and \ref{fig:angdensity_L2} compare the true and estimated angular density functions for the $L1$ and $L2$ coordinate systems, respectively. One can observe generally good agreement for both coordinate systems. We note that the KD estimation framework appears unable to fully capture the modal regions for the Frank, t and Joe copulas. We observed a marginal improvement in performance for these regions when the bandwidth parameter, $h$, was decreased; however, this significantly increased the variability in the resulting angular density estimates, and consequently, we opted to keep $h$ fixed at $1/50$. Moreover, we note that even for extremely small bandwidth parameters, the proposed KD framework was unable to approximate the modal behaviour of the Joe copula, suggesting the sample size ($n=10,000$) is not large enough to fully capture the true density function for this particular example.



\section{Additional plots for Section 7} \label{sec:sec7_additional}

\begin{figure}[H]
    \centering
    \includegraphics[width=\textwidth]{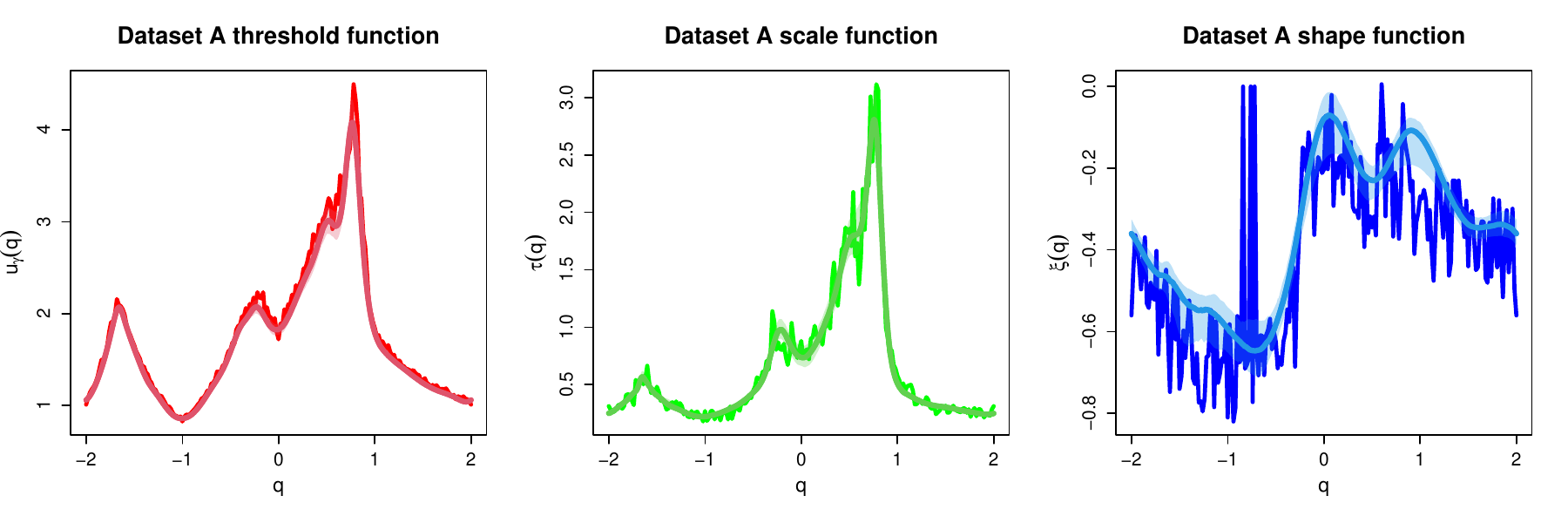}
    \caption{Comparison of estimated local (rough lines) and smooth threshold (red, left), scale (green, middle) and shape (blue, right) functions for data set A with the $L1$ coordinate system, with shaded regions denoting 95\% confidence intervals.}
    \label{fig:local_smooth_A}
\end{figure}

\begin{figure}[H]
    \centering
    \includegraphics[width=\textwidth]{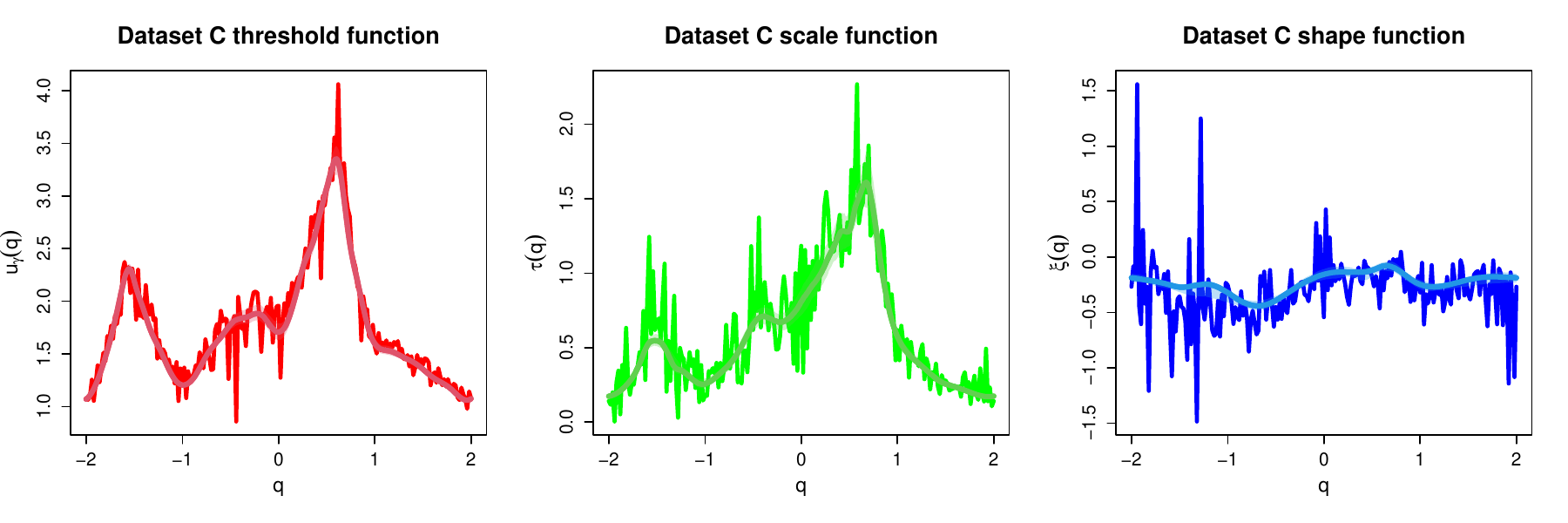}
    \caption{Comparison of estimated local (rough lines) and smooth threshold (red, left), scale (green, middle) and shape (blue, right) functions for data set C with the $L1$ coordinate system, with shaded regions denoting 95\% confidence intervals.}
    \label{fig:local_smooth_C}
\end{figure}

Figures \ref{fig:local_smooth_A} and \ref{fig:local_smooth_C} compare the smoothly and locally estimated threshold and parameters functions for data sets A and C, respectively, under the $L1$ coordinate system. One can observe generally good agreement for each component of the SPAR model.

\begin{figure}[H]
    \centering
    \includegraphics[width=\textwidth]{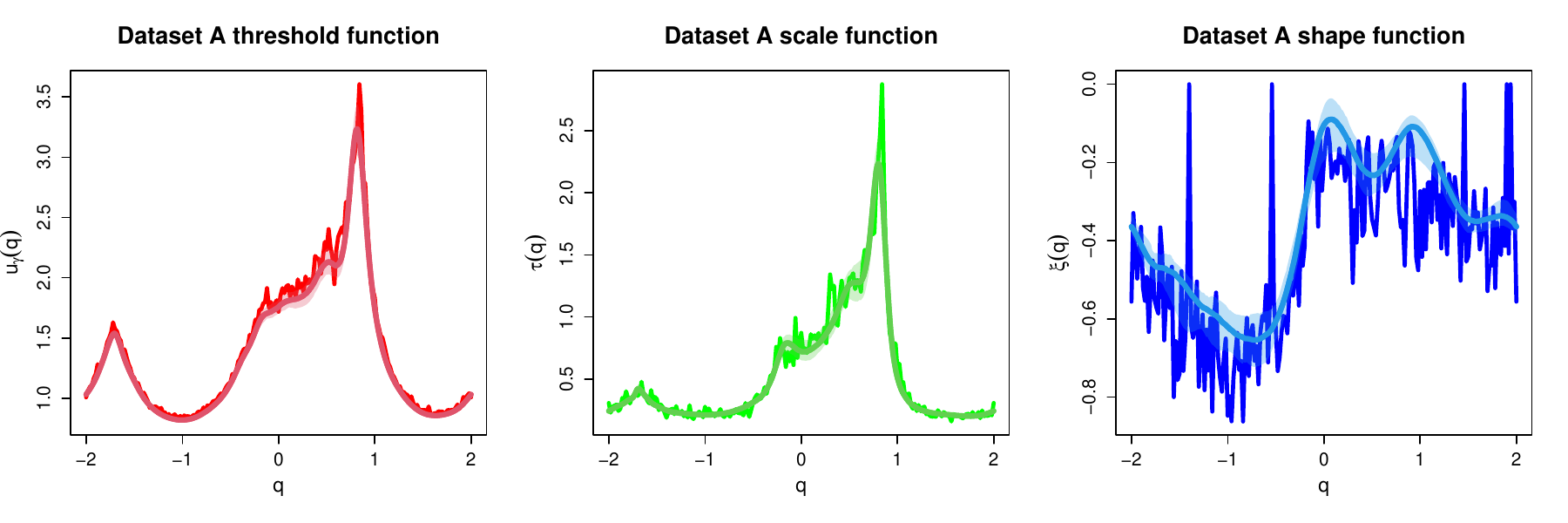}
    \caption{Comparison of estimated local (rough lines) and smooth threshold (red, left), scale (green, middle) and shape (blue, right) functions for data set A with the $L2$ coordinate system, with shaded regions denoting 95\% confidence intervals.}
    \label{fig:local_smooth_A_L2}
\end{figure}

\begin{figure}[H]
    \centering
    \includegraphics[width=\textwidth]{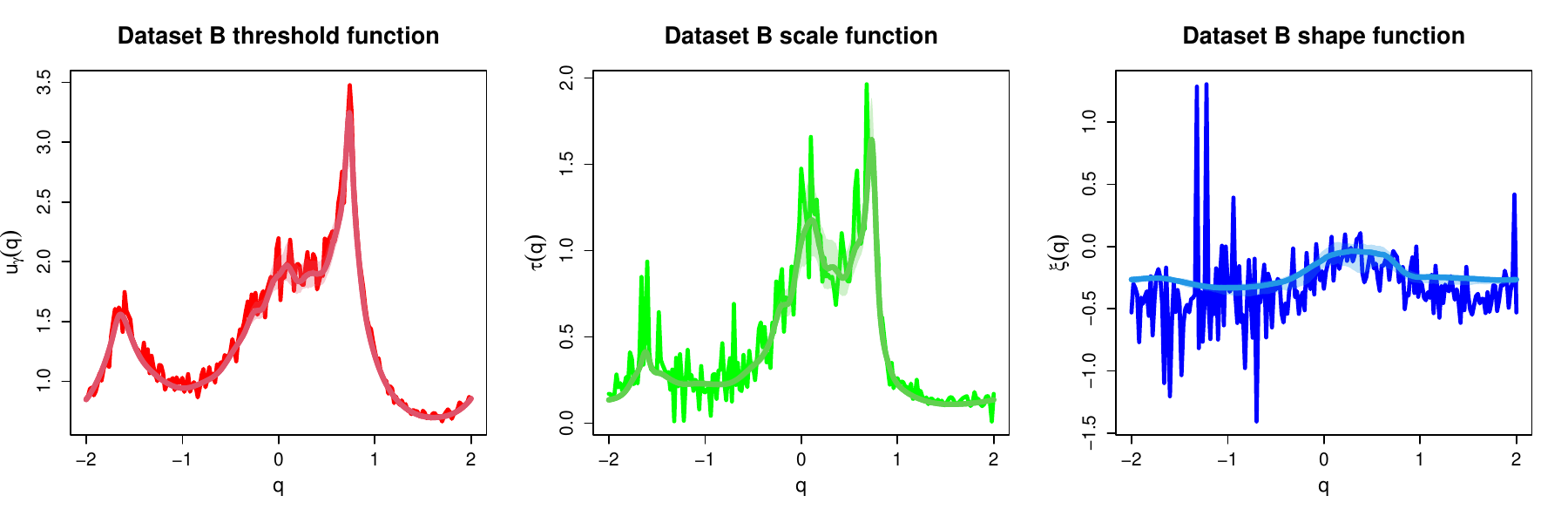}
    \caption{Comparison of estimated local (rough lines) and smooth threshold (red, left), scale (green, middle) and shape (blue, right) functions for data set B with the $L2$ coordinate system, with shaded regions denoting 95\% confidence intervals.}
    \label{fig:local_smooth_B_L2}
\end{figure}

\begin{figure}[H]
    \centering
    \includegraphics[width=\textwidth]{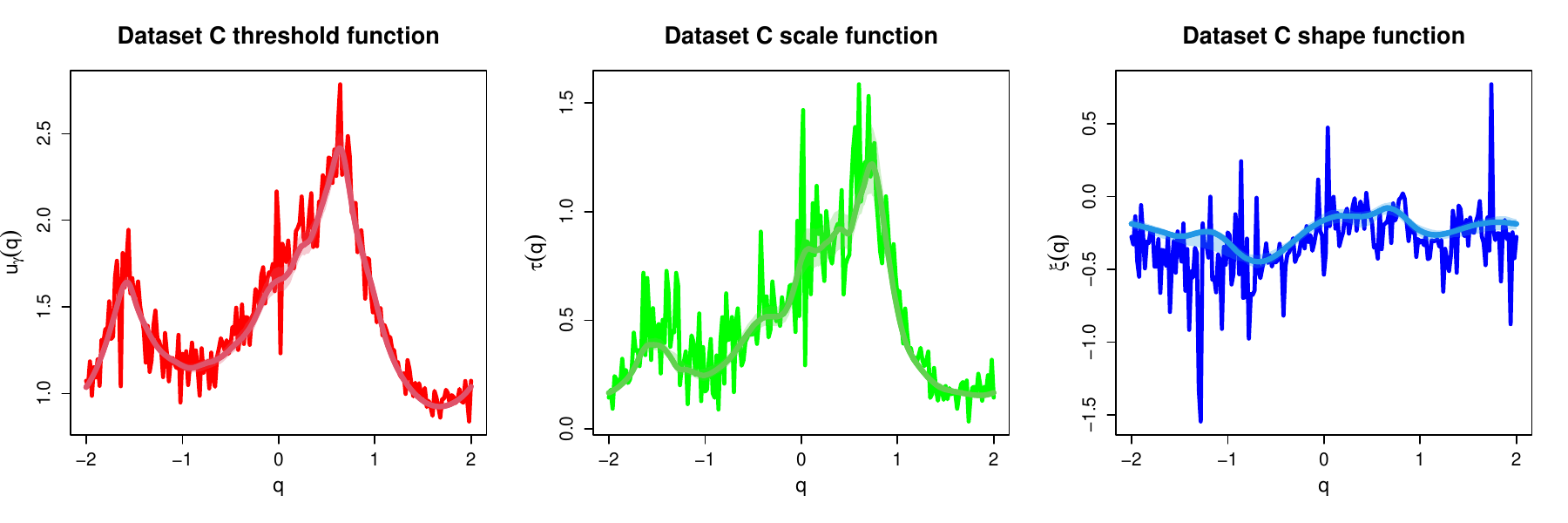}
    \caption{Comparison of estimated local (rough lines) and smooth threshold (red, left), scale (green, middle) and shape (blue, right) functions for data set C with the $L2$ coordinate system, with shaded regions denoting 95\% confidence intervals.}
    \label{fig:local_smooth_C_L2}
\end{figure}

Figures \ref{fig:local_smooth_A_L2}, \ref{fig:local_smooth_B_L2} and \ref{fig:local_smooth_C_L2} compare the smoothly and locally estimated threshold and parameters functions for data sets A, B and C, respectively, under the $L2$ coordinate system. As with the $L1$ coordinates, we obtain good agreement for each model component.

\begin{figure}[H]
    \centering
    \begin{subfigure}[b]{0.3\textwidth}
        \centering
        \includegraphics[width=\textwidth]{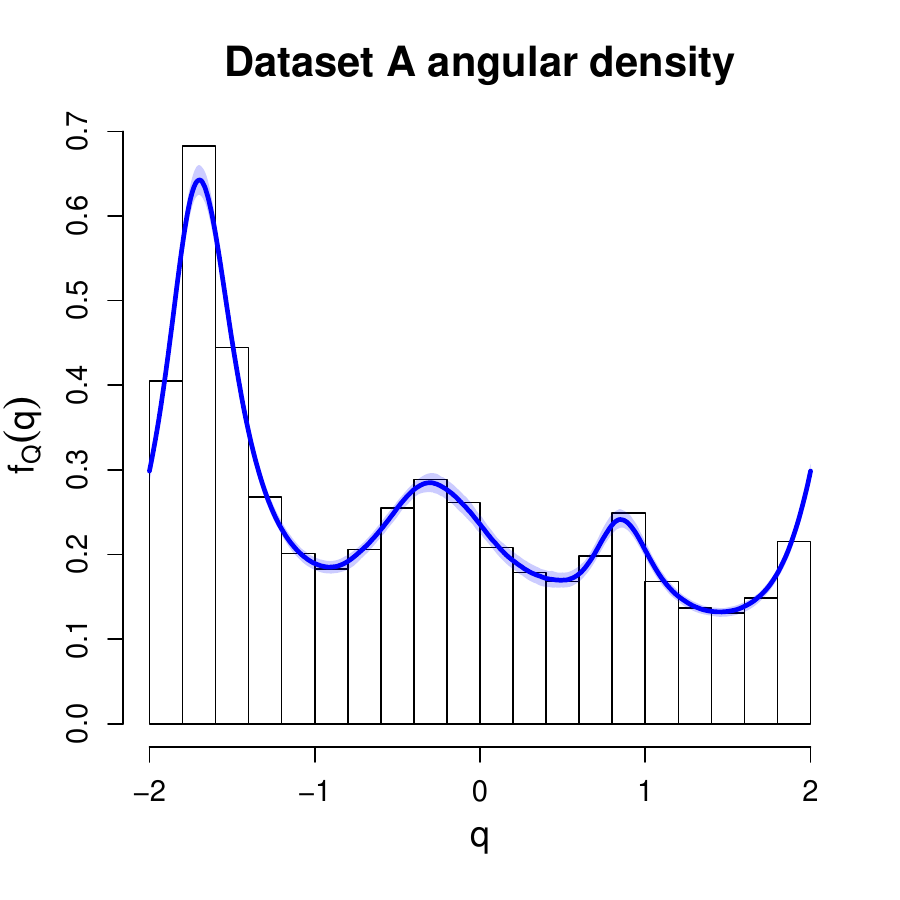}
    \end{subfigure}
    \quad
    \begin{subfigure}[b]{0.3\textwidth}  
        \centering 
        \includegraphics[width=\textwidth]{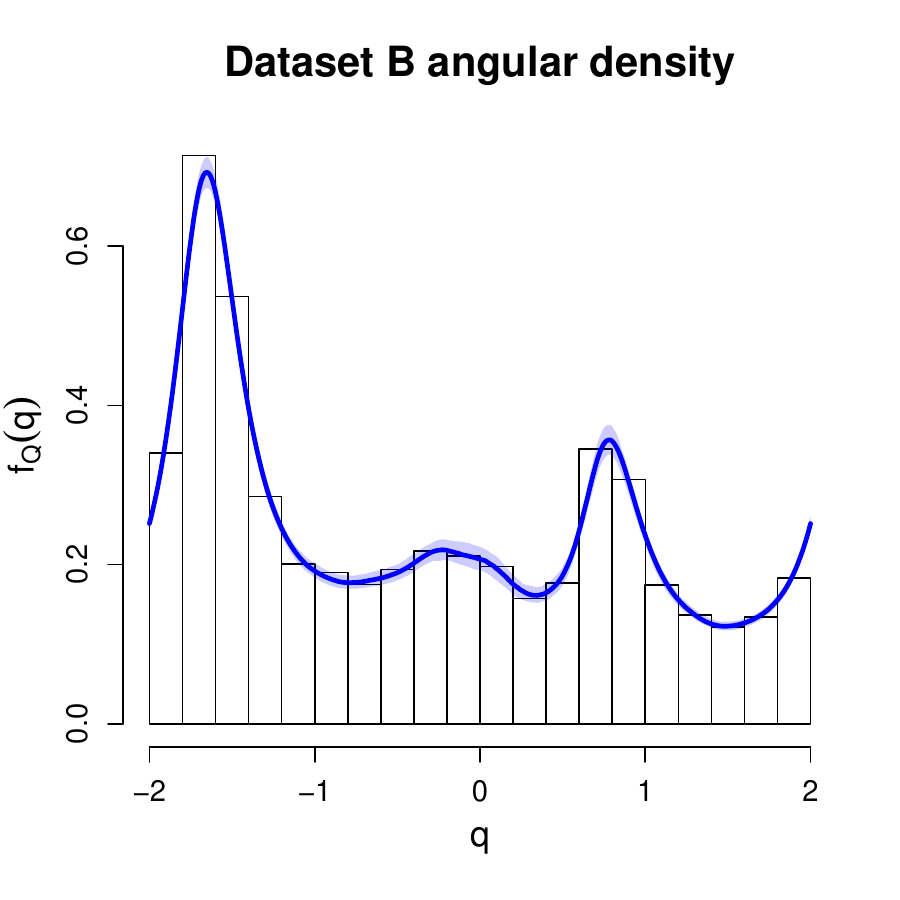}
    \end{subfigure}
    \quad
    \begin{subfigure}[b]{0.3\textwidth}  
        \centering 
        \includegraphics[width=\textwidth]{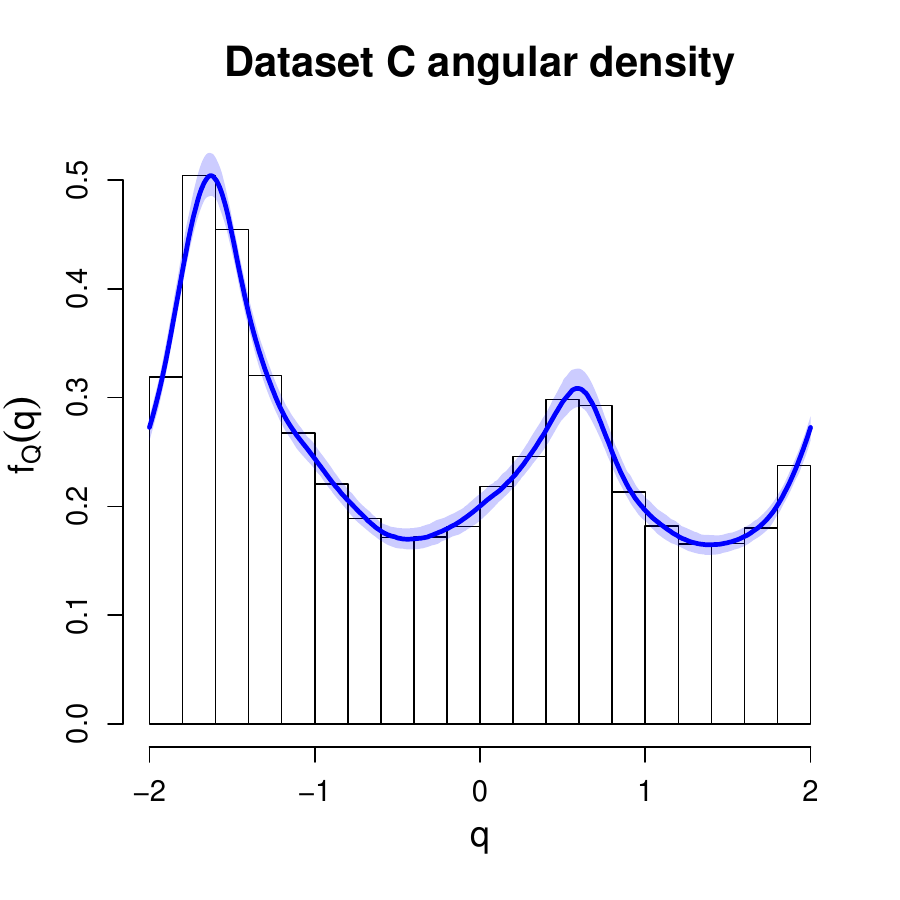}
    \end{subfigure}
    \caption{Comparison of estimated median (blue lines) angular density functions to histograms for data sets A (left), B (centre), and C (right) with the $L2$ coordinate system. The shaded regions in each plot denote the estimated 95\% confidence intervals.}
    \label{fig:angular_density_L2}
\end{figure}

Figure \ref{fig:angular_density_L2} compares the estimated median angular density functions, and corresponding confidence intervals, with the histograms for each data set under the $L2$ coordinate system. The two sets of estimates are in good agreement, providing evidence that the chosen bandwidth parameter is appropriate.

\begin{figure}[H]
    \centering
    \begin{subfigure}[b]{0.3\textwidth}
        \centering
        \includegraphics[width=\textwidth]{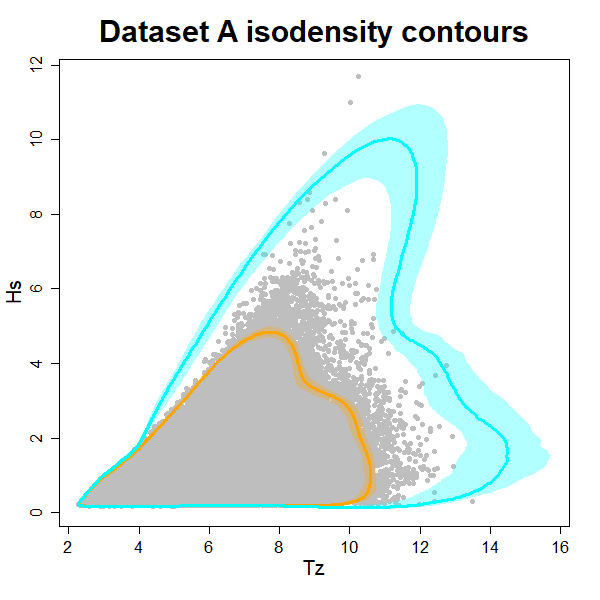}
    \end{subfigure}
    \quad
    \begin{subfigure}[b]{0.3\textwidth}  
        \centering 
        \includegraphics[width=\textwidth]{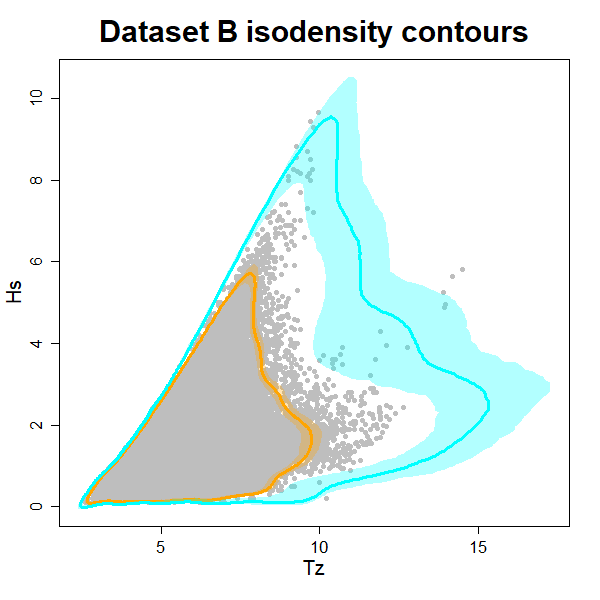}
    \end{subfigure}
    \quad
    \begin{subfigure}[b]{0.3\textwidth}  
        \centering 
        \includegraphics[width=\textwidth]{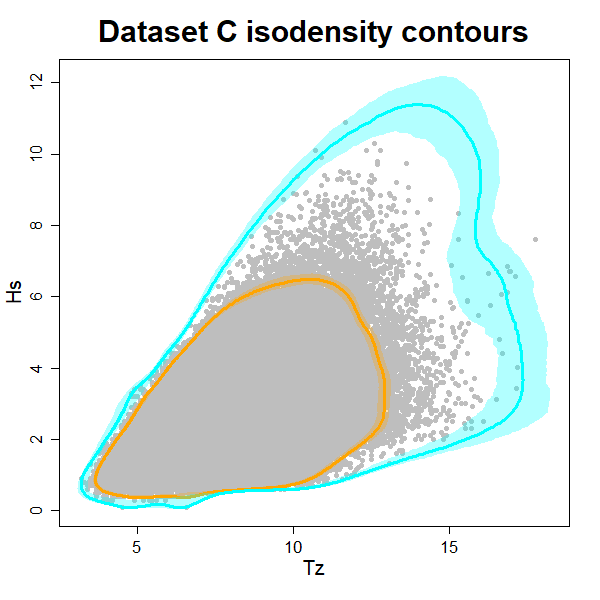}
    \end{subfigure}
    \caption{Estimated median isodensity contours at $p = 10^{-3}$ (orange lines) and $p =10^{-6}$ (cyan lines) for data sets A (left), B (centre), and C (right) with the $L2$ coordinate system. The shaded region for each contour denote the 95\% bootstrapped confidence intervals.}
    \label{fig:equidensity_contours_L2}
\end{figure}

Figure \ref{fig:equidensity_contours_L2} illustrates the estimated median isodensity contours, and corresponding uncertainty regions, for the density levels $p \in \{ 10^{-3},10^{-6}\}$ under the $L2$ coordinate system. As with the $L1$ coordinates, the estimated contours capture the shape and structure of each data set.


\begin{figure}[H]
    \centering
    \begin{subfigure}[b]{0.3\textwidth}
        \centering
        \includegraphics[width=\textwidth]{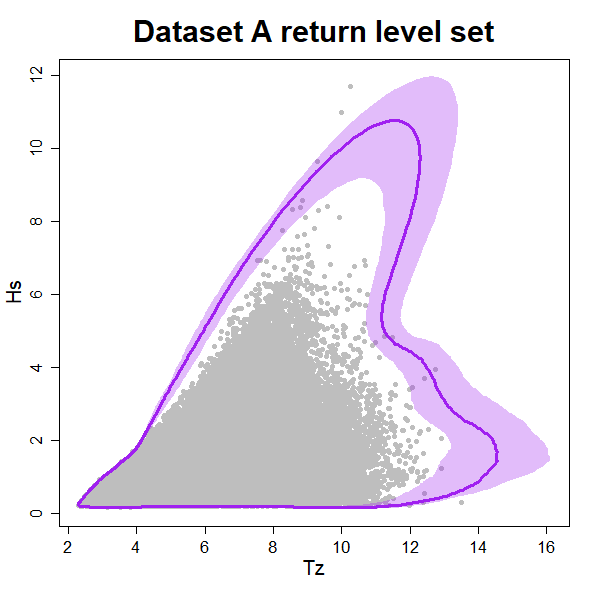}
    \end{subfigure}
    \quad
    \begin{subfigure}[b]{0.3\textwidth}  
        \centering 
        \includegraphics[width=\textwidth]{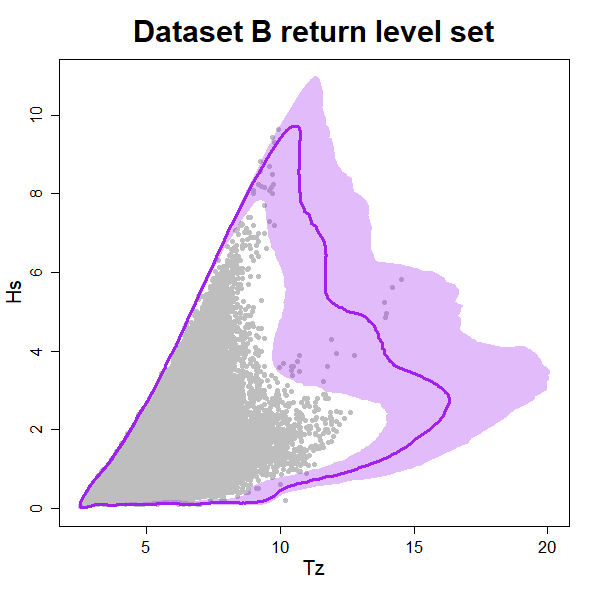}
    \end{subfigure}
    \quad
    \begin{subfigure}[b]{0.3\textwidth}  
        \centering 
        \includegraphics[width=\textwidth]{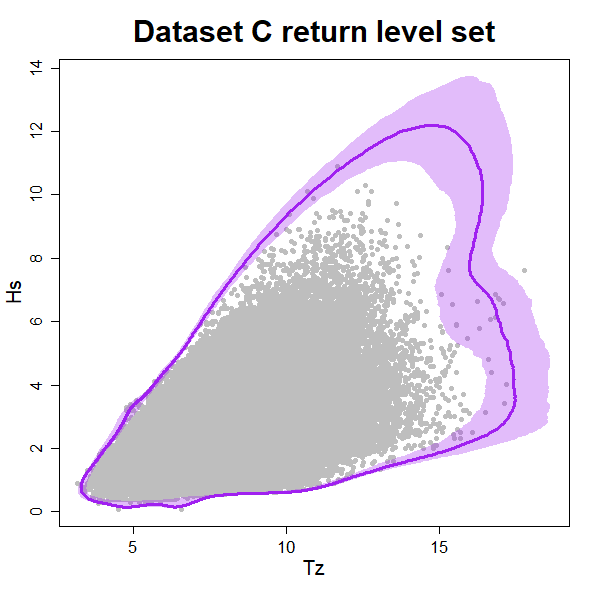}
    \end{subfigure}
    \caption{Estimated median $10$ year return level sets (purple lines) for data sets A (left), B (centre), and C (right) with the $L2$ coordinate system. The shaded region for each return level set denotes the 95\% bootstrapped confidence region.}
    \label{fig:return_level_sets_L2}
\end{figure}

Figure \ref{fig:return_level_sets_L2} illustrates the estimated median $10$ year return level sets for each data set, along with 95\% bootstrapped confidence intervals, obtained using the $L2$ coordinate system. The estimated sets appear to capture the features of the observed data, and we observe only a handful of observations outside of the return level set for each data set. 

\begin{figure}[H]
    \centering
    \begin{subfigure}[b]{0.3\textwidth}
        \centering
        \includegraphics[width=\textwidth]{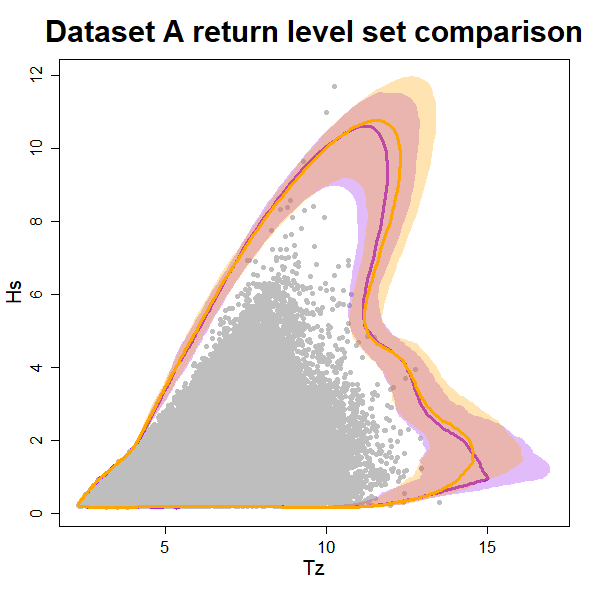}
    \end{subfigure}
    \quad
    \begin{subfigure}[b]{0.3\textwidth}  
        \centering 
        \includegraphics[width=\textwidth]{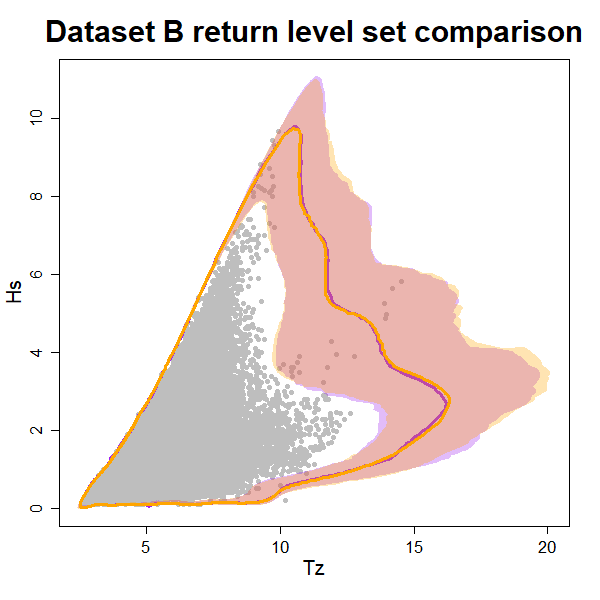}
    \end{subfigure}
    \quad
    \begin{subfigure}[b]{0.3\textwidth}  
        \centering 
        \includegraphics[width=\textwidth]{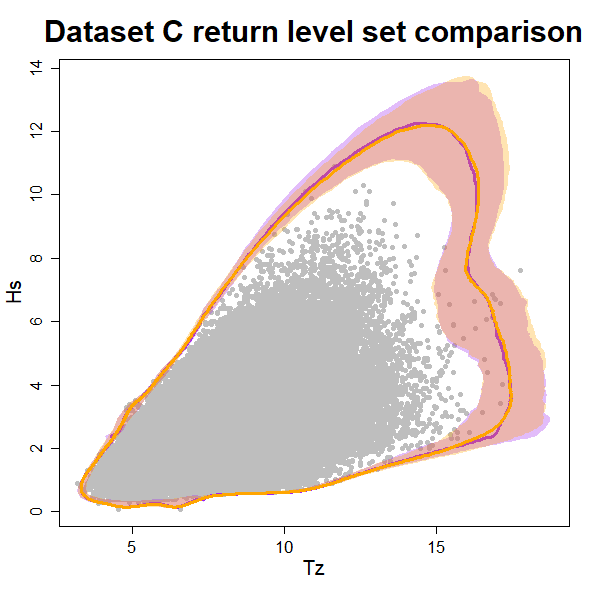}
    \end{subfigure}
    \caption{Comparison of $L1$ (purple) and $L2$ (orange) coordinate system return level sets for data sets A (left), B (centre), and C (right). The shaded region for each return level set denotes the 95\% bootstrapped confidence intervals.}
    \label{fig:return_level_set_comparison}
\end{figure}

Figure \ref{fig:return_level_set_comparison} compares the estimated $10$ year return level sets from the two coordinate systems for each of the data sets. One can observe generally good agreement between the two systems, although we note some small differences in the estimates for data set A at certain angles. The overall agreement, however, provides evidence of consistency between the two modelling approaches.  

\begin{figure}[htp]
    \centering
    \begin{subfigure}[b]{0.3\textwidth}
        \centering
        \includegraphics[width=\textwidth]{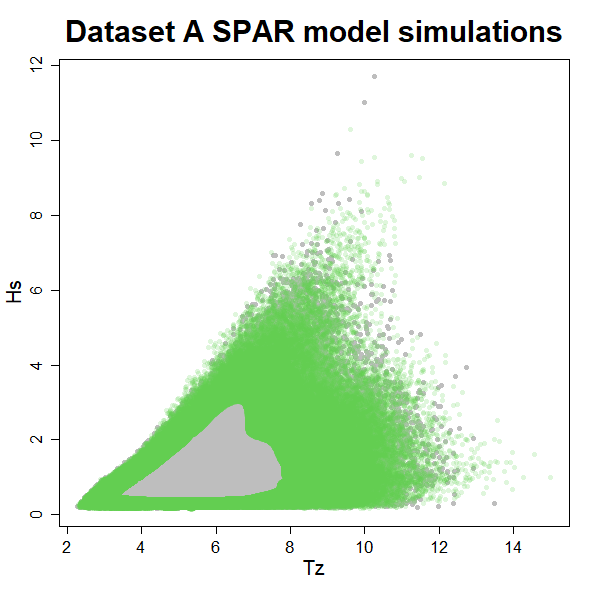}
    \end{subfigure}
    \quad
    \begin{subfigure}[b]{0.3\textwidth}  
        \centering 
        \includegraphics[width=\textwidth]{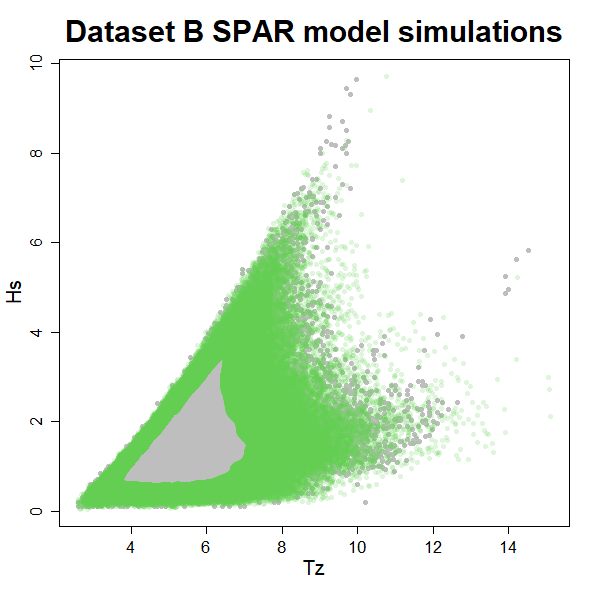}
    \end{subfigure}
    \quad
    \begin{subfigure}[b]{0.3\textwidth}  
        \centering 
        \includegraphics[width=\textwidth]{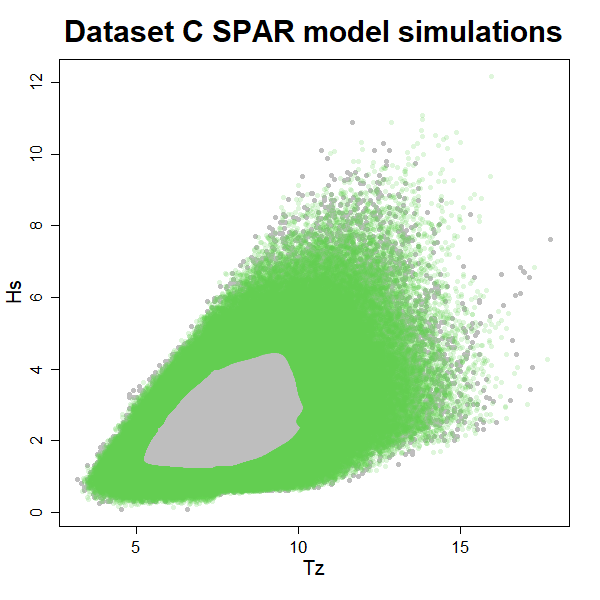}
    \end{subfigure}
    \caption{Model simulations (green) against observed time series (grey) for data sets A (left), B (centre), and C (right) with the $L1$ coordinate system.}
    \label{fig:simulated_datasets}
\end{figure}

\begin{figure}[H]
    \centering
    \begin{subfigure}[b]{0.3\textwidth}
        \centering
        \includegraphics[width=\textwidth]{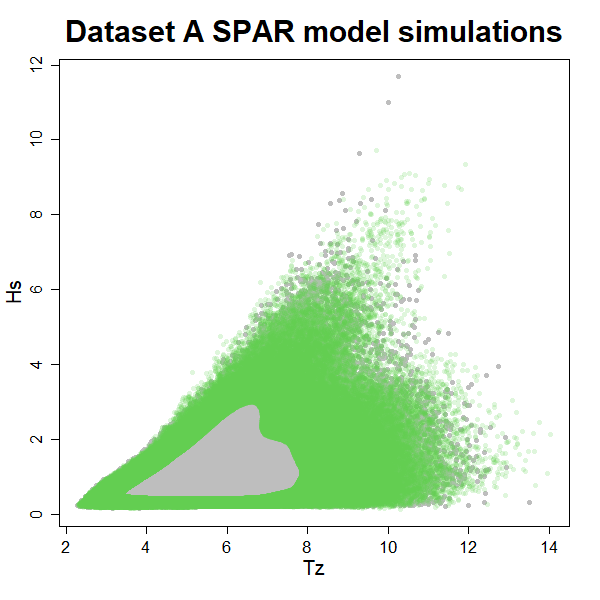}
    \end{subfigure}
    \quad
    \begin{subfigure}[b]{0.3\textwidth}  
        \centering 
        \includegraphics[width=\textwidth]{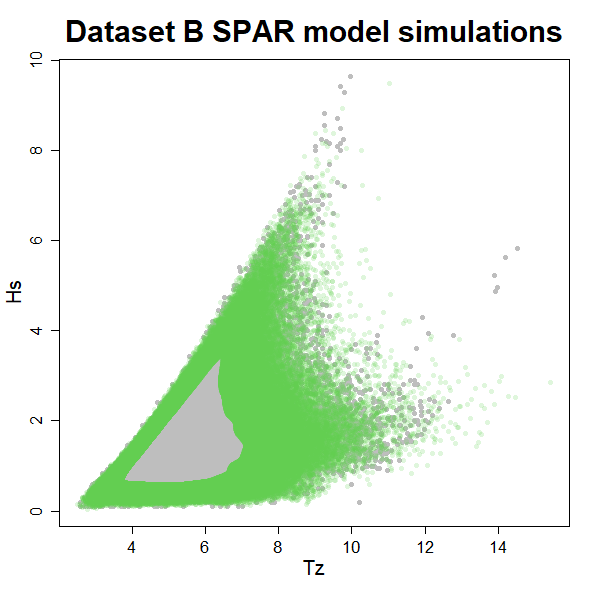}
    \end{subfigure}
    \quad
    \begin{subfigure}[b]{0.3\textwidth}  
        \centering 
        \includegraphics[width=\textwidth]{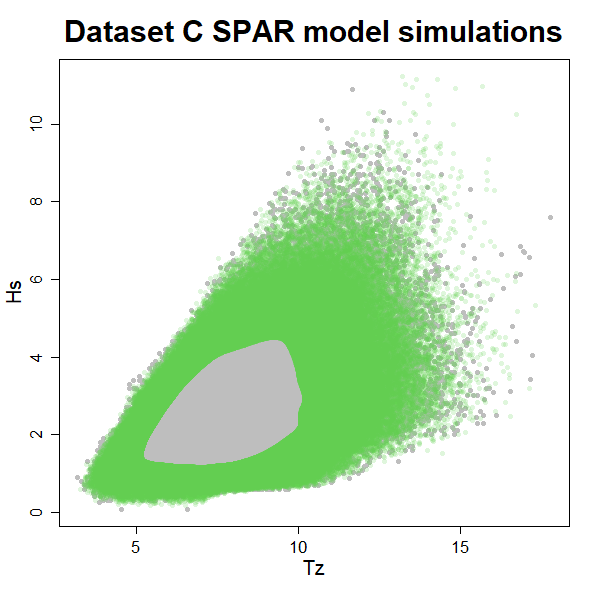}
    \end{subfigure}
    \caption{Model simulations (green) against observed time series (grey) for data sets A (left), B (centre), and C (right) with the $L2$ coordinate system.}
    \label{fig:simulated_datasets_L2}
\end{figure}

As noted in the main article, fitted SPAR models can be used to simulate new observations. This simulation is straightforward. We start by generating a random number $u^*$, uniformly distributed in $[0,1]$. A random angle $q \in (-2,2]$ can then be calculated by applying the probability integral transform so that $q = F^{-1}_{Q} (u^*)$, where $F_Q$ denotes the estimated distribution function of $Q$. A corresponding radial value $r$, is then simulated as a random value from the GP distribution with parameter vector $(u_{\gamma}(q),\xi(q),\tau(q))$. The resulting pair $(r,q)$ is then a random sample from the SPAR model. Figure \ref{fig:simulated_datasets} and \ref{fig:simulated_datasets_L2} illustrate simulated data points for each metocean data set, overlayed on top of the observed time series, for the $L1$ and $L2$ coordinate systems, respectively. For each data set, we simulated the same number of new observations as the original sample size. One can observe that the simulated data sets closely resemble the threshold exceeding observations. Moreover, these plots illustrate the regions $\mathcal{U}_{\gamma}$ for which the SPAR model is valid. 

\begin{figure}[H]
    \centering
    \includegraphics[width=\textwidth]{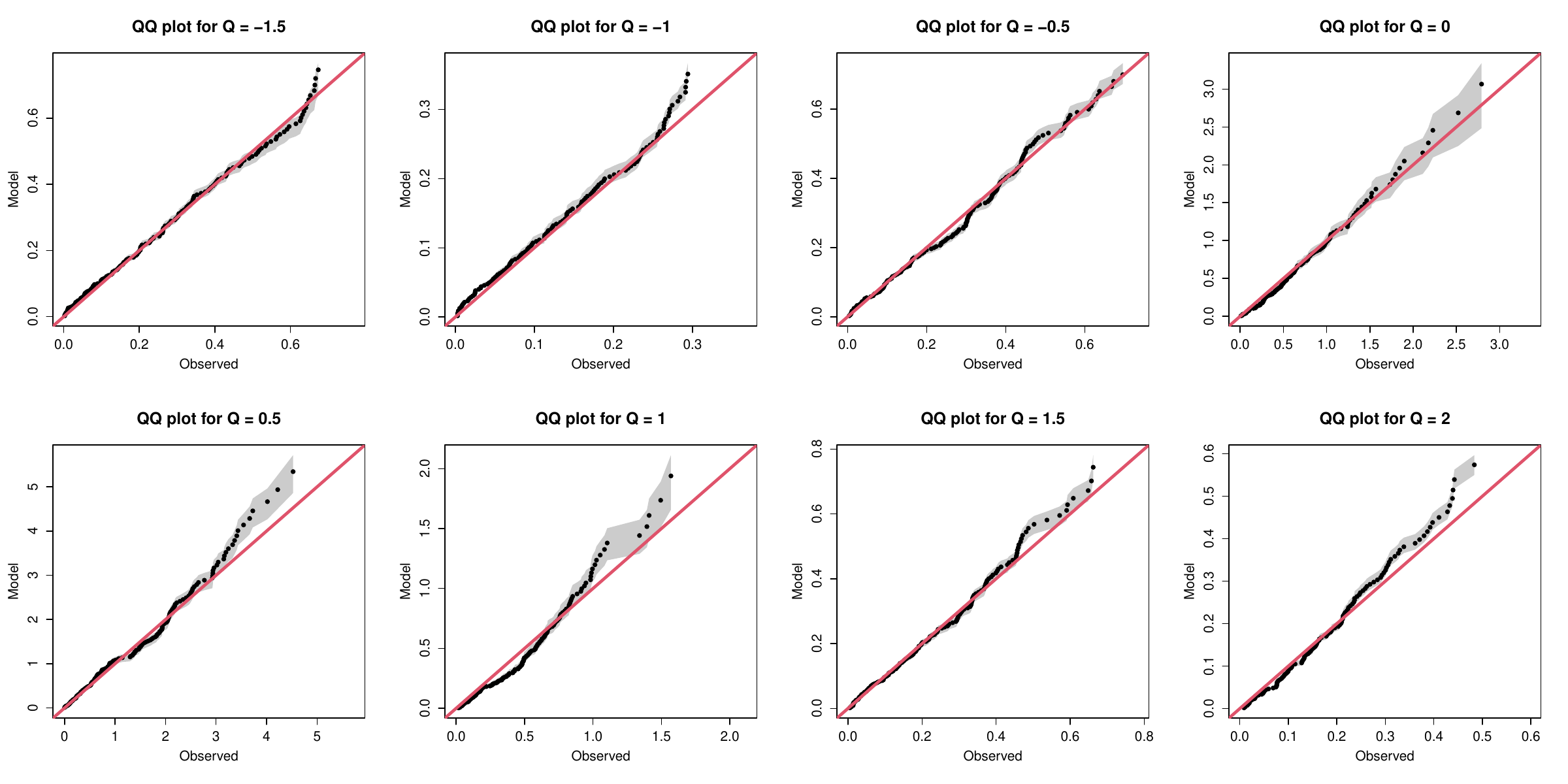}
    \caption{Local QQ plots for the fitted SPAR model on data set A with the $L1$ coordinate system. The shaded region for each plot denotes the empirical 95\% confidence region.}
    \label{fig:local_A_L1}
\end{figure}

\begin{figure}[H]
    \centering
    \includegraphics[width=\textwidth]{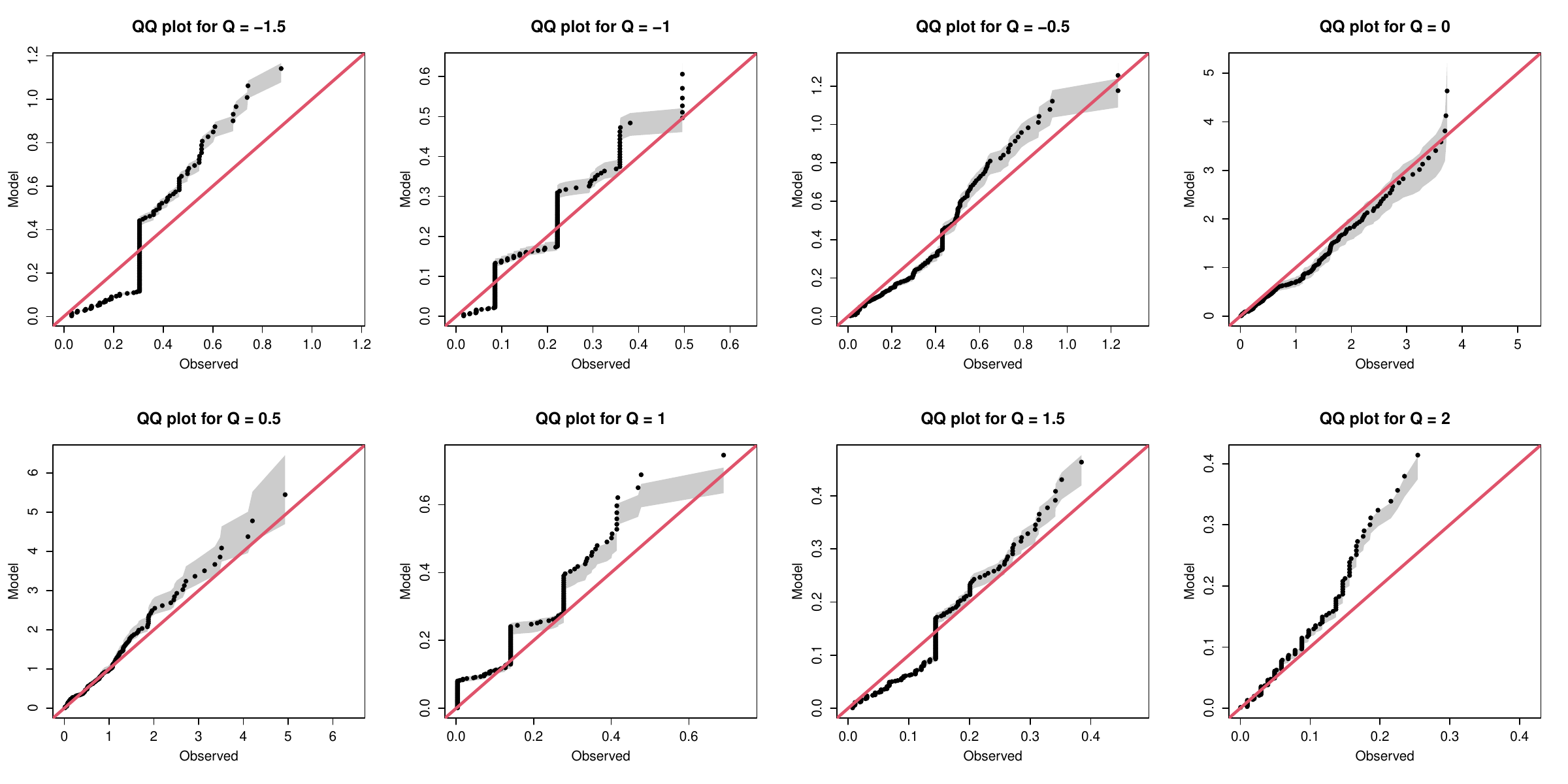}
    \caption{Local QQ plots for the fitted SPAR model on data set B with the $L1$ coordinate system. The shaded region for each plot denotes the empirical 95\% confidence region.}
    \label{fig:local_B_L1}
\end{figure}

\begin{figure}[H]
    \centering
    \includegraphics[width=\textwidth]{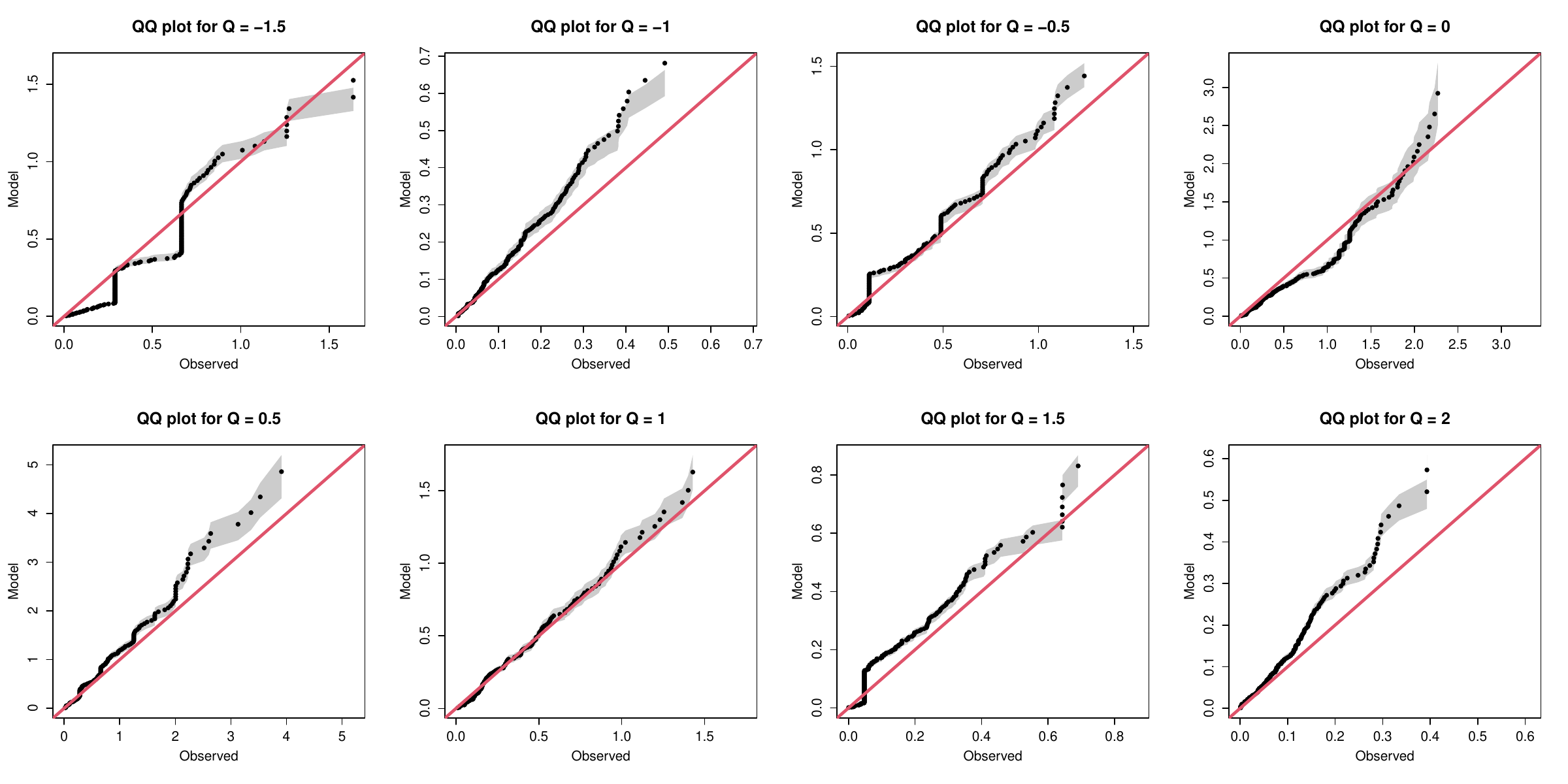}
    \caption{Local QQ plots for the fitted SPAR model on data set C with the $L1$ coordinate system. The shaded region for each plot denotes the empirical 95\% confidence region.}
    \label{fig:local_C_L1}
\end{figure}

\begin{figure}[H]
    \centering
    \includegraphics[width=\textwidth]{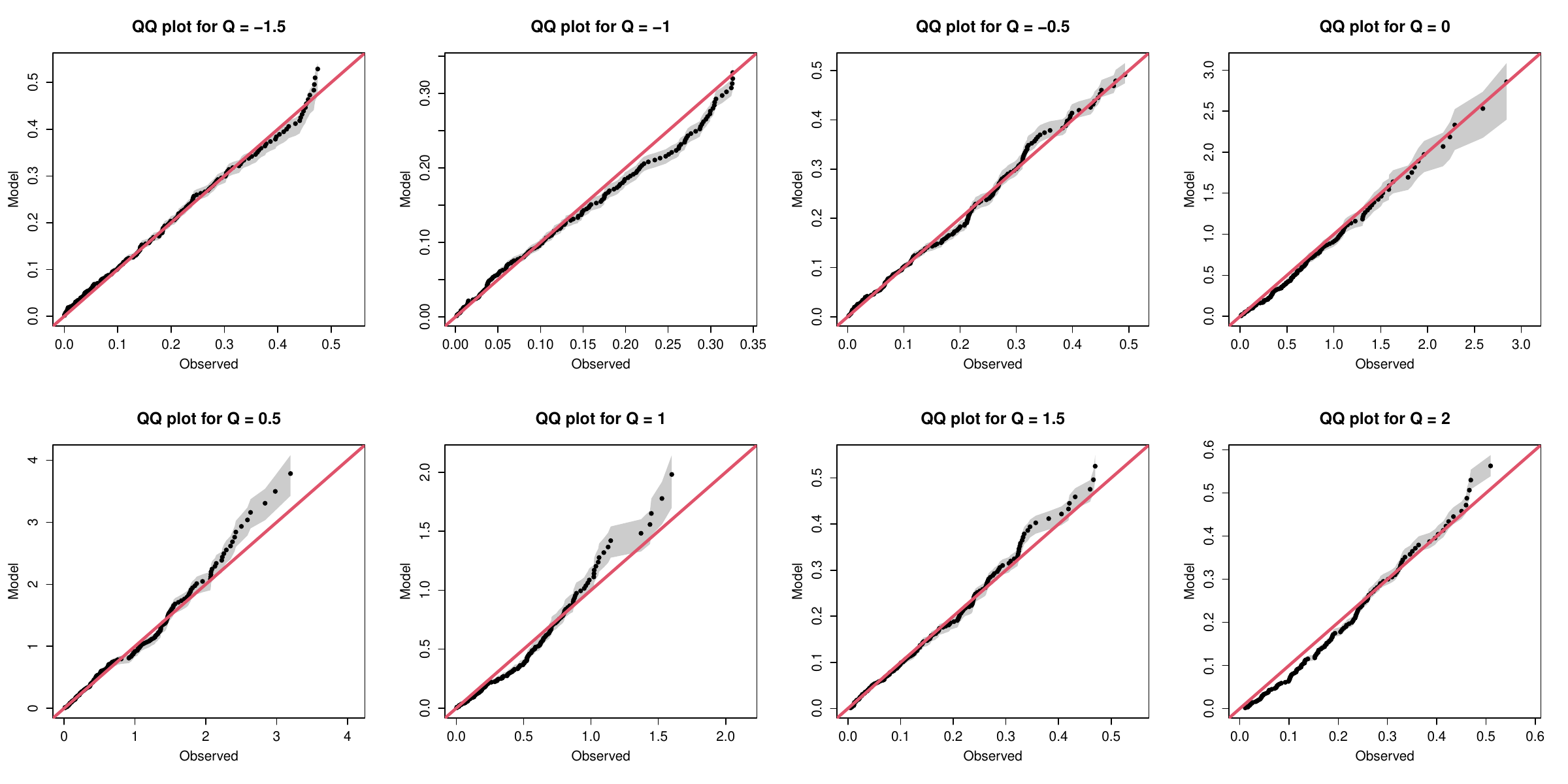}
    \caption{Local QQ plots for the fitted SPAR model on data set A with the $L2$ coordinate system. The shaded region for each plot denotes the empirical 95\% confidence region.}
    \label{fig:local_A_L2}
\end{figure}

\begin{figure}[H]
    \centering
    \includegraphics[width=\textwidth]{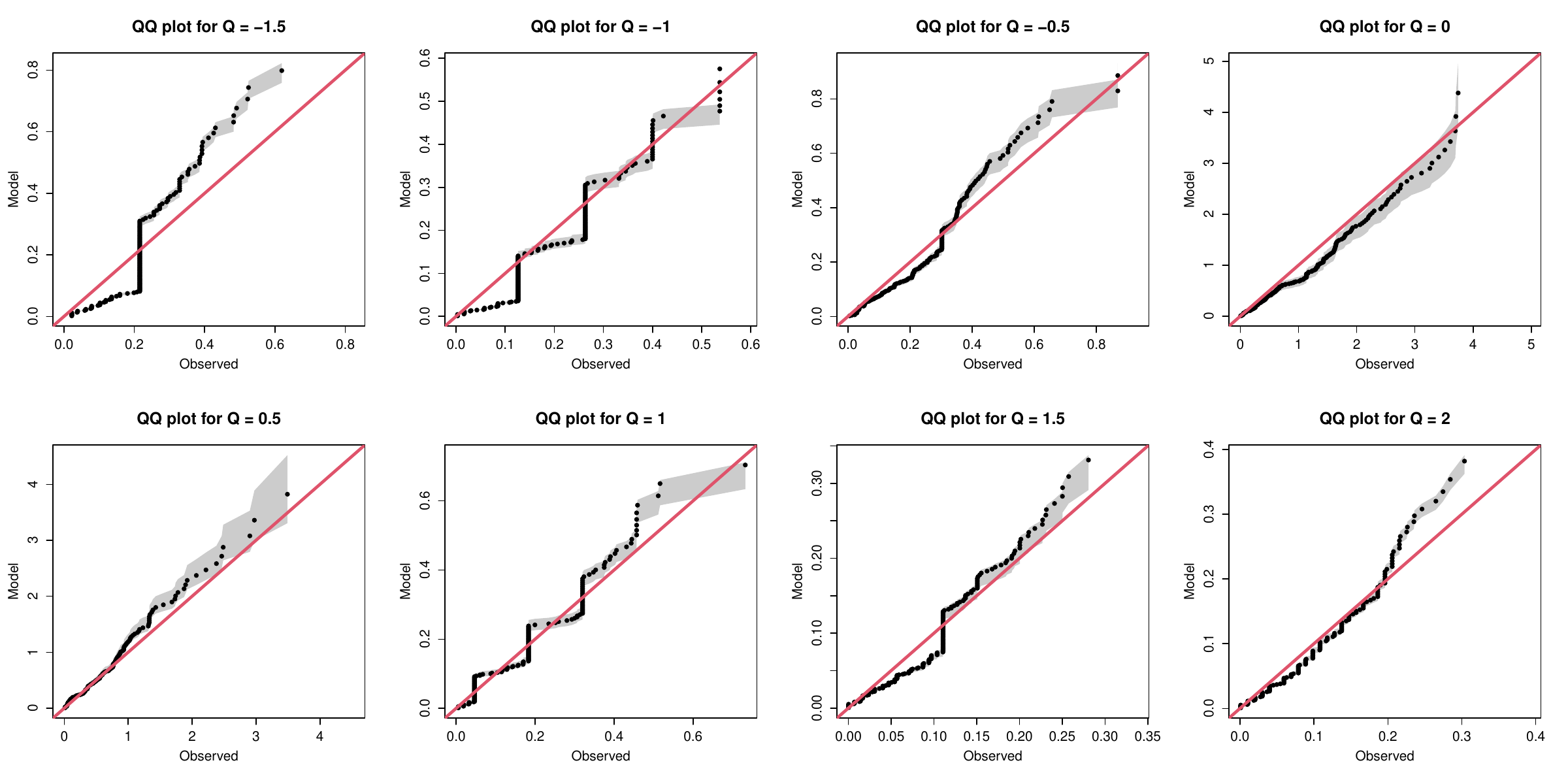}
    \caption{Local QQ plots for the fitted SPAR model on data set B with the $L2$ coordinate system. The shaded region for each plot denotes the empirical 95\% confidence region.}
    \label{fig:local_B_L2}
\end{figure}

\begin{figure}[H]
    \centering
    \includegraphics[width=\textwidth]{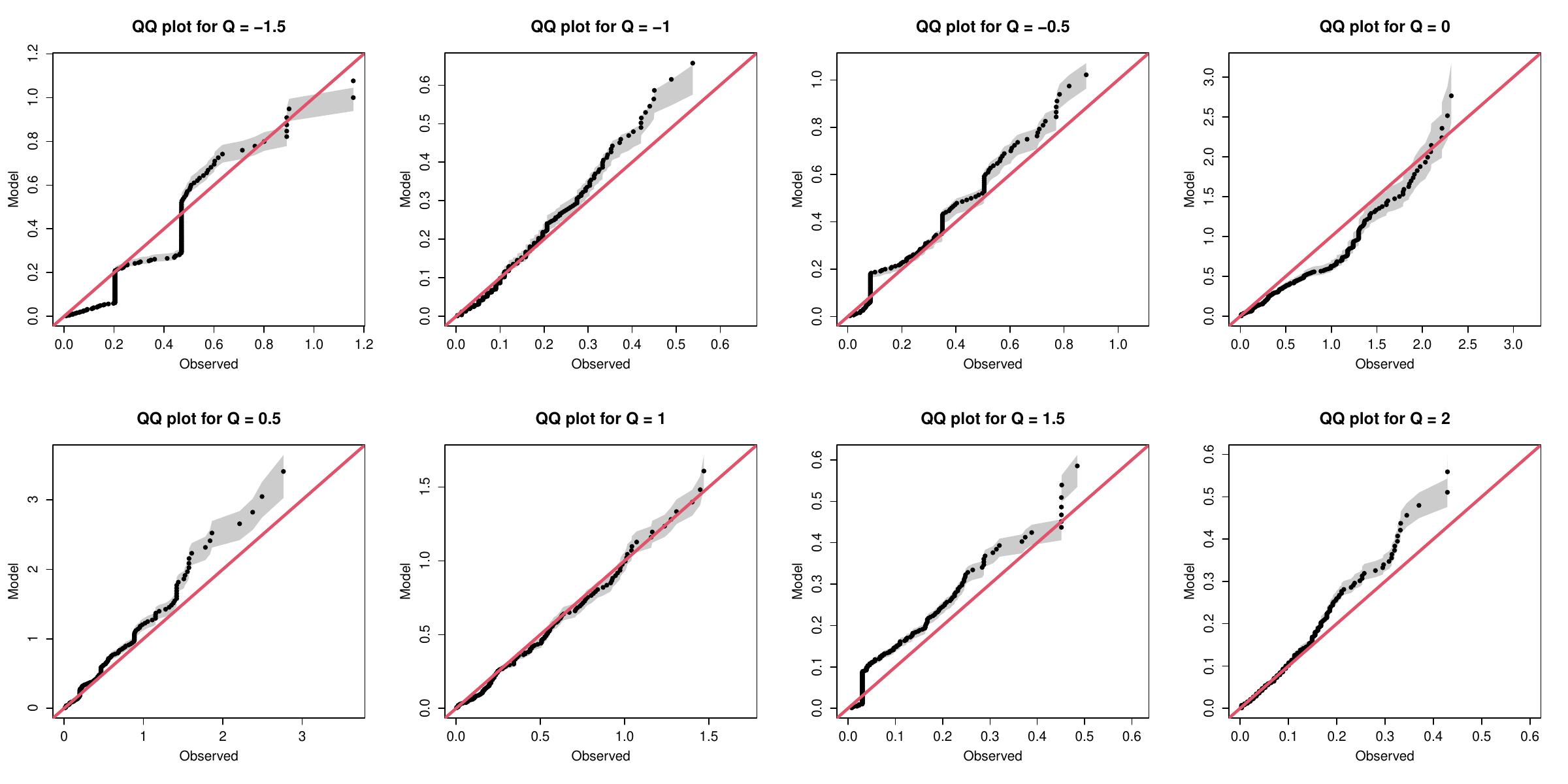}
    \caption{Local QQ plots for the fitted SPAR model on data set C with the $L2$ coordinate system. The shaded region for each plot denotes the empirical 95\% confidence region.}
    \label{fig:local_C_L2}
\end{figure}

Figures \ref{fig:local_A_L1}, \ref{fig:local_B_L1} and \ref{fig:local_C_L1} give the local window QQ plots for the SPAR model fits on data sets A, B, and C, respectively, under the $L1$ coordinate system. The corresponding plots for the $L2$ coordinate system are given in Figures \ref{fig:local_A_L2}, \ref{fig:local_B_L2} and \ref{fig:local_C_L2}. We observe generally good agreement between the estimated model and observed quantiles. Note that the step-like behaviour observed in some of the plotted quantiles occurs due to repeated observations in the time series'; these are likely a result of rounding errors in measurement equipment. 

\bibliography{LaTeX/library,LaTeX/additional}

\begin{thebibliography}{}

\bibitem[Balkema and de~Haan, 1974]{Balkema1974}
Balkema, A.~A. and de~Haan, L. (1974).
\newblock Residual Life Time at Great Age.
\newblock {\em The Annals of Probability}, 2:792--804.

\bibitem[Beirlant et~al., 2004]{Beirlant2004}
Beirlant, J., Goegebeur, Y., Teugels, J., and Segers, J. (2004).
\newblock {\em Statistics of Extremes}.
\newblock Wiley.

\bibitem[Castro-Camilo et~al., 2018]{Castro-Camilo2018}
Castro-Camilo, D., de~Carvalho, M., and Wadsworth, J. (2018).
\newblock Time-varying extreme value dependence with application to leading European stock markets.
\newblock {\em Annals of Applied Statistics}, 12:283--309.

\bibitem[Chan and Li, 2008]{chan2008tail}
Chan, Y. and Li, H. (2008).
\newblock Tail dependence for multivariate t-copulas and its monotonicity.
\newblock {\em Insurance: Mathematics and Economics}, 42(2):763--770.

\bibitem[Chaubey, 2022]{Chaubey2022}
Chaubey, Y.~P. (2022).
\newblock {\em Directional Statistics for Innovative Applications}, pages 351--378.
\newblock Springer.

\bibitem[Chavez-Demoulin and Davison, 2005]{Chavez-Demoulin2005}
Chavez-Demoulin, V. and Davison, A.~C. (2005).
\newblock Generalized additive modelling of sample extremes.
\newblock {\em Journal of the Royal Statistical Society: Series C (Applied Statistics)}, 54:207--222.

\bibitem[Chen, 2017]{Chen2017}
Chen, Y.-C. (2017).
\newblock A tutorial on kernel density estimation and recent advances.
\newblock {\em Biostatistics \& Epidemiology}, 1:161--187.

\bibitem[Davis et~al., 1988]{Davis1988}
Davis, R.~A., Mulrow, E., and Resnick, S.~I. (1988).
\newblock {Almost sure limit sets of random samples in $\mathbb{R}^d$}.
\newblock {\em Advances in Applied Probability}, 20(3):573–599.

\bibitem[Davison and Smith, 1990]{Davison1990}
Davison, A.~C. and Smith, R.~L. (1990).
\newblock Models for Exceedances Over High Thresholds.
\newblock {\em Journal of the Royal Statistical Society. Series B: Statistical Methodology}, 52:393--425.

\bibitem[de~Haan and de~Ronde, 1998]{DeHaan1998}
de~Haan, L. and de~Ronde, J. (1998).
\newblock Sea and Wind: Multivariate Extremes at Work.
\newblock {\em Extremes}, 1:7--45.

\bibitem[de~Haan and Ferreira, 2006]{haan2006extreme}
de~Haan, L. and Ferreira, A. (2006).
\newblock {\em Extreme value theory: an introduction}, volume~3.
\newblock Springer.

\bibitem[de~Haan and Resnick, 1987]{haan1987}
de~Haan, L. and Resnick, S.~I. (1987).
\newblock On regular variation of probability densities.
\newblock {\em Stochastic processes and their applications}, 25:83--93.

\bibitem[Diaconu et~al., 2021]{w13040474}
Diaconu, D.~C., Costache, R., and Popa, M.~C. (2021).
\newblock An Overview of Flood Risk Analysis Methods.
\newblock {\em Water}, 13(4).

\bibitem[Fisher, 1969]{Fisher1969}
Fisher, L. (1969).
\newblock Limiting Sets and Convex Hulls of Samples from Product Measures.
\newblock {\em The Annals of Mathematical Statistics}, 40:1824--1832.

\bibitem[García–Portugués, 2013]{García–Portugués2013}
García–Portugués, E. (2013).
\newblock Exact risk improvement of bandwidth selectors for kernel density estimation with directional data.
\newblock {\em Electronic Journal of Statistics}, 7:1655--1685.

\bibitem[Geraci and Bottai, 2007]{Geraci2007}
Geraci, M. and Bottai, M. (2007).
\newblock Quantile regression for longitudinal data using the asymmetric Laplace distribution.
\newblock {\em Biostatistics}, 8:140--154.

\bibitem[Gu, 1993]{Gu1993}
Gu, C. (1993).
\newblock Smoothing Spline Density Estimation: A Dimensionless Automatic Algorithm.
\newblock {\em Journal of the American Statistical Association}, 88:495--504.

\bibitem[Hall et~al., 1987]{Hall1987}
Hall, P., Watson, G.~S., and Cabrera, J. (1987).
\newblock Kernel Density Estimation with Spherical Data.
\newblock {\em Biometrika}, 74:751.

\bibitem[Haselsteiner et~al., 2021]{Haselsteiner2021}
Haselsteiner, A.~F., Coe, R.~G., Manuel, L., Chai, W., Leira, B., Clarindo, G., Soares, C.~G., Ásta Hannesdóttir, Dimitrov, N., Sander, A., Ohlendorf, J.~H., Thoben, K.~D., de~Hauteclocque, G., Mackay, E., Jonathan, P., Qiao, C., Myers, A., Rode, A., Hildebrandt, A., Schmidt, B., Vanem, E., and Huseby, A.~B. (2021).
\newblock A benchmarking exercise for environmental contours.
\newblock {\em Ocean Engineering}, 236:1--29.

\bibitem[Hastie et~al., 2009]{hastie2009elements}
Hastie, T., Tibshirani, R., Friedman, J.~H., and Friedman, J.~H. (2009).
\newblock {\em The elements of statistical learning: data mining, inference, and prediction}, volume~2.
\newblock Springer.

\bibitem[Hosking and Wallis, 1987]{hosking1987}
Hosking, J.~R. and Wallis, J.~R. (1987).
\newblock Parameter and quantile estimation for the generalized {Pareto} distribution.
\newblock {\em Technometrics}, 29:339--349.

\bibitem[Hua and Joe, 2011]{Hua2011}
Hua, L. and Joe, H. (2011).
\newblock {Tail order and intermediate tail dependence of multivariate copulas}.
\newblock {\em Journal of Multivariate Analysis}, 102(10):1454--1471.

\bibitem[Joe, 1997]{Joe1997}
Joe, H. (1997).
\newblock {\em Multivariate Models and Multivariate Dependence Concepts}.
\newblock Chapman and Hall/CRC.

\bibitem[Jonathan and Ewans, 2013]{Jonathan2013}
Jonathan, P. and Ewans, K. (2013).
\newblock Statistical modelling of extreme ocean environments for marine design: A review.
\newblock {\em Ocean Engineering}, 62:91--109.

\bibitem[Jones et~al., 2016]{JnsEA15}
Jones, M., Randell, D., Ewans, K., and Jonathan, P. (2016).
\newblock Statistics of extreme ocean environments: non-stationary inference for directionality and other covariate effects.
\newblock {\em Ocean Engineering}, 119:30--46.

\bibitem[Kauermann and Opsomer, 2011]{Kauermann2011}
Kauermann, G. and Opsomer, J.~D. (2011).
\newblock Data-driven selection of the spline dimension in penalized spline regression.
\newblock {\em Biometrika}, 98:225--230.

\bibitem[Keef et~al., 2013]{Keef2013a}
Keef, C., Tawn, J.~A., and Lamb, R. (2013).
\newblock Estimating the probability of widespread flood events.
\newblock {\em Environmetrics}, 24:13--21.

\bibitem[Koenker et~al., 2017]{Koenker2017}
Koenker, R., Chernozhukov, V., He, X., and Peng, L. (2017).
\newblock {\em Handbook of Quantile Regression}.
\newblock Chapman and Hall/CRC.

\bibitem[Kunsch, 1989]{Kunsch1989}
Kunsch, H.~R. (1989).
\newblock The Jackknife and the Bootstrap for General Stationary Observations.
\newblock {\em The Annals of Statistics}, 17:1217--1241.

\bibitem[Ledford and Tawn, 1996]{Ledford1996}
Ledford, A.~W. and Tawn, J.~A. (1996).
\newblock Statistics for near independence in multivariate extreme values.
\newblock {\em Biometrika}, 83:169--187.

\bibitem[Ledford and Tawn, 1997]{Ledford1997}
Ledford, A.~W. and Tawn, J.~A. (1997).
\newblock Modelling dependence within joint tail regions.
\newblock {\em Journal of the Royal Statistical Society. Series B: Statistical Methodology}, 59:475--499.

\bibitem[Mackay, 2022]{mackay2022imex}
Mackay, E. (2022).
\newblock Improved Models for Multivariate Metocean Extremes.
\newblock Technical report, Supergen ORE Hub.

\bibitem[Mackay and de~Hauteclocque, 2023]{mackay2023-DIFORM}
Mackay, E. and de~Hauteclocque, G. (2023).
\newblock Model-free environmental contours in higher dimensions.
\newblock {\em Ocean Engineering}, 273:113959.

\bibitem[Mackay et~al., 2021]{mackay2021correlation}
Mackay, E., de~Hauteclocque, G., Vanem, E., and Jonathan, P. (2021).
\newblock {The effect of serial correlation in environmental conditions on estimates of extreme events}.
\newblock {\em Ocean Engineering}, 242:110092.

\bibitem[Mackay and Haselsteiner, 2021]{Mackay2021}
Mackay, E. and Haselsteiner, A.~F. (2021).
\newblock Marginal and total exceedance probabilities of environmental contours.
\newblock {\em Marine Structures}, 75:1--24.

\bibitem[Mackay and Jonathan, 2023]{Mackay2023}
Mackay, E. and Jonathan, P. (2023).
\newblock Modelling multivariate extremes through angular-radial decomposition of the density function.
\newblock {\em arXiv}, 2310.12711.

\bibitem[Mackay et~al., 2024]{mackay2024}
Mackay, E., {Murphy-Barltrop}, C., and Jonathan, P. (2024).
\newblock {The SPAR model: a new paradigm for multivariate extremes. Application to joint distributions of metocean variables}.
\newblock In {\em 43rd International Conference on Ocean, Offshore \& Arctic Engineering}, page OMAE2024:130932, Singapore.

\bibitem[Majumder et~al., 2023]{Majumder2023}
Majumder, R., Shaby, B.~A., Reich, B.~J., and Cooley, D. (2023).
\newblock Semiparametric Estimation of the Shape of the Limiting Bivariate Point Cloud.
\newblock {\em arXiv}, 2306.13257.

\bibitem[Marzio et~al., 2011]{DiMarzia2011}
Marzio, M.~D., Panzera, A., and Taylor, C.~C. (2011).
\newblock Kernel density estimation on the torus.
\newblock {\em Journal of Statistical Planning and Inference}, 141:2156--2173.

\bibitem[Murphy et~al., 2023]{Murphy2023}
Murphy, C., Tawn, J.~A., and Varty, Z. (2023).
\newblock Automated threshold selection and associated inference uncertainty for univariate extremes.
\newblock {\em arXiv}, 2310.17999.

\bibitem[Murphy-Barltrop et~al., 2023]{Murphy-Barltrop2023}
Murphy-Barltrop, C. J.~R., Wadsworth, J.~L., and Eastoe, E.~F. (2023).
\newblock New estimation methods for extremal bivariate return curves.
\newblock {\em Environmetrics}, e2797:1--22.

\bibitem[Nolde, 2014]{Nolde2014}
Nolde, N. (2014).
\newblock Geometric interpretation of the residual dependence coefficient.
\newblock {\em Journal of Multivariate Analysis}, 123:85--95.

\bibitem[Nolde and Wadsworth, 2022]{Nolde2022}
Nolde, N. and Wadsworth, J.~L. (2022).
\newblock Linking representations for multivariate extremes via a limit set.
\newblock {\em Advances in Applied Probability}, 54:688--717.

\bibitem[Northrop and Jonathan, 2011]{Northrop2011}
Northrop, P.~J. and Jonathan, P. (2011).
\newblock Threshold modelling of spatially dependent non-stationary extremes with application to hurricane-induced wave heights.
\newblock {\em Environmetrics}, 22:799--809.

\bibitem[Oh et~al., 2011]{Oh2011}
Oh, H.-S., Lee, T. C.~M., and Nychka, D.~W. (2011).
\newblock Fast Nonparametric Quantile Regression With Arbitrary Smoothing Methods.
\newblock {\em Journal of Computational and Graphical Statistics}, 20:510--526.

\bibitem[Oliveira et~al., 2012]{Oliveira2012}
Oliveira, M., Crujeiras, R., and Rodríguez-Casal, A. (2012).
\newblock A plug-in rule for bandwidth selection in circular density estimation.
\newblock {\em Computational Statistics \& Data Analysis}, 56:3898--3908.

\bibitem[Papastathopoulos et~al., 2024]{Papastathopoulos2024}
Papastathopoulos, I., de~Monte, L., Campbell, R., and Rue, H. (2024).
\newblock Statistical inference for radially-stable generalized Pareto distributions and return level-sets in geometric extremes.
\newblock {\em arXiv}, 2310.06130.

\bibitem[Perperoglou et~al., 2019]{Perperoglou2019}
Perperoglou, A., Sauerbrei, W., Abrahamowicz, M., and Schmid, M. (2019).
\newblock A review of spline function procedures in R.
\newblock {\em BMC Medical Research Methodology}, 19:1--16.

\bibitem[Politis and Romano, 1994]{Politis1994}
Politis, D.~N. and Romano, J.~P. (1994).
\newblock The Stationary Bootstrap.
\newblock {\em Journal of the American Statistical Association}, 89:1303--1313.

\bibitem[Ramos and Ledford, 2009]{Ramos2009}
Ramos, A. and Ledford, A. (2009).
\newblock A new class of models for bivariate joint tails.
\newblock {\em Journal of the Royal Statistical Society. Series B: Statistical Methodology}, 71:219--241.

\bibitem[Randell et~al., 2016]{Randell2016}
Randell, D., Turnbull, K., Ewans, K., and Jonathan, P. (2016).
\newblock Bayesian inference for nonstationary marginal extremes.
\newblock {\em Environmetrics}, 27:439--450.

\bibitem[Resnick, 2002]{Resnick2002}
Resnick, S. (2002).
\newblock Hidden Regular Variation, Second Order Regular Variation and Asymptotic Independence.
\newblock {\em Extremes}, 5:303--336.

\bibitem[Resnick, 1987]{Resnick1987}
Resnick, S.~I. (1987).
\newblock {\em Extreme Values, Regular Variation and Point Processes}.
\newblock Springer New York.

\bibitem[Resnick, 2007]{Resnick2007}
Resnick, S.~I. (2007).
\newblock {\em {Heavy-Tail Phenomena: Probabilistic and Statistical Modeling}}.
\newblock Springer.

\bibitem[Simpson and Tawn, 2022]{Simpson2022}
Simpson, E.~S. and Tawn, J.~A. (2022).
\newblock Estimating the limiting shape of bivariate scaled sample clouds: with additional benefits of self-consistent inference for existing extremal dependence properties.
\newblock {\em arXiv}, 2207.02626.

\bibitem[Sklar, 1959]{Sklar1959}
Sklar, A. (1959).
\newblock Fonctions de repartition a n dimensions et leurs marges.
\newblock {\em Publ. Inst. Statist. Univ. Paris}, 8:229--231.

\bibitem[Southworth et~al., 2024]{texmex}
Southworth, H., Heffernan, J.~E., and Metcalfe, P.~D. (2024).
\newblock {texmex: statistical modelling of extreme values}.
\newblock \url{https://cran.r-project.org/package=texmex}.

\bibitem[Taylor, 2008]{Taylor2008}
Taylor, C.~C. (2008).
\newblock Automatic bandwidth selection for circular density estimation.
\newblock {\em Computational Statistics and Data Analysis}, 52:3493--3500.

\bibitem[Tendijck et~al., 2024]{TndEA24}
Tendijck, S., Randell, D., Feld, G., and Jonathan, P. (2024).
\newblock Uncertainties in return values from non-stationary extreme value analysis of peaks over threshold using the generalised Pareto distribution.
\newblock {\em Submitted to Ocean Engineering; pre-print at www.lancaster.ac.uk/~jonathan}.

\bibitem[Towe et~al., 2023]{Towe2023}
Towe, R., Randell, D., Kensler, J., Feld, G., and Jonathan, P. (2023).
\newblock Estimation of associated values from conditional extreme value models.
\newblock {\em Ocean Engineering}, 272:113808.

\bibitem[Towe et~al., 2024]{TowEA24}
Towe, R., Ross, E., Randell, D., and Jonathan, P. (2024).
\newblock {covXtreme: MATLAB software for non-stationary penalised piecewise constant marginal and conditional extreme value models}.
\newblock {\em Environmental Modelling \& Software}, 177:106035.

\bibitem[Vanem et~al., 2022]{Vanem2022}
Vanem, E., Zhu, T., and Babanin, A. (2022).
\newblock Statistical modelling of the ocean environment – A review of recent developments in theory and applications.
\newblock {\em Marine Structures}, 86.

\bibitem[Wadsworth and Campbell, 2024]{Wadsworth2024}
Wadsworth, J.~L. and Campbell, R. (2024).
\newblock Statistical inference for multivariate extremes via a geometric approach.
\newblock {\em Journal of the Royal Statistical Society Series B: Statistical Methodology}, 2208.14951.

\bibitem[Wadsworth and Tawn, 2013]{Wadsworth2013}
Wadsworth, J.~L. and Tawn, J.~A. (2013).
\newblock A new representation for multivariate tail probabilities.
\newblock {\em Bernoulli}, 19:2689--2714.

\bibitem[Wadsworth et~al., 2017]{Wadsworth2017}
Wadsworth, J.~L., Tawn, J.~A., Davison, A.~C., and Elton, D.~M. (2017).
\newblock Modelling across extremal dependence classes.
\newblock {\em Journal of the Royal Statistical Society. Series B: Statistical Methodology}, 79:149--175.

\bibitem[Wand, 2000]{Wand2000}
Wand, M.~P. (2000).
\newblock A Comparison of Regression Spline Smoothing Procedures.
\newblock {\em Computational Statistics}, 15:443--462.

\bibitem[Wood, 2003]{Wood2003}
Wood, S.~N. (2003).
\newblock Thin Plate Regression Splines.
\newblock {\em Journal of the Royal Statistical Society Series B: Statistical Methodology}, 65:95--114.

\bibitem[Wood, 2017]{Wood2017}
Wood, S.~N. (2017).
\newblock {\em Generalized Additive Models}.
\newblock Chapman and Hall/CRC.

\bibitem[Wood et~al., 2016]{Wood2016}
Wood, S.~N., Pya, N., and Säfken, B. (2016).
\newblock Smoothing Parameter and Model Selection for General Smooth Models.
\newblock {\em Journal of the American Statistical Association}, 111:1548--1563.

\bibitem[Youngman, 2020]{Youngman2020}
Youngman, B. (2020).
\newblock evgam: Generalised Additive Extreme Value Models.
\newblock {\em R Package}.

\bibitem[Youngman, 2019]{Youngman2019}
Youngman, B.~D. (2019).
\newblock Generalized Additive Models for Exceedances of High Thresholds With an Application to Return Level Estimation for U.S. Wind Gusts.
\newblock {\em Journal of the American Statistical Association}, 114:1865--1879.

\bibitem[Yu and Moyeed, 2001]{Yu2001}
Yu, K. and Moyeed, R.~A. (2001).
\newblock Bayesian quantile regression.
\newblock {\em Statistics \& Probability Letters}, 54:437--447.

\bibitem[Zanini et~al., 2020]{Zanini2020}
Zanini, E., Eastoe, E., Jones, M.~J., Randell, D., and Jonathan, P. (2020).
\newblock Flexible covariate representations for extremes.
\newblock {\em Environmetrics}, 31:1--28.

\end{thebibliography}

\end{document}